\documentclass[superscriptaddress,showpacs,twocolumn,prb]{revtex4-1}

\usepackage{hyperref}
\usepackage{graphicx}
\usepackage{color}
\usepackage{amsbsy}
\usepackage{pstricks}
\usepackage{multirow}

\newif\ifGPcolor
 \GPcolortrue
\newif\ifGPblacktext
 \GPblacktextfalse

\begin{document}

\title{$g$-factors and diamagnetic coefficients of electrons, holes and excitons in InAs/InP quantum dots}

\date{\today}

\author{J. van Bree}
\email{j.v.bree@tue.nl}
\author{A. Yu. Silov}
\author{P. M. Koenraad}
\affiliation{PSN, COBRA, University of Technology Eindhoven, 5600 MB Eindhoven, The Netherlands.}
\author{M. E. Flatt{\'e}}
\author{C. E. Pryor}
\affiliation{Department of Physics and Astronomy and Optical Science and Technology Center, University of Iowa, Iowa City, Iowa 52242, USA}

\begin{abstract}
The electron, hole, and exciton $g$-factors and diamagnetic coefficients have been calculated using envelope-function theory for cylindrical InAs/InP quantum dots in the presence of a magnetic field parallel to the dot symmetry axis. A clear connection is established between the electron $g$-factor and the amplitude of the those valence-state envelope functions which possess non-zero orbital momentum associated with the envelope function. The  dependence of the exciton 
diamagnetic coefficients on the quantum dot height is found to correlate with the energy dependence of the effective mass. Calculated exciton $g$-factor and diamagnetic coefficients, constructed from the values associated with the electron and hole constituents of the exciton, match experimental data well, however including the Coulomb interaction between the electron and hole states improves the agreement. Remote-band contributions to the valence-band electronic structure, included perturbatively, reduce the agreement between theory and experiment.  
\end{abstract}

\maketitle

\section{Introduction}\label{sec:intro}
The magnetic moment and Zeeman energy splittings of an electronic spin are controlled by the $g$-tensor, whose value differs from the bare electron's $g$-factor of $\sim 2$ due to the spin-orbit interaction in a solid\cite{VanVleck1932}. Thus an understanding of the $g$-tensor not only provides insight into the spin-orbit interaction in a solid,
but it also can yield insight into the effective orbital motion (and resulting angular momentum) associated with a quantum state\cite{VanVleck1932}.  The spin-orbit interaction affects the magnetic moment of an electronic state by permitting a contribution from the orbital angular momentum in addition to the spin momentum (leading to an effective Land{\'e} $g$-factor in atomic systems). In solids, in addition to the atomistic (Bloch) orbital angular momentum, the envelope wave function also contributes to the magnetic moment via the envelope orbital angular momentum. 

The $g$-factors of bulk semiconductors, which often differ substantially from the electron's bare $g$-factor, cannot be explained solely by the large orbital angular momenta of Bloch states that contribute to the orbital motion of the state within a crystal lattice unit cell. Instead the spatial extension of the wave function across many unit cells leads to the large $g$-factors through interatomic circulating currents originating from the spin-orbit potential\cite{Yafet1963}. As a wave packet in a semiconductor crystal can have a substantial extent due to the small effective mass $m^*$, the envelope wave function can sustain large extended circulating currents leading to a large envelope orbital angular momentum. In the absence of spin-orbit coupling these currents vanish and do not contribute to the modification of the $g$-factor at all. However, in most semiconductors the spin-orbit coupling is substantial and will lift the spin degeneracy, leading to a finite spin-dependent envelope orbital angular momentum. Therefore the $g$-factor is modified from the Land{\'e}-factor by means of the additional envelope orbital angular momentum generated by circulating currents in the envelope wave function.

These circulating currents are reminiscent of the origin of diamagnetism: the response of electrons to an applied magnetic field is to create circulating currents, leading to a magnetization opposed to the applied field. The underlying origin of the diamagnetic circulating currents is  different than those associated with  the $g$-tensor, which come from the spatially-periodic spin-orbit potential. However, for both phenomena the generation of orbital angular momentum by extended circulating currents is essential to understand the effect and explain the experimental results. It was Ehrenfest who initially identified the role of extended circulating currents in producing large differences in the diamagnetic susceptibilities of various materials\cite{Ehrenfest1925,Ehrenfest1929}.

In quantum dots, confinement alters the envelope orbital momentum of the envelope wave function by preventing it from extending over as many lattice unit cells as it would in the bulk. In the limit of infinite confinement potentials and vanishing dot size, the envelope orbital momentum is quenched and the $g$-factor is solely determined by the atomistic (Bloch) Land{\'e}-factor. This idea has been successfully applied to paramagnetic impurities in salts\cite{VanVleck1932}. The effect of envelope orbital momentum quenching was demonstrated in Ref.~\onlinecite{Pryor2006b} for a spherical, unstrained InAs nanocrystal, by showing that the electron $g$-factor approached the bare electron value of $\sim 2$ much more quickly than would be expected from a parametrized bulk model based on Ref.~\onlinecite{Roth1959}.  A similar quenching effect on the orbital angular momentum should reduce the diamagnetic coefficient in small quantum dots with large  barriers. For both quantities it is crucial to explore how confinement affects the envelope orbital angular momentum. An understanding of the nature of the $g$-tensor's dependence on dot composition and structure  can  thus clarify the role of orbital angular momentum quenching in determining this property in a quantum dot\cite{Pryor2006b}. Diamagnetic coefficients provide a picture of the orbital motion associated with a quantum state through a different path than the $g$-tensor; a theory which correctly describes both has passed a stringent test of its picture of the orbital motion of the electronic state. 


In addition to these fundamental concerns, a single spin in a quantum dot provides a model system for the observation and manipulation of an individual quantum system\cite{Awschalom2002,Hanson2007}, and has been proposed as a qubit for use in quantum computation\cite{Loss1998}. 
$g$-tensors (but not diamagnetic coefficients) figure prominently in spin-based proposals for quantum information processing as they provide an effective spin-manipulation method, and hence qubit-operation mechanism. For example, proposals for single-spin control via an electric field include electric field modification of the $g$-tensor to bring the spin into and out of resonance with a global RF field\cite{Loss1998,Kane1998,Xiao2004}, or to tilt the spin's precessional axis\cite{Kato2003,Pingenot2008,De2009,Andlauer2009,Pingenot2011}. They also provide energy-level structures favorable for optical pumping of spin polarization of an electron\cite{Chen2004,Economou2006,Dutt2006}, or a hole\cite{Gerardot2008,Brunner2009} in a quantum dot.  Due to energy-conservation constraints the $g$-tensor also can control the  spin lifetime ($T_1$) for a dot\cite{Khaetskii2001,Kroutvar2004}. 

\begin{figure}
\begin{center}
\includegraphics[width=0.75\columnwidth]{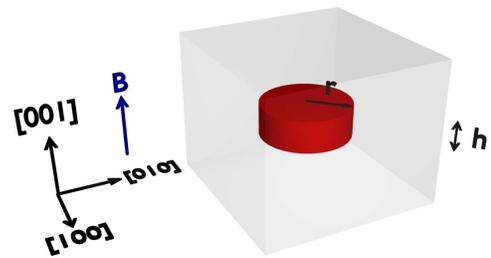}
\caption{(Color online) The quantum dots are made out of pure InAs (red) embedded in InP (grey), and have a cylindrical shape with a radius $r$ and height $h$.\label{fig:QD}}
\end{center}
\end{figure}

Compositionally-defined  quantum dots\cite{Bimberg1998} permit much larger barrier heights than gate-defined dots, as well as efficient optical access with simultaneous control of charging\cite{Heiss2008}. The larger confinement energies in such dots for both electrons and holes from these barriers may in principle permit single-spin behavior at higher operating temperature than in lithographic dots\cite{Hanson2007}, however the intrinsic strain and dot shape asymmetry provide a challenge to theory. Shown  in Fig.~\ref{fig:QD} is a schematic dot, with height $h$ and radius $r$, which can be formed through metal-organic vapor-phase epitaxy (MOVPE) of InAs on an InP substrate. Details of the growth of these roughly cylindrical InAs dots can be found in Ref.~\onlinecite{Anantathanasarn2005}.  Recent experimental work\cite{Kleemans2009} has investigated the magneto-optical properties of such dots by probing the component of the exciton $g$-tensor along the $[001]$-direction (this component is from now on referred as the $g$-factor). A strong dependence of the exciton $g$-factor ($g_{ex}$) on the dot height was shown, whereas the same work found the exciton diamagnetic coefficient ($\alpha_{ex}$) was mostly independent of the height. The quantum dot radii had little influence on either $g_{ex}$ or $\alpha_{ex}$.  Theoretical calculations are needed to provide insight into the details of the observed dependences of $g_{ex}$ and $\alpha_{ex}$ and to provide clarity about the ability to tune the $g_{ex}$-factor through control of the size and shape of the quantum dot. 

Here we describe an extensive series of multi-band envelope-function calculations of electron, hole, and exciton $g$-factors and diamagnetic coefficients based on an eight-band strain-dependent ${\bf k}\cdot{\bf p}$ model of the bulk semiconductor constituents for a series of quantum dots similar to the ones reported in Ref.~\onlinecite{Kleemans2009}. The calculational method is detailed in Sec.~\ref{sec:methods}. The calculated $g$-factors and diamagnetic coefficients of the electron and hole are discussed separately in sections \ref{sec:g-factor} and \ref{sec:alpha}. 
The exciton $g_{ex}$-factor and $\alpha_{ex}$, which are constructed from the values for the electron and hole, are compared to experimental data in section \ref{sec:comparison}. The Appendices provide a detailed comparative analysis with our results of other schemes to calculate the $g$-factor and diamagnetic coefficient. 

\section{Methods} \label{sec:methods}

The $g$-factors and diamagnetic coefficients of quantum dots are calculated using eight-band strain-dependent ${\bf k}~\cdot~{\bf p}$-theory on a real space grid \cite{Pryor1998}. The strain is calculated using linear elasticity continuum theory. All calculations are performed at $T=0$~K and material parameters are taken from Ref. \onlinecite{Vurgaftman2001}. The magnetic field is coupled both to the envelope wave function and the spin of the Bloch functions\cite{Pryor2006b}, where the latter is taken into account by the Zeeman Hamiltonian:
\begin{eqnarray} 
{\cal H}_{\text{Zeeman}} = \frac{1}{2} \mu_B {\bf B} \cdot \left(
\begin{array}{ccc}
2 \boldsymbol{\sigma} & 0 & 0 \\
0 & \frac{4}{3} {\bf J} & 0 \\
0 & 0 & \frac{2}{3} \boldsymbol{\sigma}
\end{array}
\right),
\end{eqnarray}
where $\mu_B$ is the Bohr magneton and $\boldsymbol{\sigma}$ and ${\bf J}$ are the spin matrices of spins $\frac{1}{2}$ and $\frac{3}{2}$ respectively. The $g$-factors for the conduction, valence and spin-orbit bands are respectively $2$, $\frac{4}{3}$ and $\frac{2}{3}$ ({\it i.e.} the Land{\'e}-factors). Note that we have explicitly left out the remote band contributions. This approximation is validated in appendix \ref{app:remote}.

\begin{figure*}
\begin{center}

\begingroup
  \makeatletter
  \providecommand\color[2][]{%
    \GenericError{(gnuplot) \space\space\space\@spaces}{%
      Package color not loaded in conjunction with
      terminal option `colourtext'%
    }{See the gnuplot documentation for explanation.%
    }{Either use 'blacktext' in gnuplot or load the package
      color.sty in LaTeX.}%
    \renewcommand\color[2][]{}%
  }%
  \providecommand\includegraphics[2][]{%
    \GenericError{(gnuplot) \space\space\space\@spaces}{%
      Package graphicx or graphics not loaded%
    }{See the gnuplot documentation for explanation.%
    }{The gnuplot epslatex terminal needs graphicx.sty or graphics.sty.}%
    \renewcommand\includegraphics[2][]{}%
  }%
  \providecommand\rotatebox[2]{#2}%
  \@ifundefined{ifGPcolor}{%
    \newif\ifGPcolor
    \GPcolortrue
  }{}%
  \@ifundefined{ifGPblacktext}{%
    \newif\ifGPblacktext
    \GPblacktexttrue
  }{}%
  \let\gplgaddtomacro\g@addto@macro
  \gdef\gplbacktext{}%
  \gdef\gplfronttext{}%
  \makeatother
  \ifGPblacktext
    \def\colorrgb#1{}%
    \def\colorgray#1{}%
  \else
    \ifGPcolor
      \def\colorrgb#1{\color[rgb]{#1}}%
      \def\colorgray#1{\color[gray]{#1}}%
      \expandafter\def\csname LTw\endcsname{\color{white}}%
      \expandafter\def\csname LTb\endcsname{\color{black}}%
      \expandafter\def\csname LTa\endcsname{\color{black}}%
      \expandafter\def\csname LT0\endcsname{\color[rgb]{1,0,0}}%
      \expandafter\def\csname LT1\endcsname{\color[rgb]{0,1,0}}%
      \expandafter\def\csname LT2\endcsname{\color[rgb]{0,0,1}}%
      \expandafter\def\csname LT3\endcsname{\color[rgb]{1,0,1}}%
      \expandafter\def\csname LT4\endcsname{\color[rgb]{0,1,1}}%
      \expandafter\def\csname LT5\endcsname{\color[rgb]{1,1,0}}%
      \expandafter\def\csname LT6\endcsname{\color[rgb]{0,0,0}}%
      \expandafter\def\csname LT7\endcsname{\color[rgb]{1,0.3,0}}%
      \expandafter\def\csname LT8\endcsname{\color[rgb]{0.5,0.5,0.5}}%
    \else
      \def\colorrgb#1{\color{black}}%
      \def\colorgray#1{\color[gray]{#1}}%
      \expandafter\def\csname LTw\endcsname{\color{white}}%
      \expandafter\def\csname LTb\endcsname{\color{black}}%
      \expandafter\def\csname LTa\endcsname{\color{black}}%
      \expandafter\def\csname LT0\endcsname{\color{black}}%
      \expandafter\def\csname LT1\endcsname{\color{black}}%
      \expandafter\def\csname LT2\endcsname{\color{black}}%
      \expandafter\def\csname LT3\endcsname{\color{black}}%
      \expandafter\def\csname LT4\endcsname{\color{black}}%
      \expandafter\def\csname LT5\endcsname{\color{black}}%
      \expandafter\def\csname LT6\endcsname{\color{black}}%
      \expandafter\def\csname LT7\endcsname{\color{black}}%
      \expandafter\def\csname LT8\endcsname{\color{black}}%
    \fi
  \fi
  \setlength{\unitlength}{0.0500bp}%
  \begin{picture}(10368.00,7761.60)%
    \gplgaddtomacro\gplbacktext{%
      \csname LTb\endcsname%
      \put(688,4392){\makebox(0,0)[r]{\strut{}-10}}%
      \put(688,4911){\makebox(0,0)[r]{\strut{}-8}}%
      \put(688,5431){\makebox(0,0)[r]{\strut{}-6}}%
      \put(688,5950){\makebox(0,0)[r]{\strut{}-4}}%
      \put(688,6469){\makebox(0,0)[r]{\strut{}-2}}%
      \put(688,6989){\makebox(0,0)[r]{\strut{} 0}}%
      \put(688,7508){\makebox(0,0)[r]{\strut{} 2}}%
      \put(784,4232){\makebox(0,0){\strut{} 0}}%
      \put(1788,4232){\makebox(0,0){\strut{} 5}}%
      \put(2792,4232){\makebox(0,0){\strut{} 10}}%
      \put(3795,4232){\makebox(0,0){\strut{} 15}}%
      \put(4799,4232){\makebox(0,0){\strut{} 20}}%
      \csname LTb\endcsname%
      \put(224,5950){\rotatebox{-270}{\makebox(0,0){\strut{}g-factor $g_c$}}}%
      \put(2791,3992){\makebox(0,0){\strut{}Height [nm]}}%
    }%
    \gplgaddtomacro\gplfronttext{%
      \csname LTb\endcsname%
      \put(4498,7228){\makebox(0,0)[r]{\strut{}Electron - unstrained (A)}}%
      \put(4217,5483){\makebox(0,0){\strut{}Radius}}%
      \csname LT0\endcsname%
      \put(4458,5296){\makebox(0,0)[r]{\strut{}7 nm}}%
      \csname LT1\endcsname%
      \put(4458,5140){\makebox(0,0)[r]{\strut{}9 nm}}%
      \csname LT2\endcsname%
      \put(4458,4984){\makebox(0,0)[r]{\strut{}11 nm}}%
      \csname LT3\endcsname%
      \put(4458,4828){\makebox(0,0)[r]{\strut{}13 nm}}%
      \csname LT4\endcsname%
      \put(4458,4672){\makebox(0,0)[r]{\strut{}15 nm}}%
    }%
    \gplgaddtomacro\gplbacktext{%
      \csname LTb\endcsname%
      \put(688,512){\makebox(0,0)[r]{\strut{}-10}}%
      \put(688,1031){\makebox(0,0)[r]{\strut{}-8}}%
      \put(688,1551){\makebox(0,0)[r]{\strut{}-6}}%
      \put(688,2070){\makebox(0,0)[r]{\strut{}-4}}%
      \put(688,2589){\makebox(0,0)[r]{\strut{}-2}}%
      \put(688,3109){\makebox(0,0)[r]{\strut{} 0}}%
      \put(688,3628){\makebox(0,0)[r]{\strut{} 2}}%
      \put(784,352){\makebox(0,0){\strut{} 0}}%
      \put(1230,352){\makebox(0,0){\strut{} 0.01}}%
      \put(1676,352){\makebox(0,0){\strut{} 0.02}}%
      \put(2122,352){\makebox(0,0){\strut{} 0.03}}%
      \put(2568,352){\makebox(0,0){\strut{} 0.04}}%
      \put(3015,352){\makebox(0,0){\strut{} 0.05}}%
      \put(3461,352){\makebox(0,0){\strut{} 0.06}}%
      \put(3907,352){\makebox(0,0){\strut{} 0.07}}%
      \put(4353,352){\makebox(0,0){\strut{} 0.08}}%
      \put(4799,352){\makebox(0,0){\strut{} 0.09}}%
      \csname LTb\endcsname%
      \put(224,2070){\rotatebox{-270}{\makebox(0,0){\strut{}g-factor $g_c$}}}%
      \put(2791,112){\makebox(0,0){\strut{}Sum of all components having $L^{\text{env}}_z = \pm1$}}%
    }%
    \gplgaddtomacro\gplfronttext{%
      \csname LTb\endcsname%
      \put(4498,3348){\makebox(0,0)[r]{\strut{}Electron - unstrained (B)}}%
      \put(4217,1603){\makebox(0,0){\strut{}Radius}}%
      \csname LT0\endcsname%
      \put(4458,1416){\makebox(0,0)[r]{\strut{}7 nm}}%
      \csname LT1\endcsname%
      \put(4458,1260){\makebox(0,0)[r]{\strut{}9 nm}}%
      \csname LT2\endcsname%
      \put(4458,1104){\makebox(0,0)[r]{\strut{}11 nm}}%
      \csname LT3\endcsname%
      \put(4458,948){\makebox(0,0)[r]{\strut{}13 nm}}%
      \csname LT4\endcsname%
      \put(4458,792){\makebox(0,0)[r]{\strut{}15 nm}}%
    }%
    \gplgaddtomacro\gplbacktext{%
      \csname LTb\endcsname%
      \put(5872,4392){\makebox(0,0)[r]{\strut{}-10}}%
      \put(5872,4911){\makebox(0,0)[r]{\strut{}-8}}%
      \put(5872,5431){\makebox(0,0)[r]{\strut{}-6}}%
      \put(5872,5950){\makebox(0,0)[r]{\strut{}-4}}%
      \put(5872,6469){\makebox(0,0)[r]{\strut{}-2}}%
      \put(5872,6989){\makebox(0,0)[r]{\strut{} 0}}%
      \put(5872,7508){\makebox(0,0)[r]{\strut{} 2}}%
      \put(5968,4232){\makebox(0,0){\strut{} 0}}%
      \put(6972,4232){\makebox(0,0){\strut{} 5}}%
      \put(7975,4232){\makebox(0,0){\strut{} 10}}%
      \put(8979,4232){\makebox(0,0){\strut{} 15}}%
      \put(9982,4232){\makebox(0,0){\strut{} 20}}%
      \csname LTb\endcsname%
      \put(5408,5950){\rotatebox{-270}{\makebox(0,0){\strut{}g-factor $g_c$}}}%
      \put(7975,3992){\makebox(0,0){\strut{}Height [nm]}}%
    }%
    \gplgaddtomacro\gplfronttext{%
      \csname LTb\endcsname%
      \put(9681,7228){\makebox(0,0)[r]{\strut{}Electron - strained (C)}}%
      \put(9400,5483){\makebox(0,0){\strut{}Radius}}%
      \csname LT0\endcsname%
      \put(9641,5296){\makebox(0,0)[r]{\strut{}7 nm}}%
      \csname LT1\endcsname%
      \put(9641,5140){\makebox(0,0)[r]{\strut{}9 nm}}%
      \csname LT2\endcsname%
      \put(9641,4984){\makebox(0,0)[r]{\strut{}11 nm}}%
      \csname LT3\endcsname%
      \put(9641,4828){\makebox(0,0)[r]{\strut{}13 nm}}%
      \csname LT4\endcsname%
      \put(9641,4672){\makebox(0,0)[r]{\strut{}15 nm}}%
    }%
    \gplgaddtomacro\gplbacktext{%
      \csname LTb\endcsname%
      \put(6064,512){\makebox(0,0)[r]{\strut{} 0}}%
      \put(6064,824){\makebox(0,0)[r]{\strut{} 0.01}}%
      \put(6064,1135){\makebox(0,0)[r]{\strut{} 0.02}}%
      \put(6064,1447){\makebox(0,0)[r]{\strut{} 0.03}}%
      \put(6064,1758){\makebox(0,0)[r]{\strut{} 0.04}}%
      \put(6064,2070){\makebox(0,0)[r]{\strut{} 0.05}}%
      \put(6064,2382){\makebox(0,0)[r]{\strut{} 0.06}}%
      \put(6064,2693){\makebox(0,0)[r]{\strut{} 0.07}}%
      \put(6064,3005){\makebox(0,0)[r]{\strut{} 0.08}}%
      \put(6064,3316){\makebox(0,0)[r]{\strut{} 0.09}}%
      \put(6064,3628){\makebox(0,0)[r]{\strut{} 0.1}}%
      \put(6160,352){\makebox(0,0){\strut{} 0}}%
      \put(7116,352){\makebox(0,0){\strut{} 5}}%
      \put(8071,352){\makebox(0,0){\strut{} 10}}%
      \put(9027,352){\makebox(0,0){\strut{} 15}}%
      \put(9982,352){\makebox(0,0){\strut{} 20}}%
      \csname LTb\endcsname%
      \put(5408,2070){\rotatebox{-270}{\makebox(0,0){\strut{}Sum of all components having certain $L^{\text{env}}_z$}}}%
      \put(8071,112){\makebox(0,0){\strut{}Height [nm]}}%
    }%
    \gplgaddtomacro\gplfronttext{%
      \csname LTb\endcsname%
      \put(9695,3348){\makebox(0,0)[r]{\strut{}Electron - unstrained (D)}}%
      \put(9428,1603){\makebox(0,0){\strut{}Radius}}%
      \csname LT0\endcsname%
      \put(9657,1416){\makebox(0,0)[r]{\strut{}7 nm}}%
      \csname LT1\endcsname%
      \put(9657,1260){\makebox(0,0)[r]{\strut{}9 nm}}%
      \csname LT2\endcsname%
      \put(9657,1104){\makebox(0,0)[r]{\strut{}11 nm}}%
      \csname LT3\endcsname%
      \put(9657,948){\makebox(0,0)[r]{\strut{}13 nm}}%
      \csname LT4\endcsname%
      \put(9657,792){\makebox(0,0)[r]{\strut{}15 nm}}%
      \csname LTb\endcsname%
      \put(6351,2756){\makebox(0,0)[l]{\strut{}$L^{\text{env}}_z = 0$}}%
      \put(7116,979){\makebox(0,0)[l]{\strut{}$L^{\text{env}}_z = \pm1$}}%
    }%
    \gplbacktext
    \put(0,0){\includegraphics{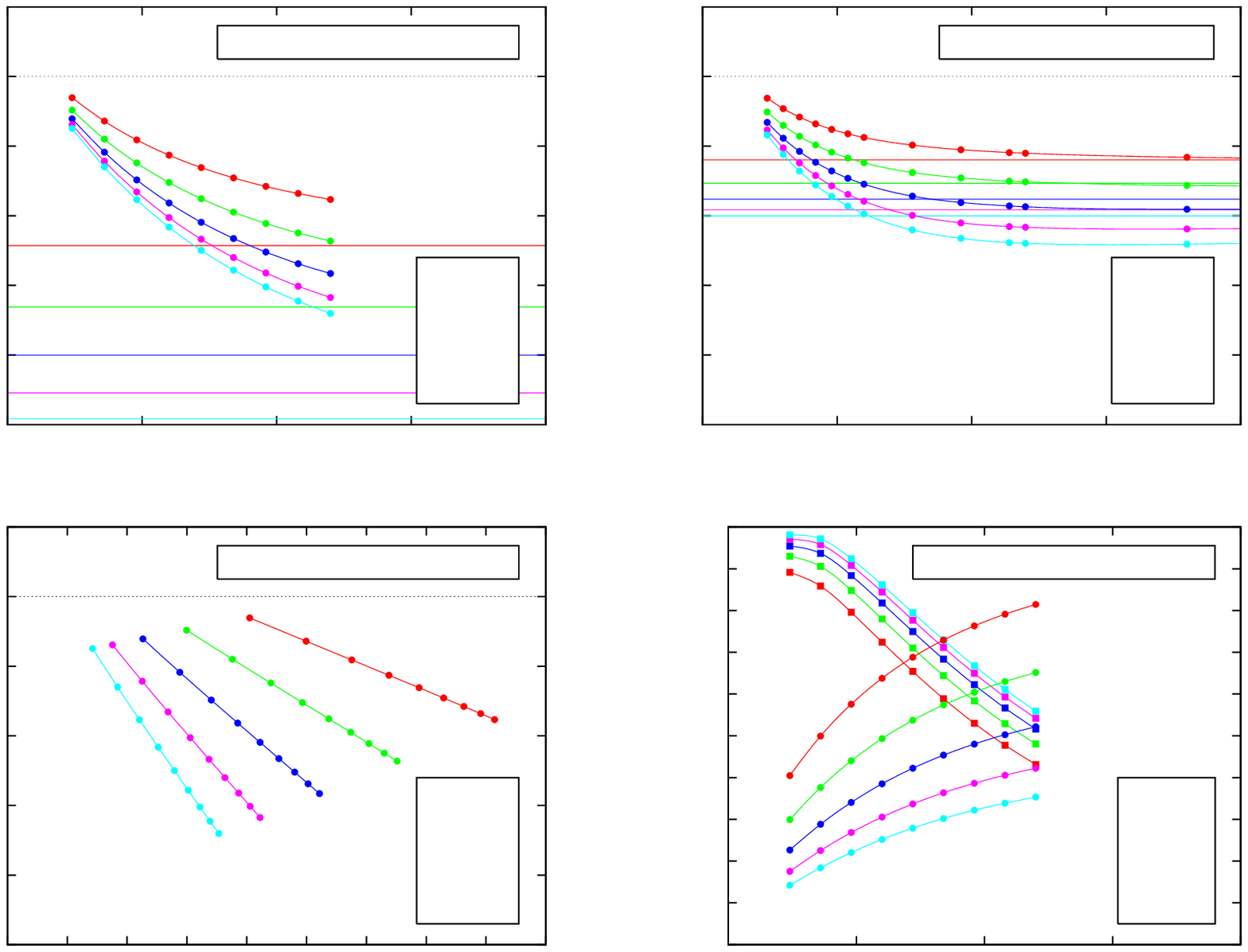}}%
    \gplfronttext
  \end{picture}%
\endgroup

\caption{(Color online) Height dependence of the electron $g_c$-factor of (a) unstrained and (c) strained InAs/InP quantum dots, respectively. (b)  the correlation between the $g_c$-factor and the sum of all components having $L_z^{\text{env}}=\pm 1$ for unstrained QDs.  (d)  the height dependence of the sum of all components having envelope angular momenta $L_z^{\text{env}}=\pm 1$ (dots) and $L_z^{\text{env}}=0$ (squares). The different colors indicate different radii of QDs. Horizontal lines are the $g_c$-factors of quantum wires having different radii.}
\label{fig:g_c}
\end{center}
\end{figure*}

The calculated quantum dots have a cylindrical shape and are assumed to be pure InAs (see Fig.~\ref{fig:QD}), as was inferred from cross-sectional scanning tunneling microscopy in Ref.~\onlinecite{Kleemans2009}. Calculations for various heights $h$ and radii $r$ for the quantum dots have been performed to study the size dependence of the $g$-factor and diamagnetic coefficient. The energy levels and wave functions of these dots are calculated with magnetic fields of $0-10$~T applied in the growth $[001]$-direction. The sizes of the quantum dots are small enough compared to the magnitude of the applied magnetic field to preclude any transition to complete magnetic confinement, which has been studied elsewhere\cite{Maes2004}. The spin-dependent energies  of both the electron ($c$) and hole states ($v$) can be parametrized as
\begin{eqnarray}
E_c\left(B\right) &=& E_{c,0} \pm \frac{1}{2} |\mu_B| g_c B + \alpha_c B^2, \\
E_v\left(B\right) &=& E_{v,0} \mp \frac{1}{2} |\mu_B| \left(g_v^0 + g_v^2 B^2 \right) B + \alpha_v B^2, \label{eq:g_v_def}
\end{eqnarray}
where $E_{(c,v),0}$ is the ground state energy, $g_{(c,v)}$ the $g$-factor, $\mu_B$ the Bohr magneton, and $\alpha_{(c,v)}$ the diamagnetic coefficient. Note that the hole (but not the electron) Zeeman energy is observed to possess significant nonlinearity; the hole $g_v$-factor has a zero-order component $g_v^0$ and a second-order component $g_v^2$. The upper (lower) signs in front of the Zeeman energy correspond to spin states parallel (anti-parallel) to the applied magnetic field. Using these definitions for the energy levels, we can define the magneto-optical properties of the exciton as:
\begin{eqnarray}
g_{ex} &=& - g_c - g_v^0, \label{eq:g_ex} \\
\alpha_{ex} &=& |\alpha_c| + |\alpha_v|, \label{eq:alpha_ex} \\
E_{em} &=& |E_{c,0}| + |E_{v,0}|, \label{eq:E_ex}
\end{eqnarray}
where $g_{ex}$ is the exciton $g$-factor, $\alpha_{ex}$ the exciton diamagnetic coefficient, and $E_{em}$ the emission energy of the QD. These definitions are consistent with the experimental definitions of Ref. \onlinecite{Kleemans2009}.


\section{$g$-factors} \label{sec:g-factor}
\begin{figure*}
\begin{center}
\begingroup
  \makeatletter
  \providecommand\color[2][]{%
    \GenericError{(gnuplot) \space\space\space\@spaces}{%
      Package color not loaded in conjunction with
      terminal option `colourtext'%
    }{See the gnuplot documentation for explanation.%
    }{Either use 'blacktext' in gnuplot or load the package
      color.sty in LaTeX.}%
    \renewcommand\color[2][]{}%
  }%
  \providecommand\includegraphics[2][]{%
    \GenericError{(gnuplot) \space\space\space\@spaces}{%
      Package graphicx or graphics not loaded%
    }{See the gnuplot documentation for explanation.%
    }{The gnuplot epslatex terminal needs graphicx.sty or graphics.sty.}%
    \renewcommand\includegraphics[2][]{}%
  }%
  \providecommand\rotatebox[2]{#2}%
  \@ifundefined{ifGPcolor}{%
    \newif\ifGPcolor
    \GPcolortrue
  }{}%
  \@ifundefined{ifGPblacktext}{%
    \newif\ifGPblacktext
    \GPblacktexttrue
  }{}%
  \let\gplgaddtomacro\g@addto@macro
  \gdef\gplbacktext{}%
  \gdef\gplfronttext{}%
  \makeatother
  \ifGPblacktext
    \def\colorrgb#1{}%
    \def\colorgray#1{}%
  \else
    \ifGPcolor
      \def\colorrgb#1{\color[rgb]{#1}}%
      \def\colorgray#1{\color[gray]{#1}}%
      \expandafter\def\csname LTw\endcsname{\color{white}}%
      \expandafter\def\csname LTb\endcsname{\color{black}}%
      \expandafter\def\csname LTa\endcsname{\color{black}}%
      \expandafter\def\csname LT0\endcsname{\color[rgb]{1,0,0}}%
      \expandafter\def\csname LT1\endcsname{\color[rgb]{0,1,0}}%
      \expandafter\def\csname LT2\endcsname{\color[rgb]{0,0,1}}%
      \expandafter\def\csname LT3\endcsname{\color[rgb]{1,0,1}}%
      \expandafter\def\csname LT4\endcsname{\color[rgb]{0,1,1}}%
      \expandafter\def\csname LT5\endcsname{\color[rgb]{1,1,0}}%
      \expandafter\def\csname LT6\endcsname{\color[rgb]{0,0,0}}%
      \expandafter\def\csname LT7\endcsname{\color[rgb]{1,0.3,0}}%
      \expandafter\def\csname LT8\endcsname{\color[rgb]{0.5,0.5,0.5}}%
    \else
      \def\colorrgb#1{\color{black}}%
      \def\colorgray#1{\color[gray]{#1}}%
      \expandafter\def\csname LTw\endcsname{\color{white}}%
      \expandafter\def\csname LTb\endcsname{\color{black}}%
      \expandafter\def\csname LTa\endcsname{\color{black}}%
      \expandafter\def\csname LT0\endcsname{\color{black}}%
      \expandafter\def\csname LT1\endcsname{\color{black}}%
      \expandafter\def\csname LT2\endcsname{\color{black}}%
      \expandafter\def\csname LT3\endcsname{\color{black}}%
      \expandafter\def\csname LT4\endcsname{\color{black}}%
      \expandafter\def\csname LT5\endcsname{\color{black}}%
      \expandafter\def\csname LT6\endcsname{\color{black}}%
      \expandafter\def\csname LT7\endcsname{\color{black}}%
      \expandafter\def\csname LT8\endcsname{\color{black}}%
    \fi
  \fi
  \setlength{\unitlength}{0.0500bp}%
  \begin{picture}(10368.00,7761.60)%
    \gplgaddtomacro\gplbacktext{%
      \csname LTb\endcsname%
      \put(688,4392){\makebox(0,0)[r]{\strut{}-10}}%
      \put(688,4738){\makebox(0,0)[r]{\strut{}-8}}%
      \put(688,5084){\makebox(0,0)[r]{\strut{}-6}}%
      \put(688,5431){\makebox(0,0)[r]{\strut{}-4}}%
      \put(688,5777){\makebox(0,0)[r]{\strut{}-2}}%
      \put(688,6123){\makebox(0,0)[r]{\strut{} 0}}%
      \put(688,6469){\makebox(0,0)[r]{\strut{} 2}}%
      \put(688,6816){\makebox(0,0)[r]{\strut{} 4}}%
      \put(688,7162){\makebox(0,0)[r]{\strut{} 6}}%
      \put(688,7508){\makebox(0,0)[r]{\strut{} 8}}%
      \put(784,4232){\makebox(0,0){\strut{} 0}}%
      \put(1788,4232){\makebox(0,0){\strut{} 5}}%
      \put(2792,4232){\makebox(0,0){\strut{} 10}}%
      \put(3795,4232){\makebox(0,0){\strut{} 15}}%
      \put(4799,4232){\makebox(0,0){\strut{} 20}}%
      \csname LTb\endcsname%
      \put(224,5950){\rotatebox{-270}{\makebox(0,0){\strut{}Zero order g-factor $g_v^0$}}}%
      \put(2791,3992){\makebox(0,0){\strut{}Height [nm]}}%
    }%
    \gplgaddtomacro\gplfronttext{%
      \csname LTb\endcsname%
      \put(4498,7228){\makebox(0,0)[r]{\strut{}Hole - unstrained (A)}}%
      \put(4217,5483){\makebox(0,0){\strut{}Radius}}%
      \csname LT0\endcsname%
      \put(4458,5296){\makebox(0,0)[r]{\strut{}7 nm}}%
      \csname LT1\endcsname%
      \put(4458,5140){\makebox(0,0)[r]{\strut{}9 nm}}%
      \csname LT2\endcsname%
      \put(4458,4984){\makebox(0,0)[r]{\strut{}11 nm}}%
      \csname LT3\endcsname%
      \put(4458,4828){\makebox(0,0)[r]{\strut{}13 nm}}%
      \csname LT4\endcsname%
      \put(4458,4672){\makebox(0,0)[r]{\strut{}15 nm}}%
    }%
    \gplgaddtomacro\gplbacktext{%
      \csname LTb\endcsname%
      \put(688,512){\makebox(0,0)[r]{\strut{}-10}}%
      \put(688,858){\makebox(0,0)[r]{\strut{}-8}}%
      \put(688,1204){\makebox(0,0)[r]{\strut{}-6}}%
      \put(688,1551){\makebox(0,0)[r]{\strut{}-4}}%
      \put(688,1897){\makebox(0,0)[r]{\strut{}-2}}%
      \put(688,2243){\makebox(0,0)[r]{\strut{} 0}}%
      \put(688,2589){\makebox(0,0)[r]{\strut{} 2}}%
      \put(688,2936){\makebox(0,0)[r]{\strut{} 4}}%
      \put(688,3282){\makebox(0,0)[r]{\strut{} 6}}%
      \put(688,3628){\makebox(0,0)[r]{\strut{} 8}}%
      \put(784,352){\makebox(0,0){\strut{} 0}}%
      \put(1453,352){\makebox(0,0){\strut{} 0.05}}%
      \put(2122,352){\makebox(0,0){\strut{} 0.1}}%
      \put(2792,352){\makebox(0,0){\strut{} 0.15}}%
      \put(3461,352){\makebox(0,0){\strut{} 0.2}}%
      \put(4130,352){\makebox(0,0){\strut{} 0.25}}%
      \put(4799,352){\makebox(0,0){\strut{} 0.3}}%
      \csname LTb\endcsname%
      \put(224,2070){\rotatebox{-270}{\makebox(0,0){\strut{}Zero-order g-factor $g_v^0$}}}%
      \put(2791,112){\makebox(0,0){\strut{}LH-component}}%
    }%
    \gplgaddtomacro\gplfronttext{%
      \csname LTb\endcsname%
      \put(4498,3348){\makebox(0,0)[r]{\strut{}Hole - unstrained (B)}}%
      \put(4217,1603){\makebox(0,0){\strut{}Radius}}%
      \csname LT0\endcsname%
      \put(4458,1416){\makebox(0,0)[r]{\strut{}7 nm}}%
      \csname LT1\endcsname%
      \put(4458,1260){\makebox(0,0)[r]{\strut{}9 nm}}%
      \csname LT2\endcsname%
      \put(4458,1104){\makebox(0,0)[r]{\strut{}11 nm}}%
      \csname LT3\endcsname%
      \put(4458,948){\makebox(0,0)[r]{\strut{}13 nm}}%
      \csname LT4\endcsname%
      \put(4458,792){\makebox(0,0)[r]{\strut{}15 nm}}%
    }%
    \gplgaddtomacro\gplbacktext{%
      \csname LTb\endcsname%
      \put(5872,4392){\makebox(0,0)[r]{\strut{}-10}}%
      \put(5872,4738){\makebox(0,0)[r]{\strut{}-8}}%
      \put(5872,5084){\makebox(0,0)[r]{\strut{}-6}}%
      \put(5872,5431){\makebox(0,0)[r]{\strut{}-4}}%
      \put(5872,5777){\makebox(0,0)[r]{\strut{}-2}}%
      \put(5872,6123){\makebox(0,0)[r]{\strut{} 0}}%
      \put(5872,6469){\makebox(0,0)[r]{\strut{} 2}}%
      \put(5872,6816){\makebox(0,0)[r]{\strut{} 4}}%
      \put(5872,7162){\makebox(0,0)[r]{\strut{} 6}}%
      \put(5872,7508){\makebox(0,0)[r]{\strut{} 8}}%
      \put(5968,4232){\makebox(0,0){\strut{} 0}}%
      \put(6972,4232){\makebox(0,0){\strut{} 5}}%
      \put(7975,4232){\makebox(0,0){\strut{} 10}}%
      \put(8979,4232){\makebox(0,0){\strut{} 15}}%
      \put(9982,4232){\makebox(0,0){\strut{} 20}}%
      \csname LTb\endcsname%
      \put(5408,5950){\rotatebox{-270}{\makebox(0,0){\strut{}Zero order g-factor $g_v^0$}}}%
      \put(7975,3992){\makebox(0,0){\strut{}Height [nm]}}%
    }%
    \gplgaddtomacro\gplfronttext{%
      \csname LTb\endcsname%
      \put(9681,7228){\makebox(0,0)[r]{\strut{}Hole - strained (C)}}%
      \put(9400,5483){\makebox(0,0){\strut{}Radius}}%
      \csname LT0\endcsname%
      \put(9641,5296){\makebox(0,0)[r]{\strut{}7 nm}}%
      \csname LT1\endcsname%
      \put(9641,5140){\makebox(0,0)[r]{\strut{}9 nm}}%
      \csname LT2\endcsname%
      \put(9641,4984){\makebox(0,0)[r]{\strut{}11 nm}}%
      \csname LT3\endcsname%
      \put(9641,4828){\makebox(0,0)[r]{\strut{}13 nm}}%
      \csname LT4\endcsname%
      \put(9641,4672){\makebox(0,0)[r]{\strut{}15 nm}}%
    }%
    \gplgaddtomacro\gplbacktext{%
      \csname LTb\endcsname%
      \put(5872,512){\makebox(0,0)[r]{\strut{}-12}}%
      \put(5872,957){\makebox(0,0)[r]{\strut{}-10}}%
      \put(5872,1402){\makebox(0,0)[r]{\strut{}-8}}%
      \put(5872,1847){\makebox(0,0)[r]{\strut{}-6}}%
      \put(5872,2293){\makebox(0,0)[r]{\strut{}-4}}%
      \put(5872,2738){\makebox(0,0)[r]{\strut{}-2}}%
      \put(5872,3183){\makebox(0,0)[r]{\strut{} 0}}%
      \put(5872,3628){\makebox(0,0)[r]{\strut{} 2}}%
      \put(5968,352){\makebox(0,0){\strut{} 0}}%
      \put(6972,352){\makebox(0,0){\strut{} 5}}%
      \put(7975,352){\makebox(0,0){\strut{} 10}}%
      \put(8979,352){\makebox(0,0){\strut{} 15}}%
      \put(9982,352){\makebox(0,0){\strut{} 20}}%
      \csname LTb\endcsname%
      \put(5408,2070){\rotatebox{-270}{\makebox(0,0){\strut{}Second-order g-factor $g_v^2$ $[10^{-3}/\text{T}^2]$}}}%
      \put(7975,112){\makebox(0,0){\strut{}Height [nm]}}%
    }%
    \gplgaddtomacro\gplfronttext{%
      \csname LTb\endcsname%
      \put(9681,3348){\makebox(0,0)[r]{\strut{}Hole - strained (D)}}%
      \put(9400,1603){\makebox(0,0){\strut{}Radius}}%
      \csname LT0\endcsname%
      \put(9641,1416){\makebox(0,0)[r]{\strut{}7 nm}}%
      \csname LT1\endcsname%
      \put(9641,1260){\makebox(0,0)[r]{\strut{}9 nm}}%
      \csname LT2\endcsname%
      \put(9641,1104){\makebox(0,0)[r]{\strut{}11 nm}}%
      \csname LT3\endcsname%
      \put(9641,948){\makebox(0,0)[r]{\strut{}13 nm}}%
      \csname LT4\endcsname%
      \put(9641,792){\makebox(0,0)[r]{\strut{}15 nm}}%
    }%
    \gplbacktext
    \put(0,0){\includegraphics{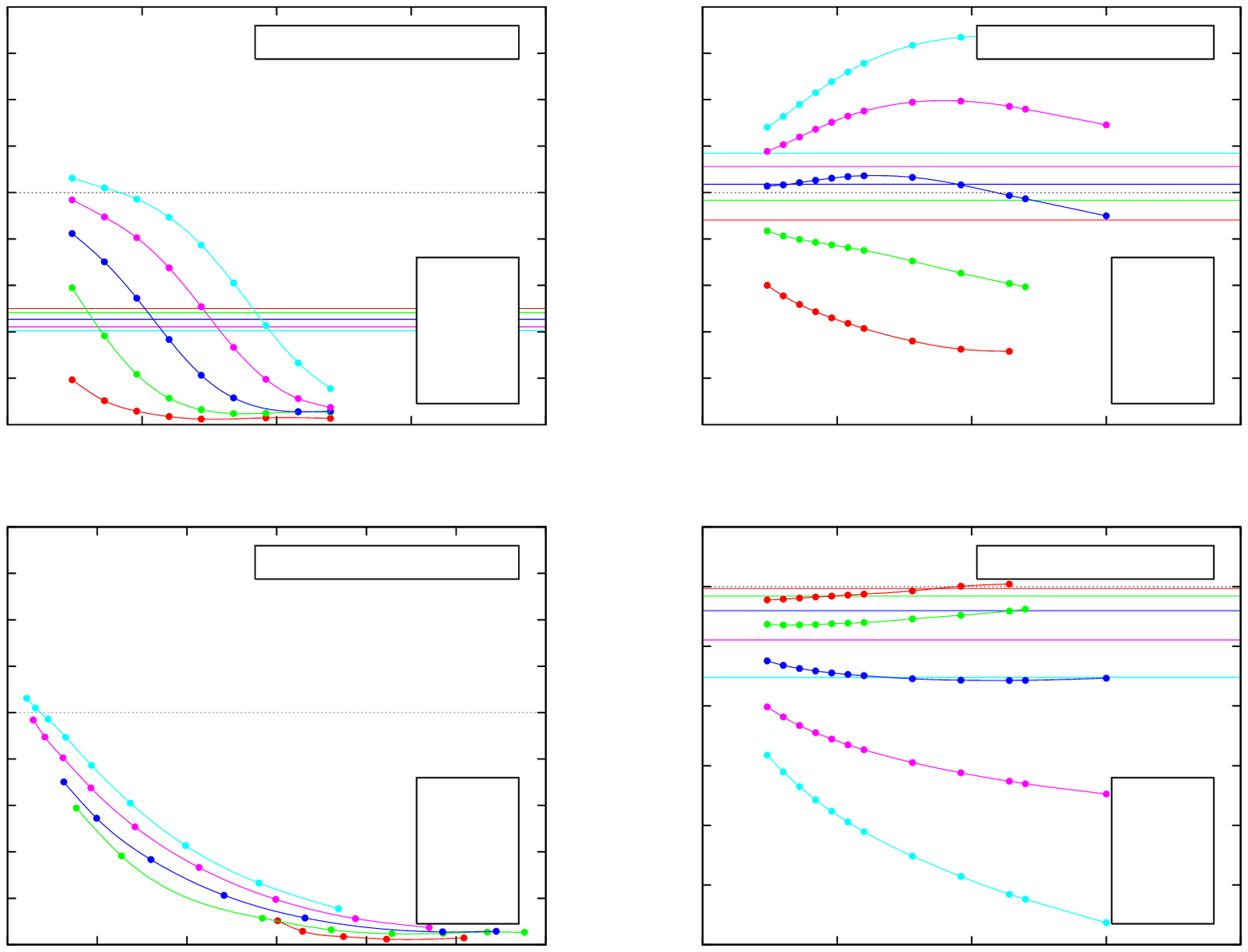}}%
    \gplfronttext
  \end{picture}%
\endgroup

\caption{(Color online)  The height dependence of the zero-order hole $g_v^0$-factor of (a) unstrained and (c) strained InAs/InP quantum dots.  (b)  the dependence of the zero-order hole $g_v^0$-factor on the LH-component for unstrained and strained InAs/InP quantum dots.  (d)  the height dependence of the second-order hole $g_v^2$-factor of strained InAs/InP quantum dots. The different colors indicate different radii of QDs. The horizontal lines are the $g_v^{0,2}$-factors of quantum wires having different radii.}
\label{fig:g_v_strain}
\end{center}
\end{figure*}

The calculated electron $g_c$-factors and hole $g_v$-factors are shown in Figs.~\ref{fig:g_c} and \ref{fig:g_v_strain}. The different colors indicate different radii $r$ of the QDs. The continuous horizontal lines are the $g$-factors calculated for quantum wires (i.e. quantum dots having infinite height). 

\subsection{Electron state}

\begin{table*}[p]
\caption{Combinations of $J^{\text{Bloch}}$ and $L^{\text{env}}$ leading to $|F,F_z\rangle=|\frac{1}{2},\pm\frac{1}{2}\rangle$, for each of the different components $i$ of $\Psi_c^i\left({\bf r}\right)$, of a QD having height of $6$~nm and radius of $11$~nm. The color scale is different for each plot.\label{table:F12}}
\begin{ruledtabular}
\begin{tabular}{cccccc}
\multirow{3}{*}{$i$} & \multirow{3}{*}{$|J^{\text{Bloch}},J^{\text{Bloch}}_z,L^{\text{env}},L^{\text{env}}_z\rangle$} & Cut of $|\Psi_c^i\left({\bf r}\right)|^2$ at $B=0$~T& \multicolumn{3}{c}{Cut of $|\Psi_c^i\left({\bf r}\right)|^2$ at $B=2$~T} \\
 & & $|F=\frac{1}{2},F_z=\pm\frac{1}{2}\rangle$ & $|J^{\text{Bloch}},J^{\text{Bloch}}_z\rangle$ & $|F=\frac{1}{2},F_z=+\frac{1}{2}\rangle$ & $|F=\frac{1}{2},F_z=-\frac{1}{2}\rangle$ \vspace{1mm} \\
\hline\hline \vspace{1mm}
CB
 & \begin{tabular}{l}
$|\frac{1}{2},\pm\frac{1}{2},0,0\rangle$
 \end{tabular}
  & \parbox{0.3\columnwidth}{\vspace{1mm}\includegraphics[width=0.4\columnwidth]{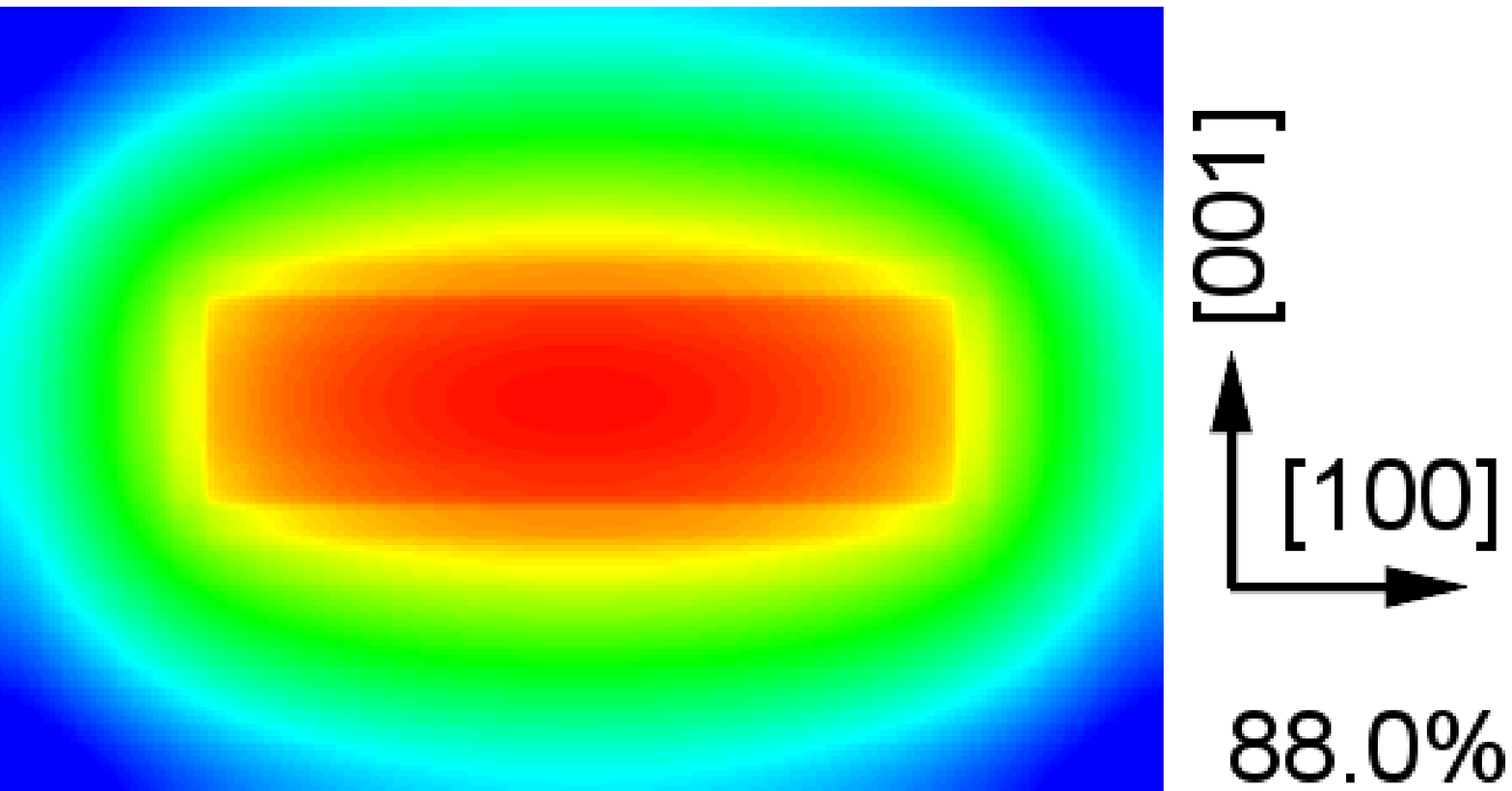}}
  & \begin{tabular}{l} $|\frac{1}{2},-\frac{1}{2}\rangle$ \\ \vspace{3mm} \\ $|\frac{1}{2},+\frac{1}{2}\rangle$ \end{tabular}
  & \parbox{0.2\columnwidth}{\vspace{1mm}\includegraphics[width=0.21\columnwidth]{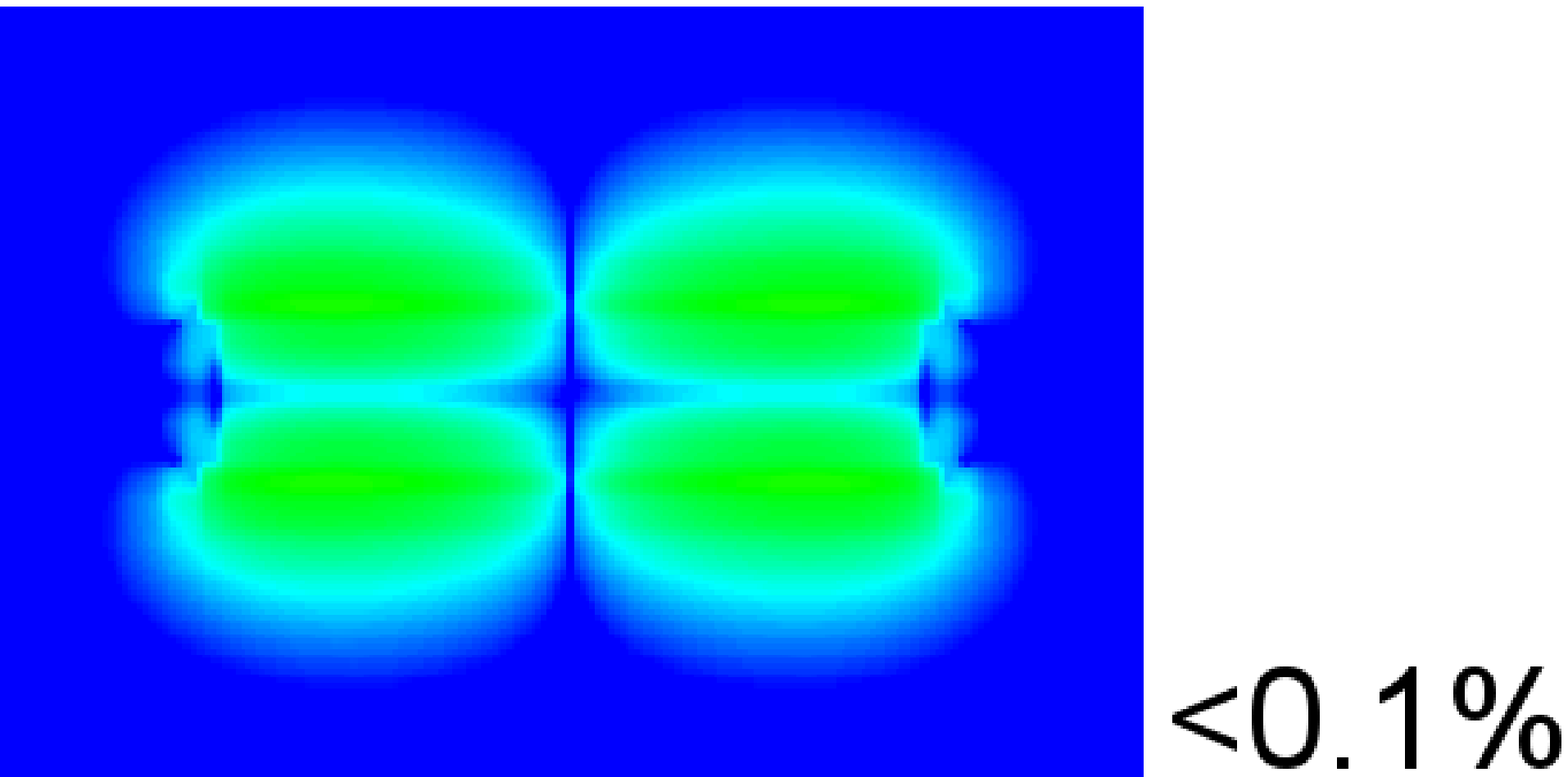} \\ \vspace{1mm}\includegraphics[width=0.21\columnwidth]{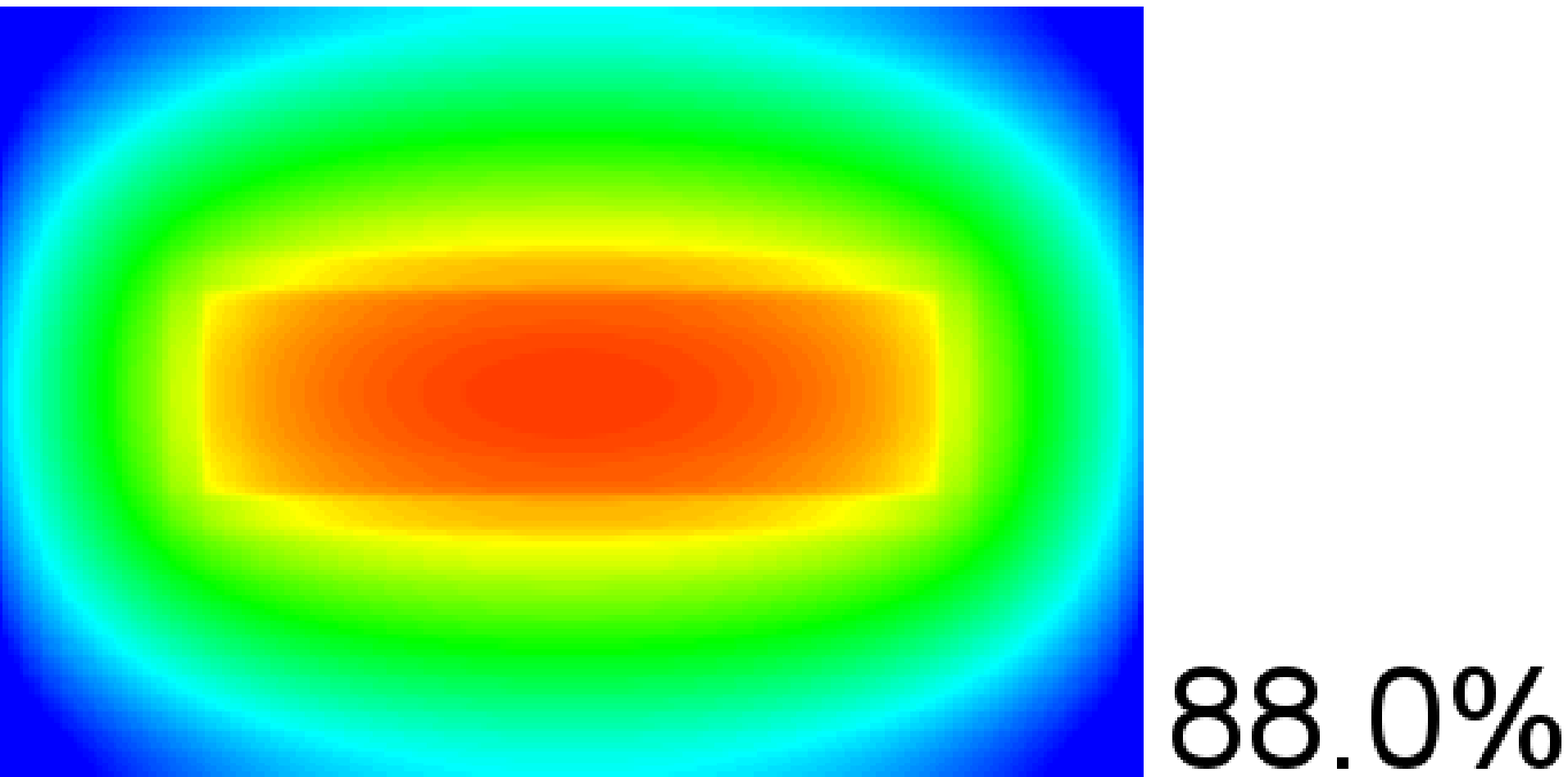}}
  & \parbox{0.2\columnwidth}{\vspace{1mm}\includegraphics[width=0.21\columnwidth]{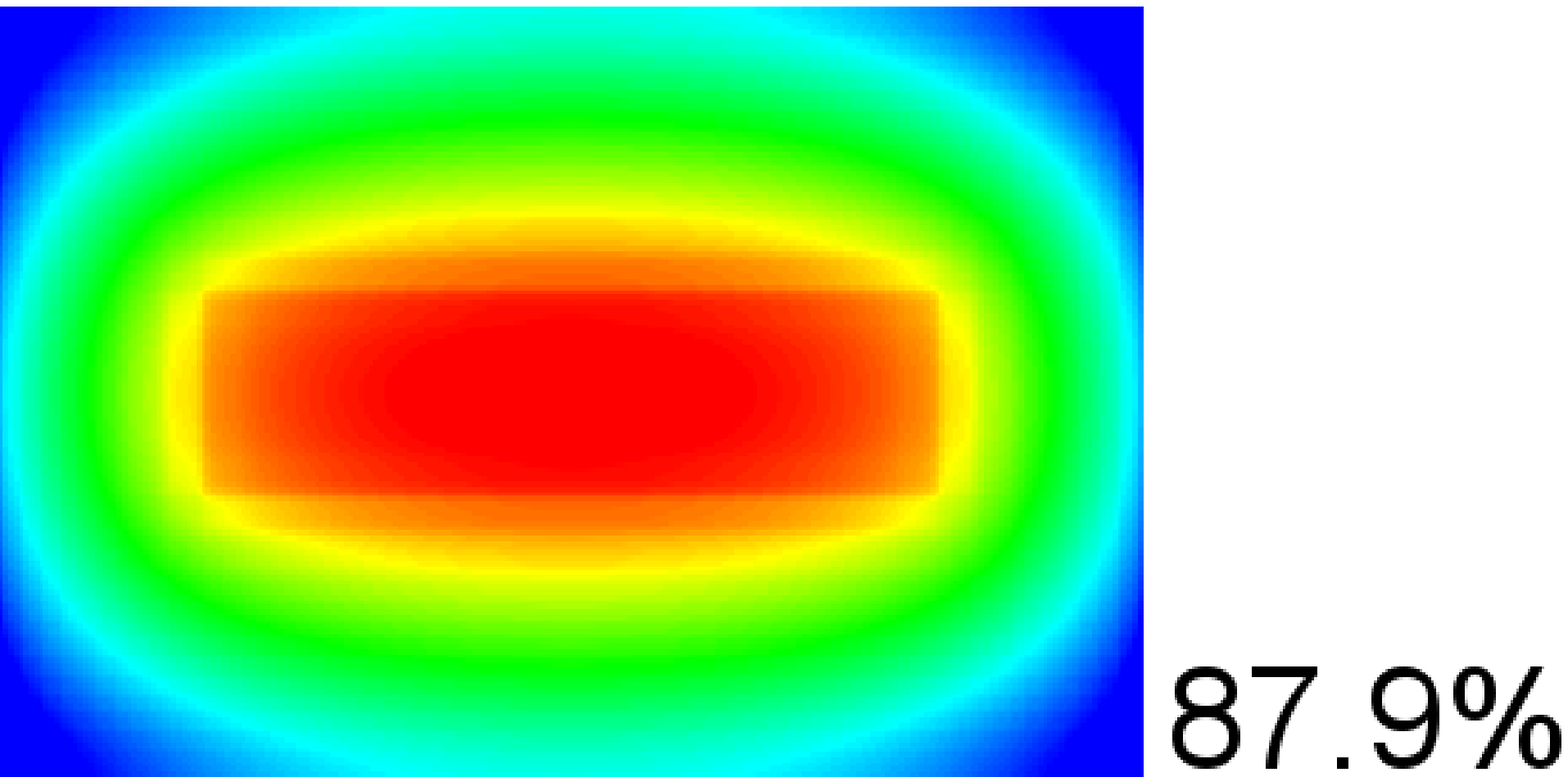} \\ \vspace{1mm}\includegraphics[width=0.21\columnwidth]{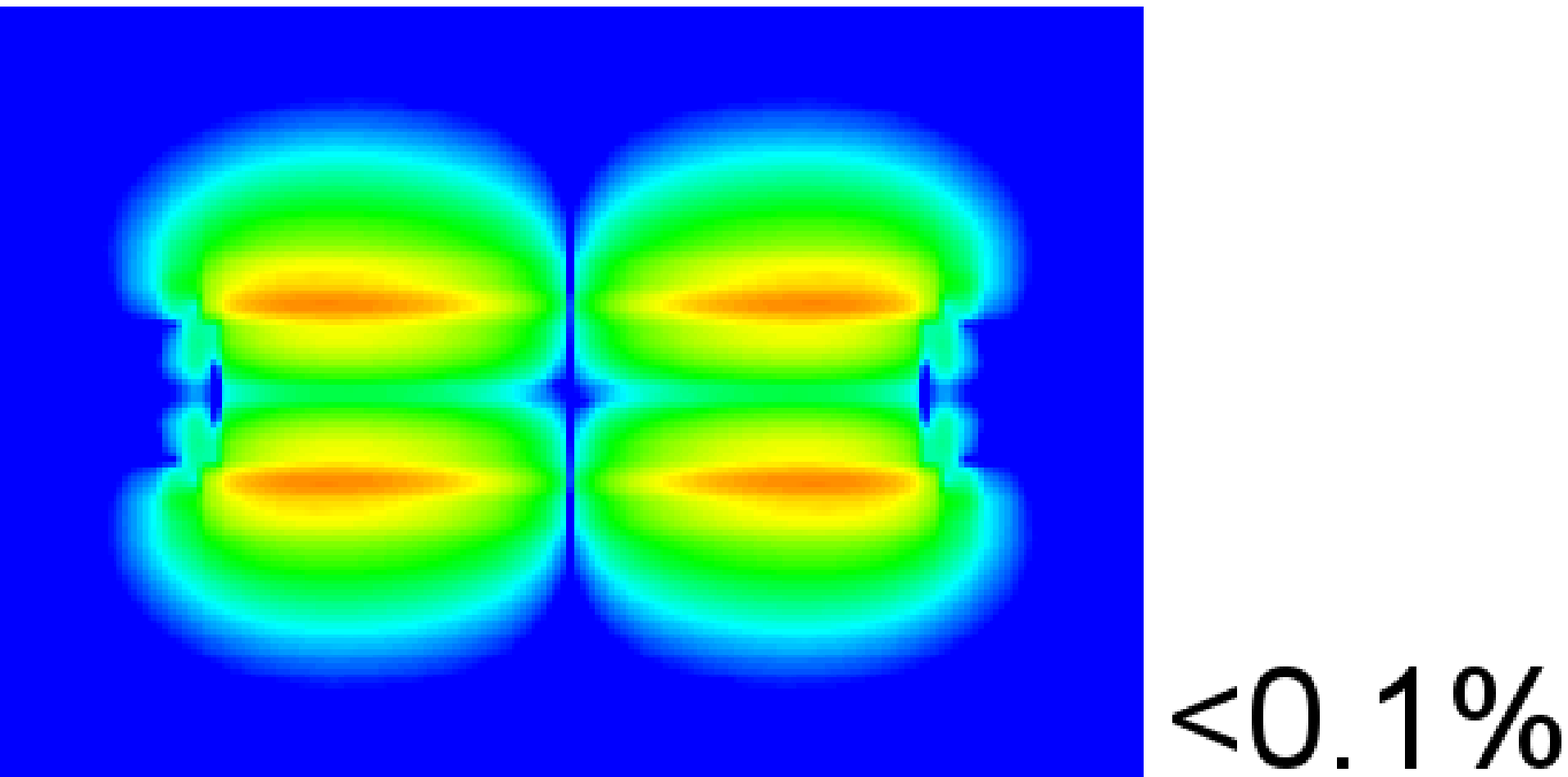}} \\
\hline \vspace{1mm}
HH
 & \begin{tabular}{l}
$|\frac{3}{2},\pm\frac{3}{2},1,\mp1\rangle$
 \end{tabular}
  & \parbox{0.3\columnwidth}{\vspace{1mm}\includegraphics[width=0.4\columnwidth]{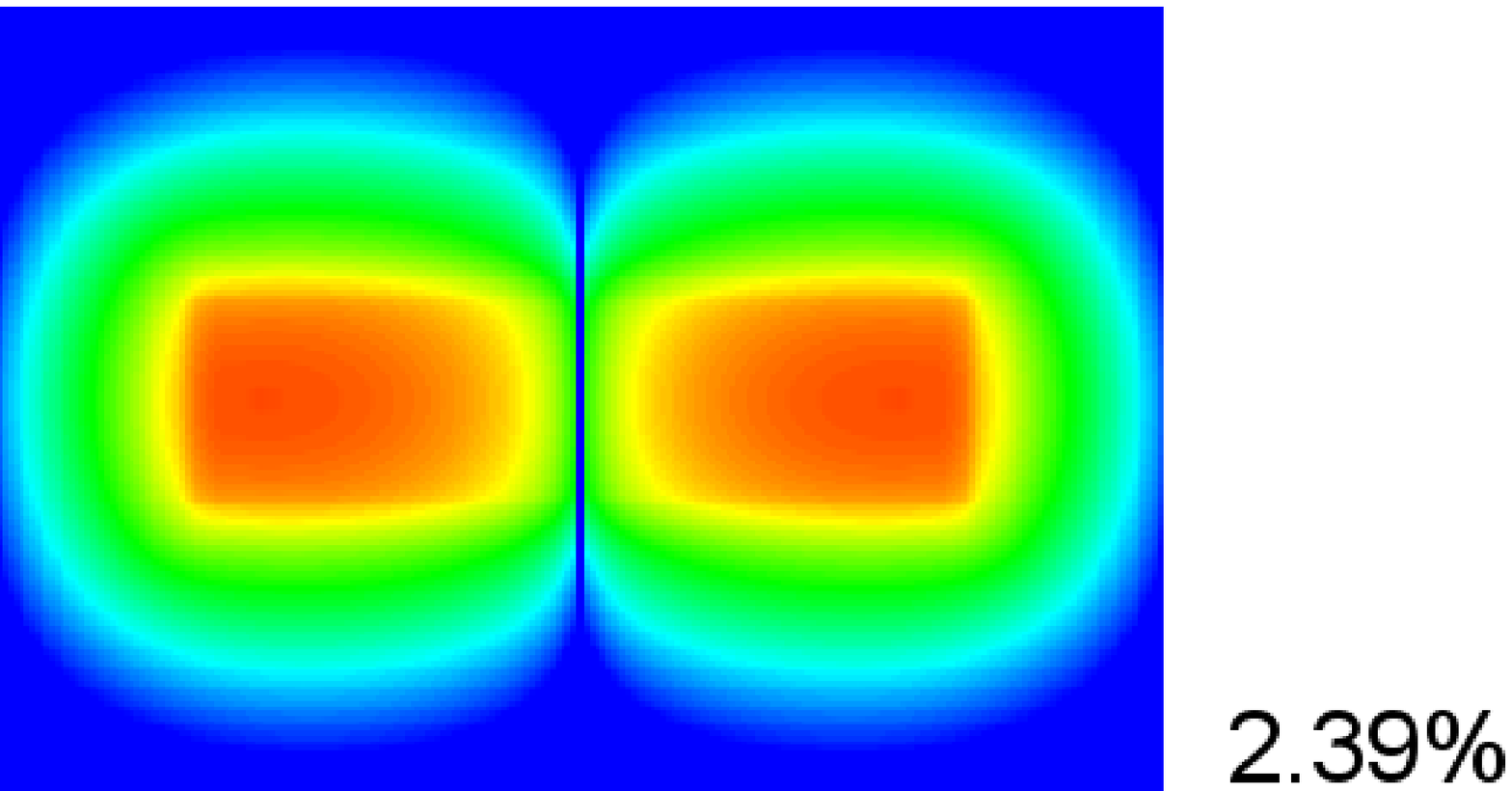}}
  & \begin{tabular}{l} $|\frac{3}{2},+\frac{3}{2}\rangle$ \\ \vspace{3mm} \\ $|\frac{3}{2},-\frac{3}{2}\rangle$ \end{tabular}
  & \parbox{0.2\columnwidth}{\vspace{1mm}\includegraphics[width=0.21\columnwidth]{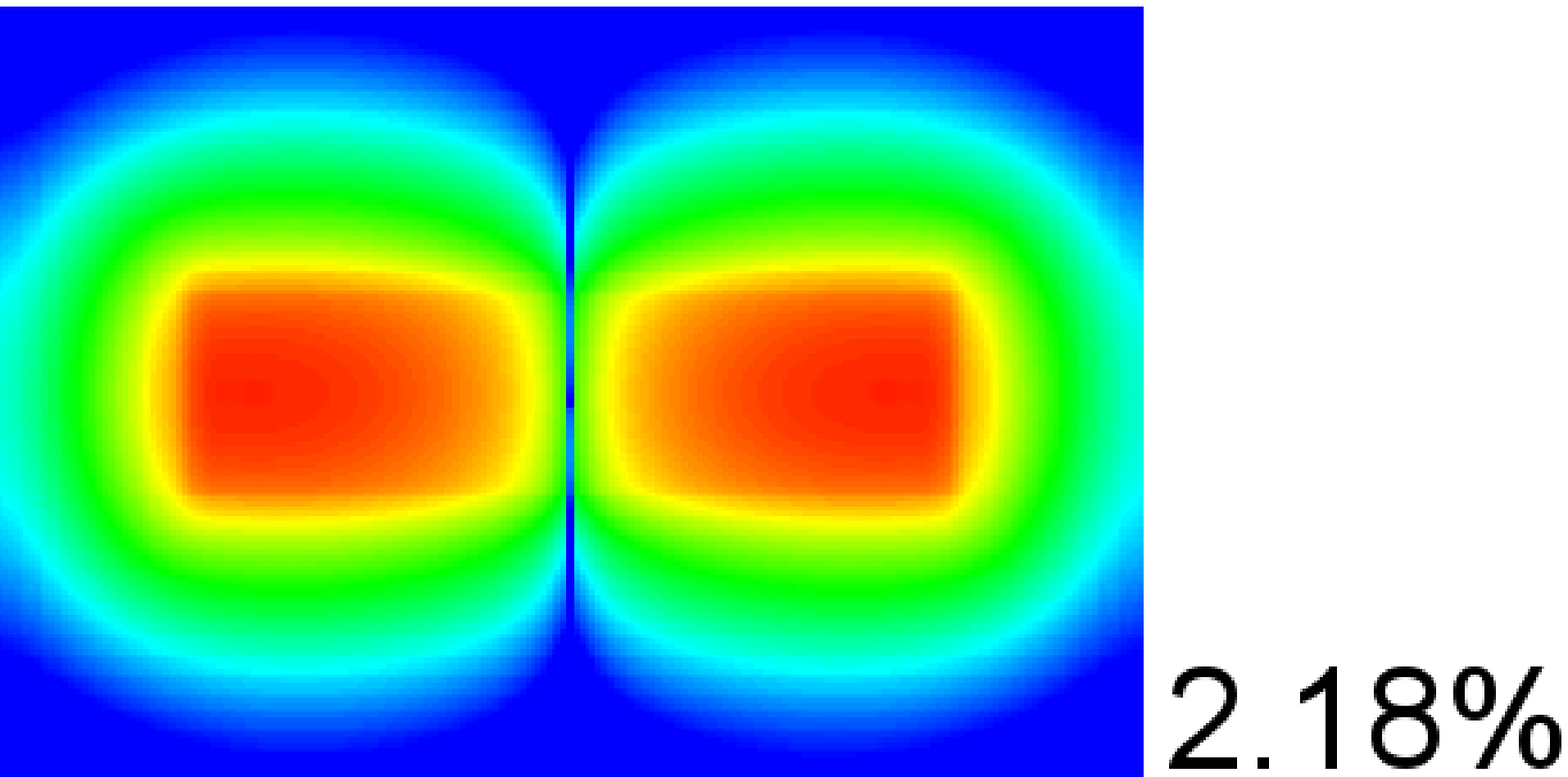} \\ \vspace{1mm}\includegraphics[width=0.21\columnwidth]{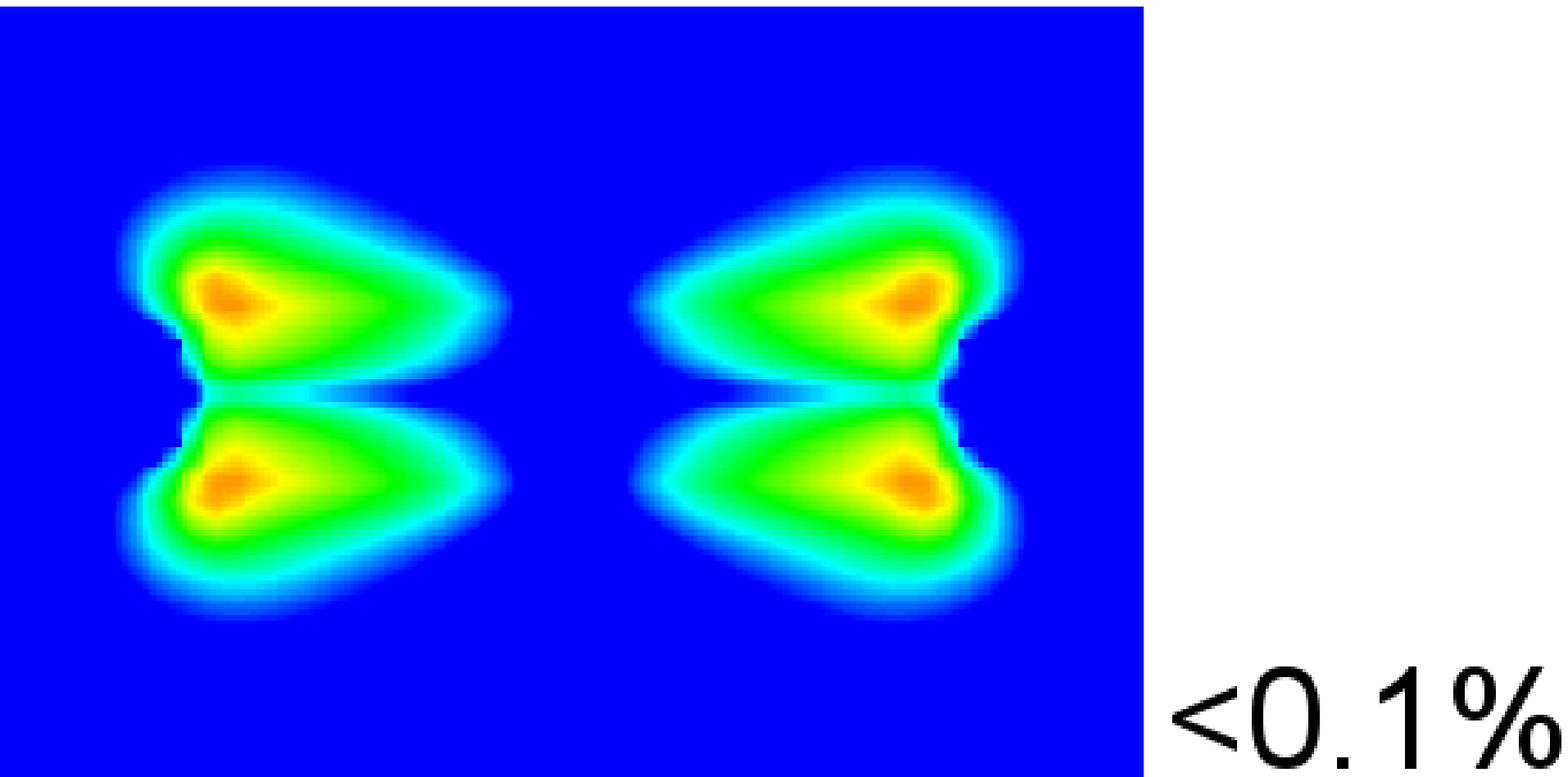}}
  & \parbox{0.2\columnwidth}{\vspace{1mm}\includegraphics[width=0.21\columnwidth]{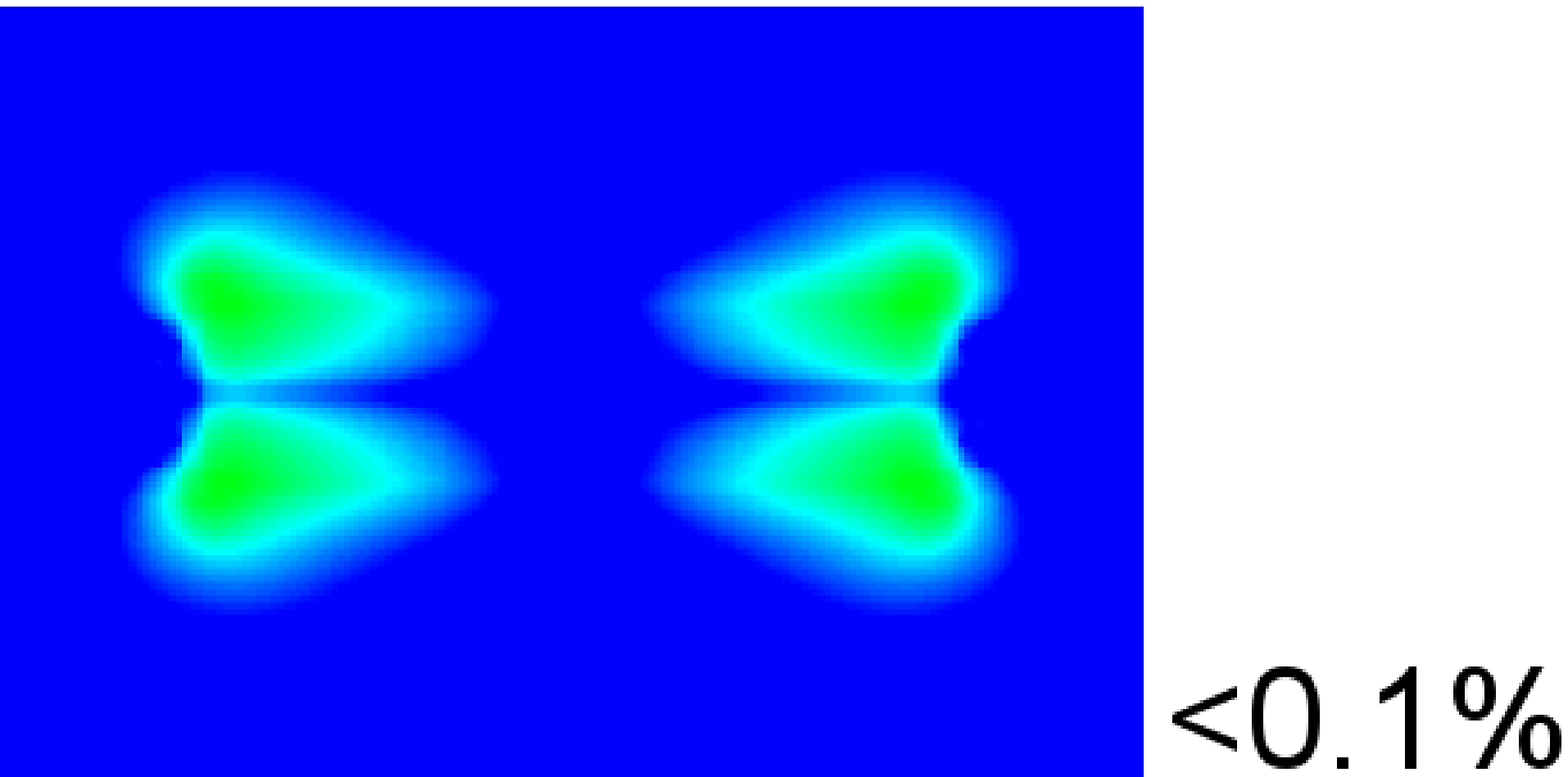} \\ \vspace{1mm}\includegraphics[width=0.21\columnwidth]{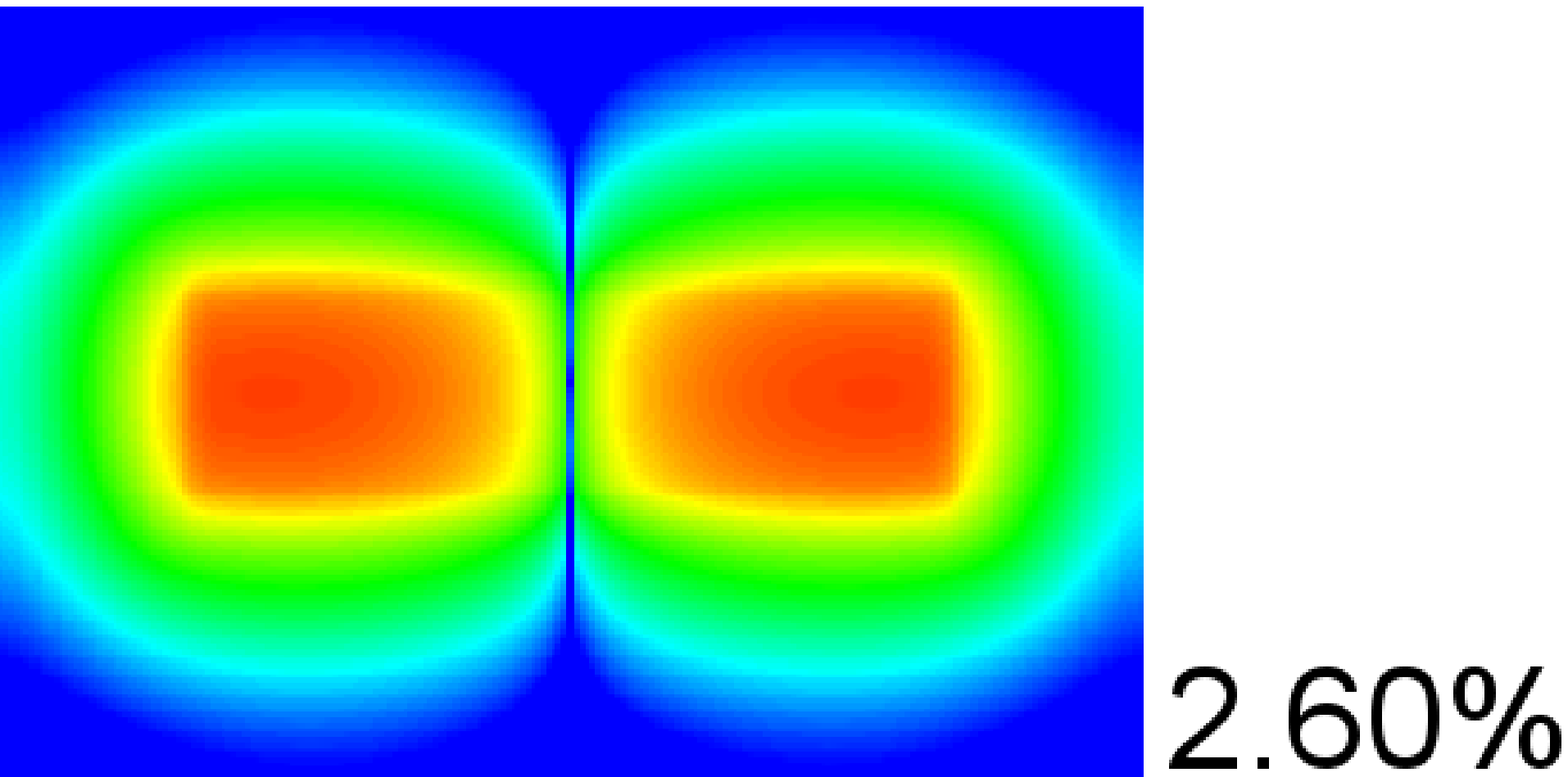}} \\
\hline \vspace{1mm}
LH
 & \begin{tabular}{l}
$|\frac{3}{2},\mp\frac{1}{2},1,\pm1\rangle$ \\
$|\frac{3}{2},\pm\frac{1}{2},1,0\rangle$
  \end{tabular}
 & \parbox{0.3\columnwidth}{\vspace{1mm}\includegraphics[width=0.4\columnwidth]{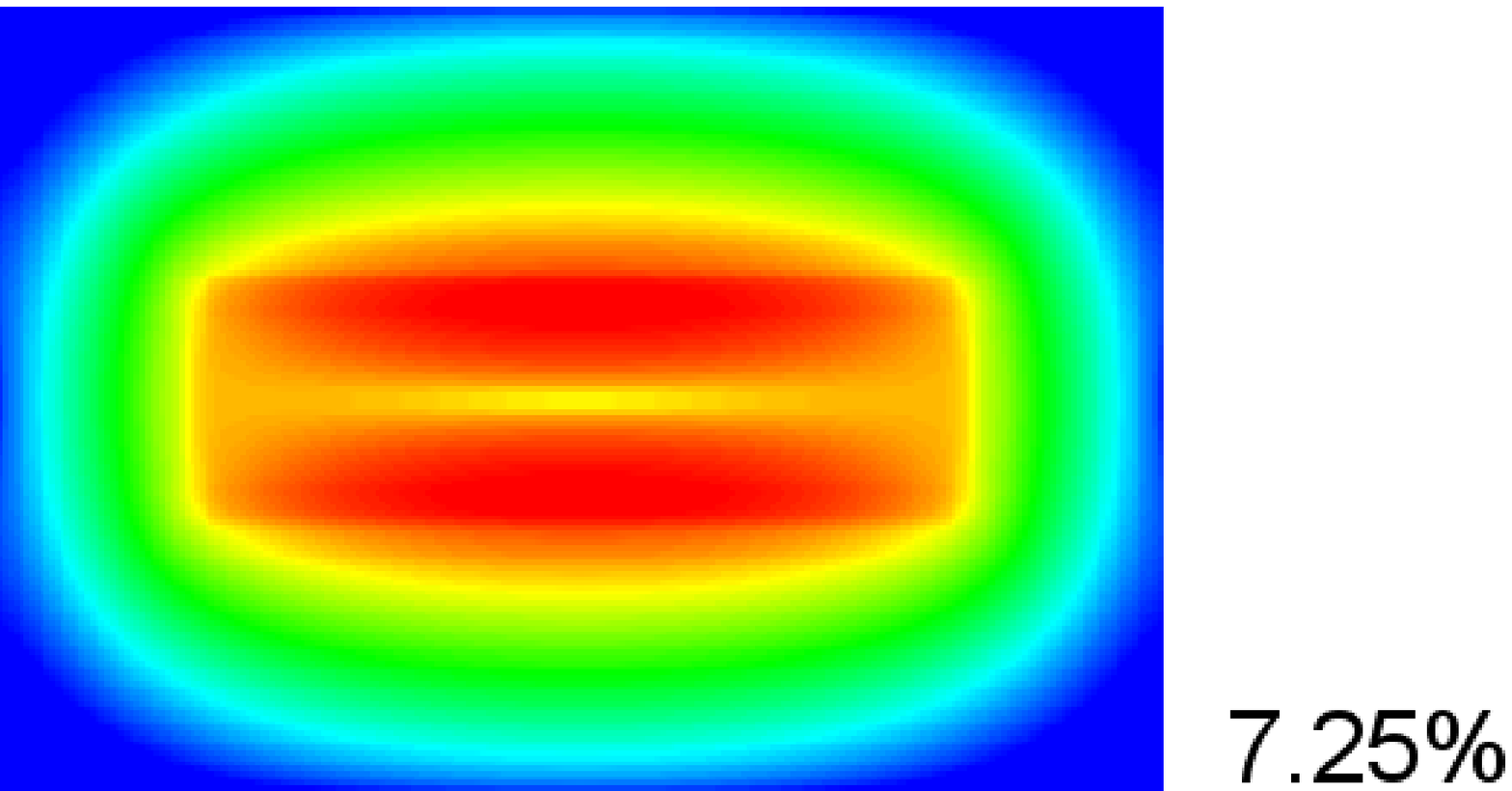}}
 & \begin{tabular}{l} $|\frac{3}{2},+\frac{1}{2}\rangle$ \\ \vspace{3mm} \\ $|\frac{3}{2},-\frac{1}{2}\rangle$ \end{tabular}
 & \parbox{0.2\columnwidth}{\vspace{1mm}\includegraphics[width=0.21\columnwidth]{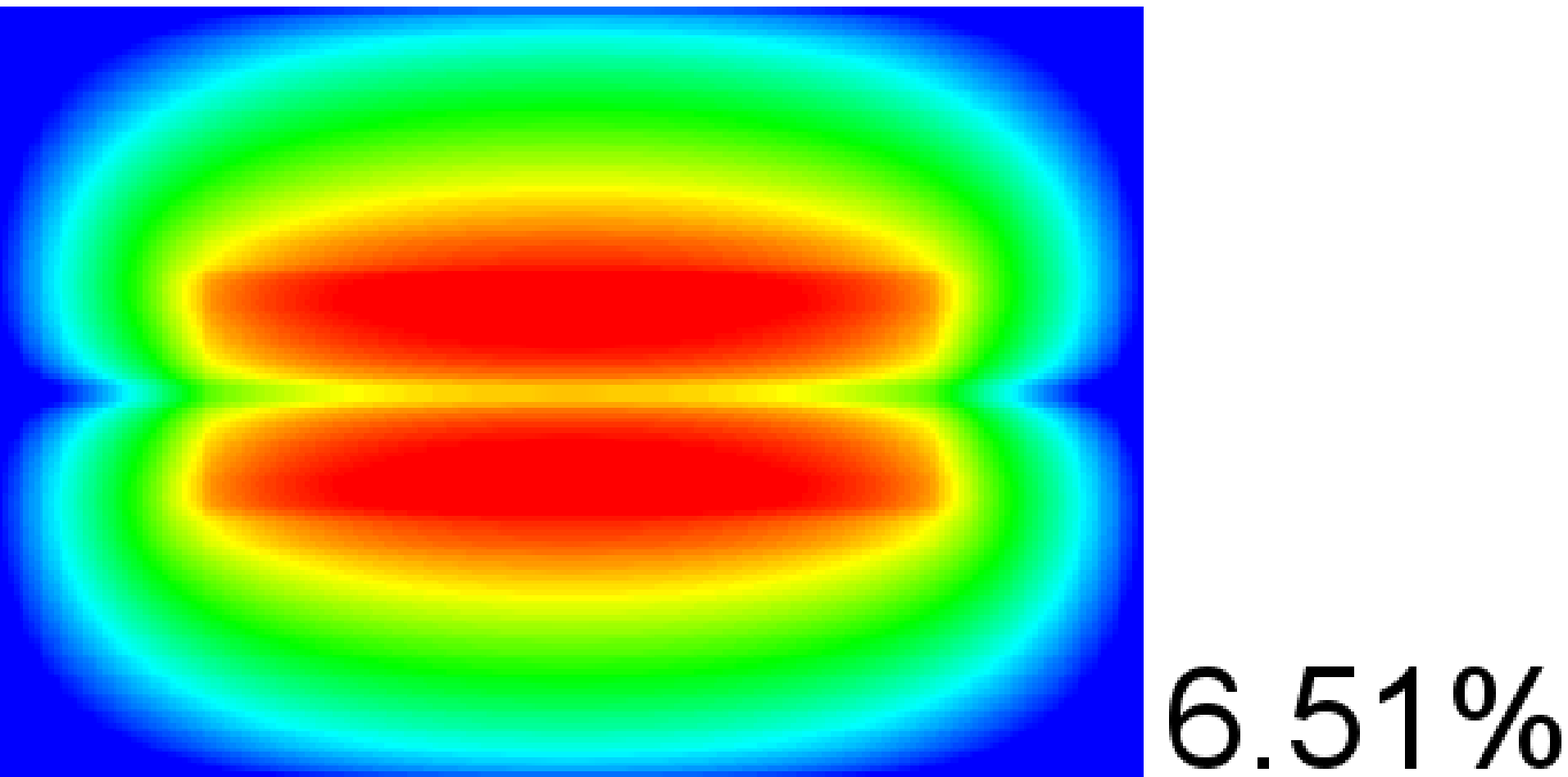} \\ \vspace{1mm}\includegraphics[width=0.21\columnwidth]{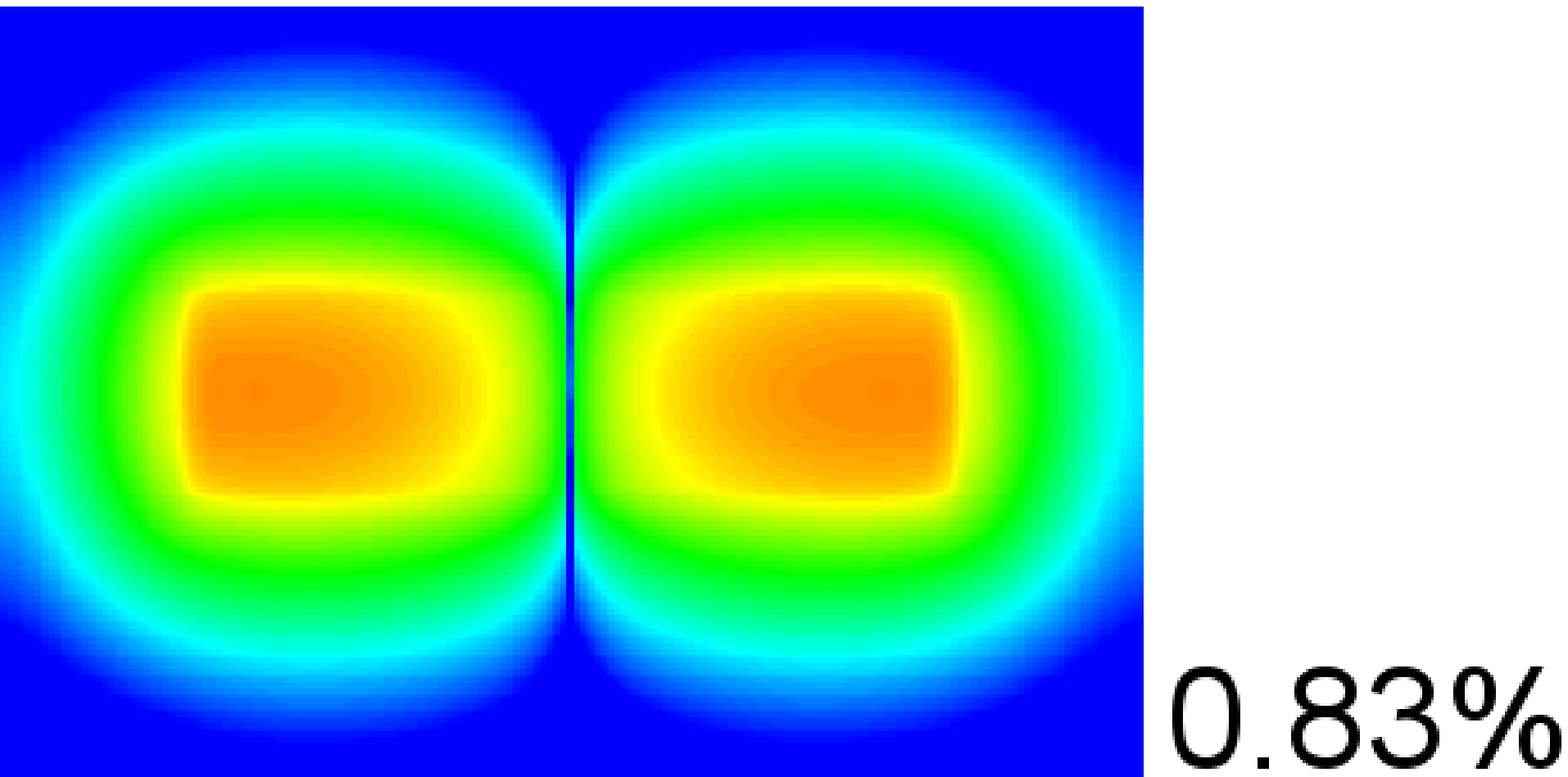}}
 & \parbox{0.2\columnwidth}{\vspace{1mm}\includegraphics[width=0.21\columnwidth]{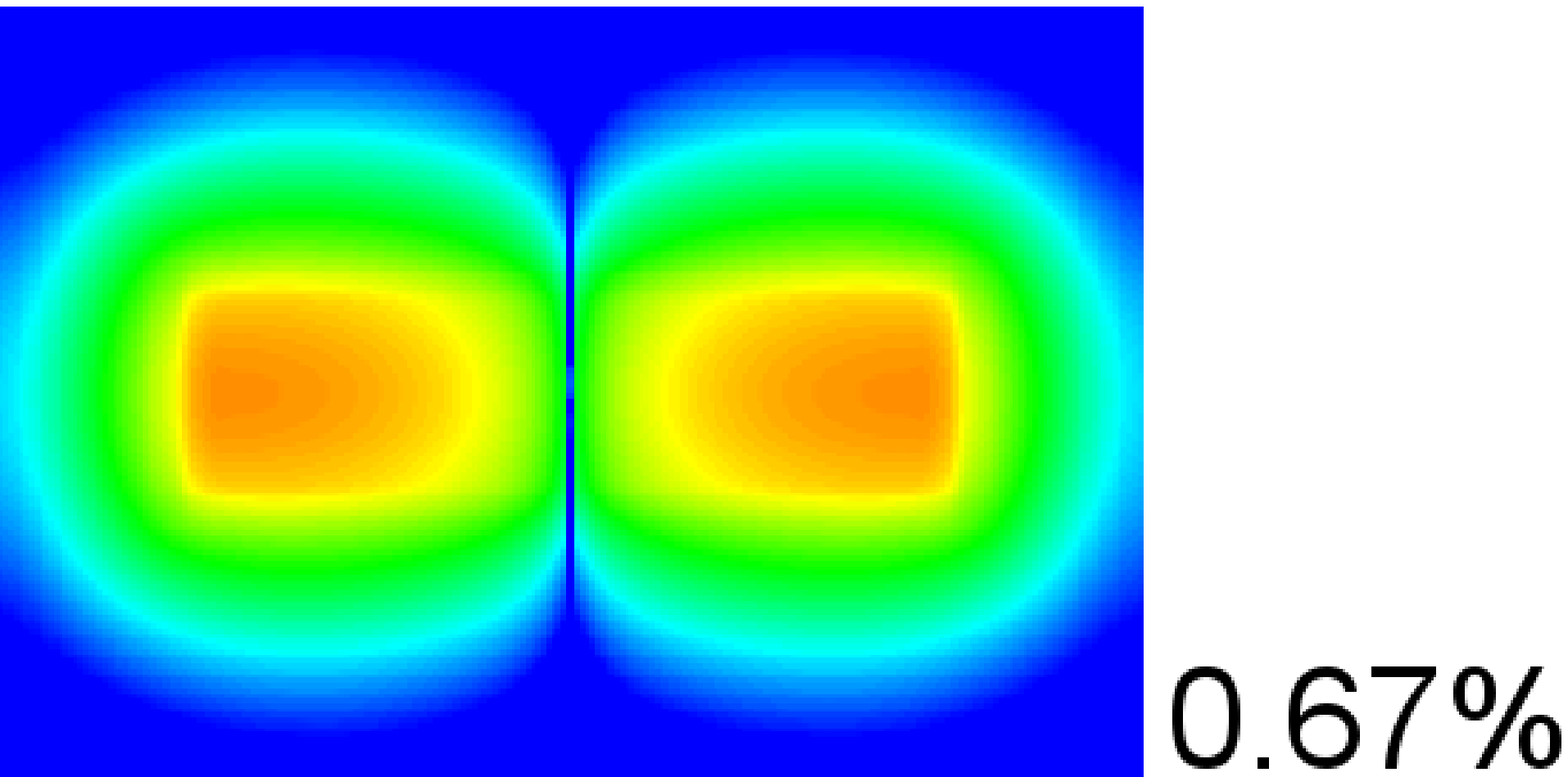} \\ \vspace{1mm}\includegraphics[width=0.21\columnwidth]{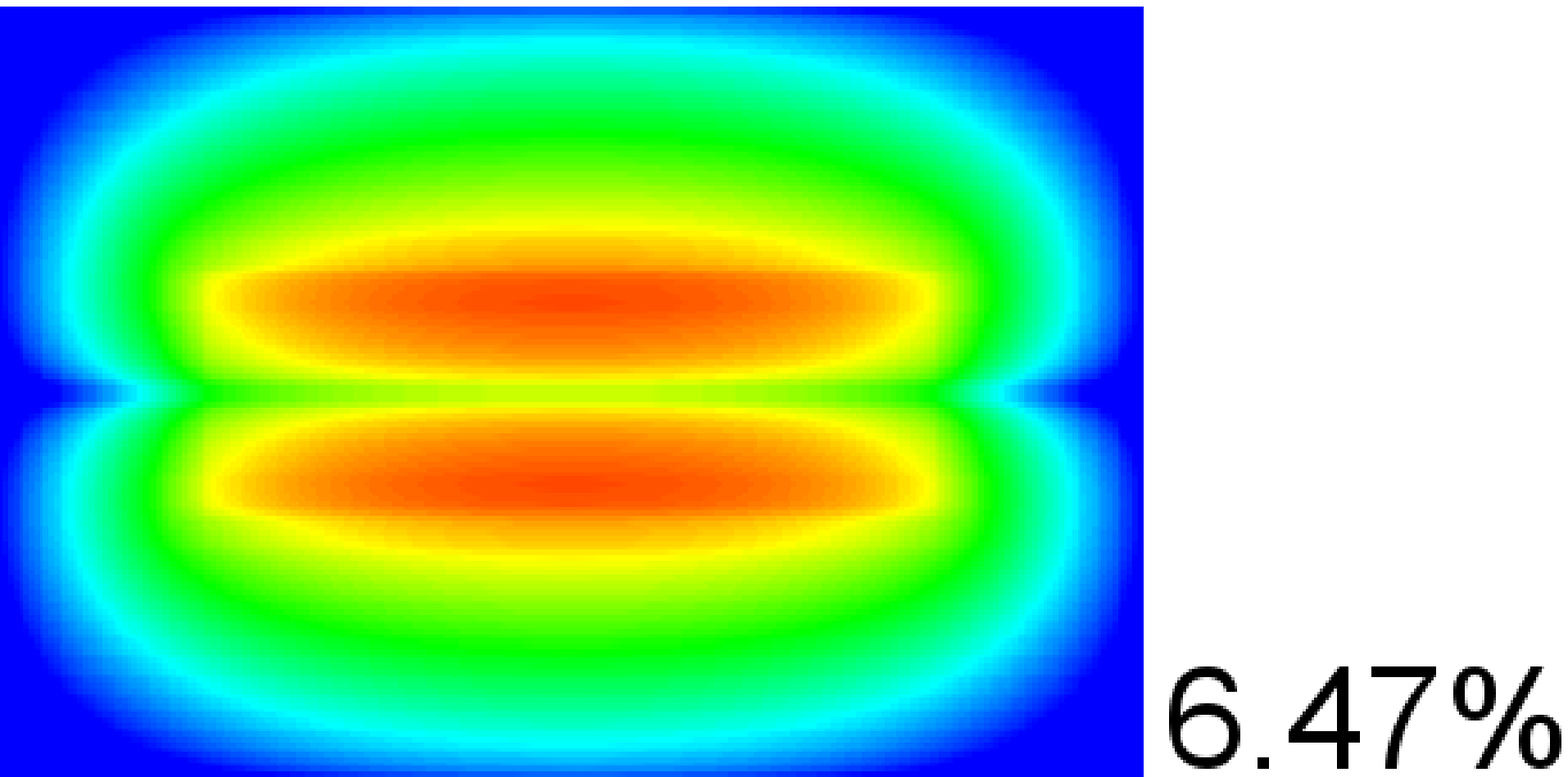}} \\
\hline \vspace{1mm}
SO
 & \begin{tabular}{l}
$|\frac{1}{2},\mp\frac{1}{2},1,\pm 1\rangle$ \\
$|\frac{1}{2},\pm\frac{1}{2},1,0 \rangle$
  \end{tabular}
 & \parbox{0.3\columnwidth}{\vspace{1mm}\includegraphics[width=0.4\columnwidth]{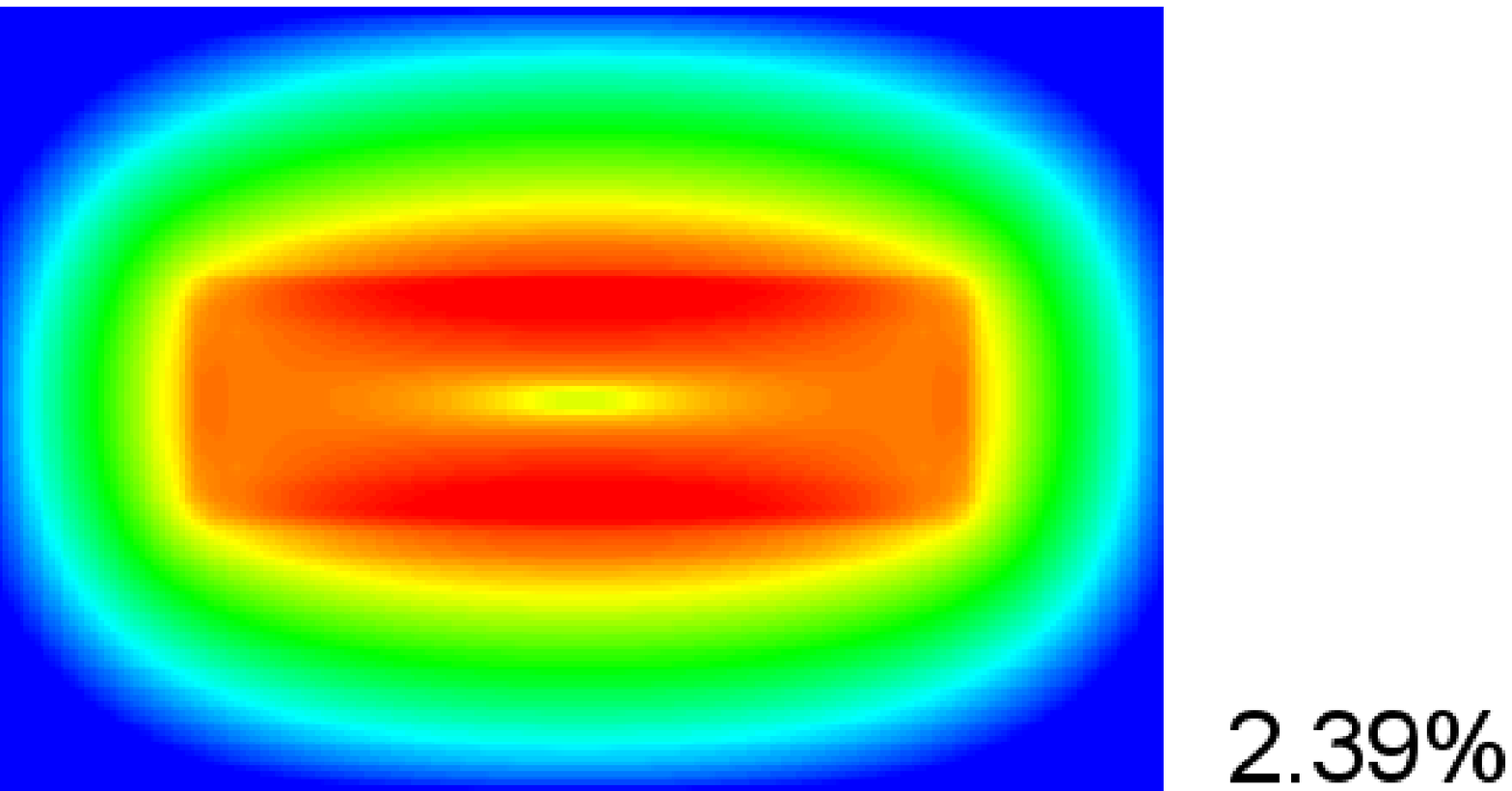}}
 & \begin{tabular}{l} $|\frac{1}{2},-\frac{1}{2}\rangle$ \\ \vspace{3mm} \\ $|\frac{1}{2},+\frac{1}{2}\rangle$ \end{tabular}
 & \parbox{0.2\columnwidth}{\vspace{1mm}\includegraphics[width=0.21\columnwidth]{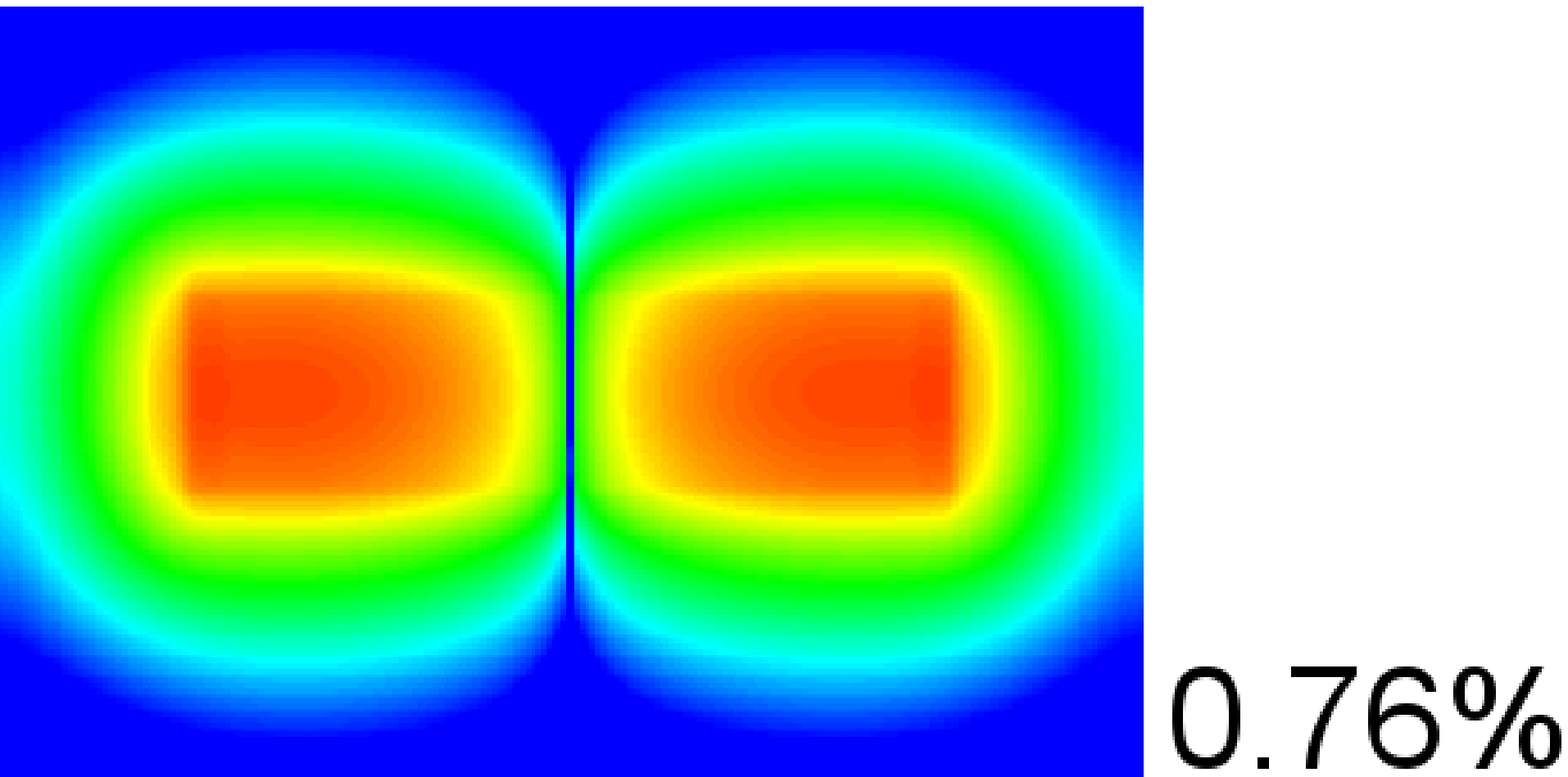} \\ \vspace{1mm}\includegraphics[width=0.21\columnwidth]{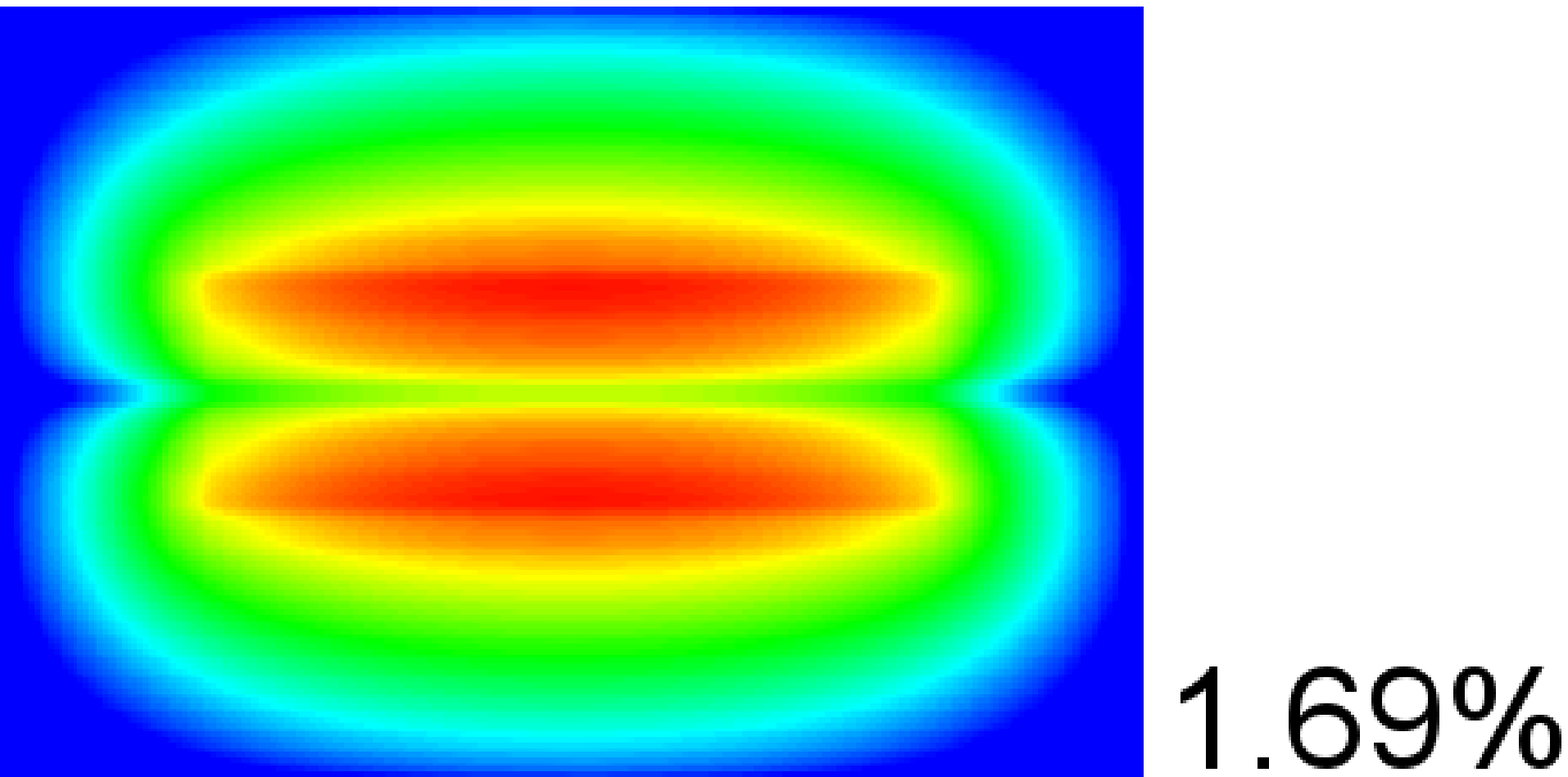}}
 & \parbox{0.2\columnwidth}{\vspace{1mm}\includegraphics[width=0.21\columnwidth]{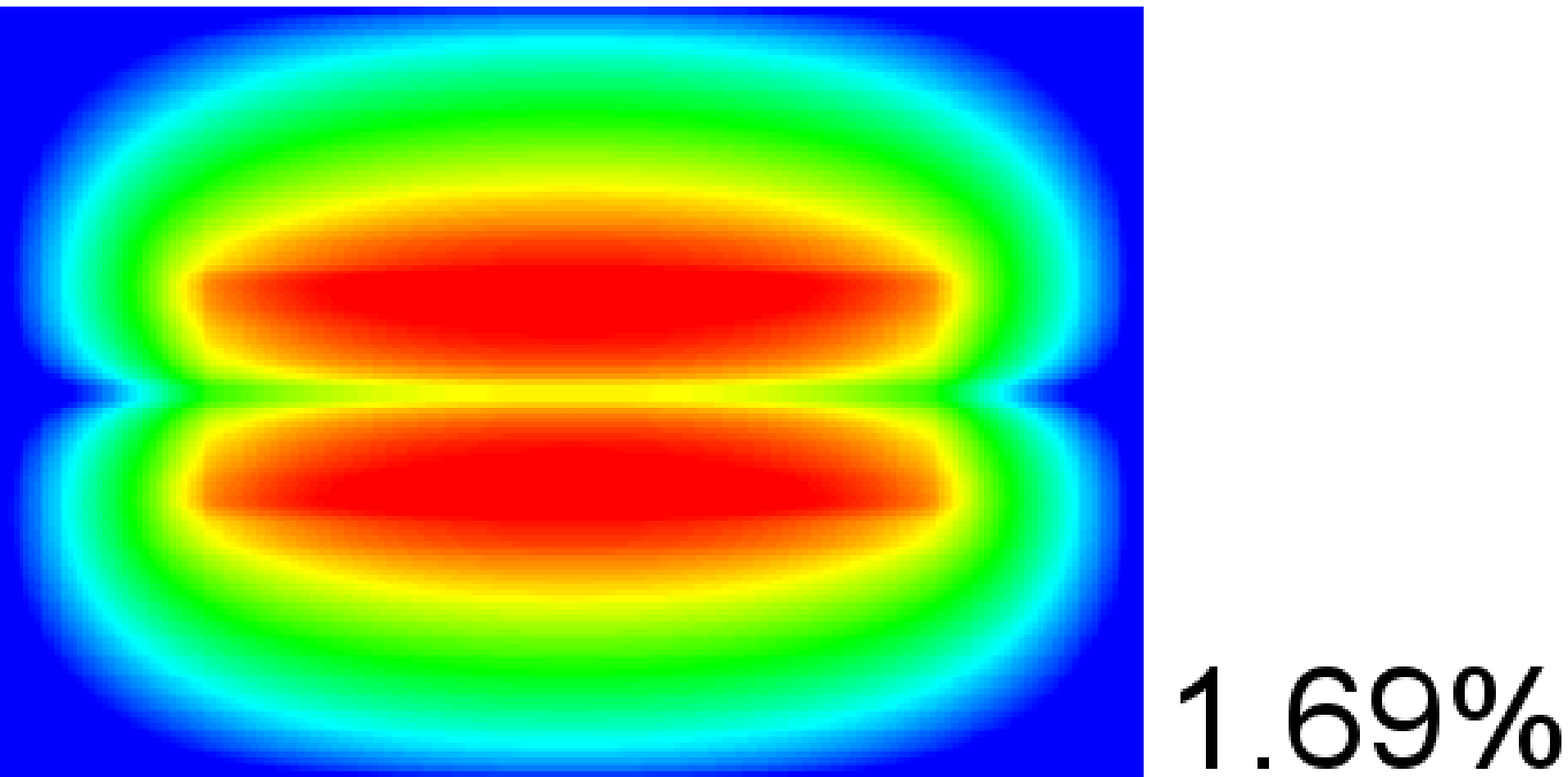} \\ \vspace{1mm}\includegraphics[width=0.21\columnwidth]{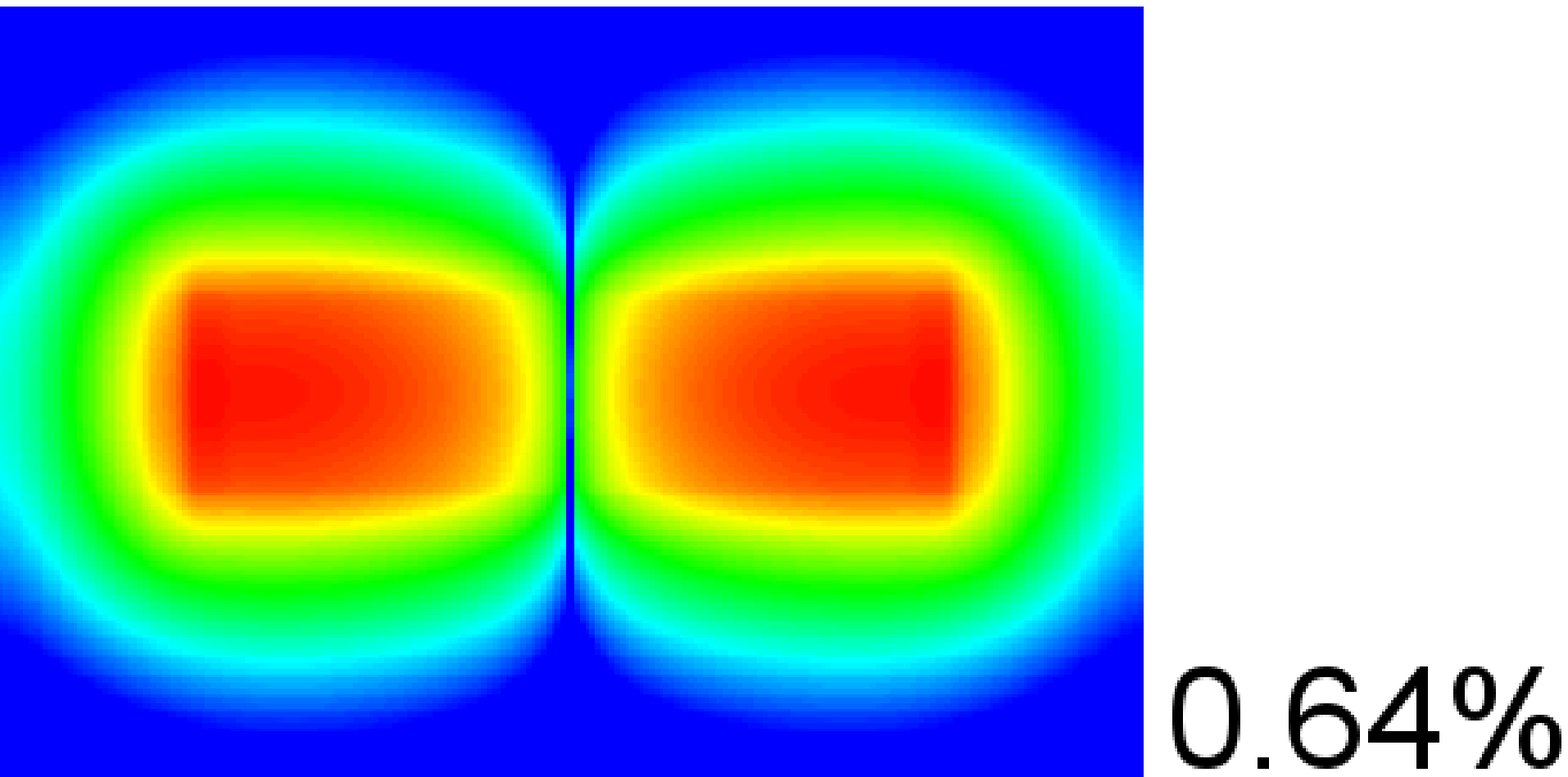}}
\end{tabular}
\end{ruledtabular}
\end{table*}

\begin{table*}[p]
\caption{Combinations of $J^{\text{Bloch}}$ and $L^{\text{env}}$ leading to $|F,F_z\rangle=|\frac{3}{2},\pm\frac{3}{2}\rangle$, for each of the different components $i$ of $\Psi_v^i\left({\bf r}\right)$, of a QD having height of $6$~nm and radius of $11$~nm. The color scale is different for each plot.\label{table:F32}}
\begin{ruledtabular}
\begin{tabular}{cccccc}
\multirow{3}{*}{$i$} & \multirow{3}{*}{$|J^{\text{Bloch}},J^{\text{Bloch}}_z,L^{\text{env}},L^{\text{env}}_z\rangle$} & Cut of $|\Psi_v^i\left({\bf r}\right)|^2$ at $B=0$~T & \multicolumn{3}{c}{Cut of $|\Psi_v^i\left({\bf r}\right)|^2$ at $B=2$~T} \\
 & & $|F=\frac{3}{2},F_z=\pm\frac{3}{2}\rangle$ & $|J^{\text{Bloch}},J^{\text{Bloch}}_z\rangle$ & $|F=\frac{3}{2},F_z=+\frac{3}{2}\rangle$ & $|F=\frac{3}{2},F_z=-\frac{3}{2}\rangle$ \vspace{1mm} \\
\hline\hline \vspace{1mm}
CB
 & \begin{tabular}{l}
$|\frac{1}{2},\pm\frac{1}{2},1,\pm 1\rangle$
\end{tabular}
 & \parbox{0.3\columnwidth}{\vspace{1mm}\includegraphics[width=0.4\columnwidth]{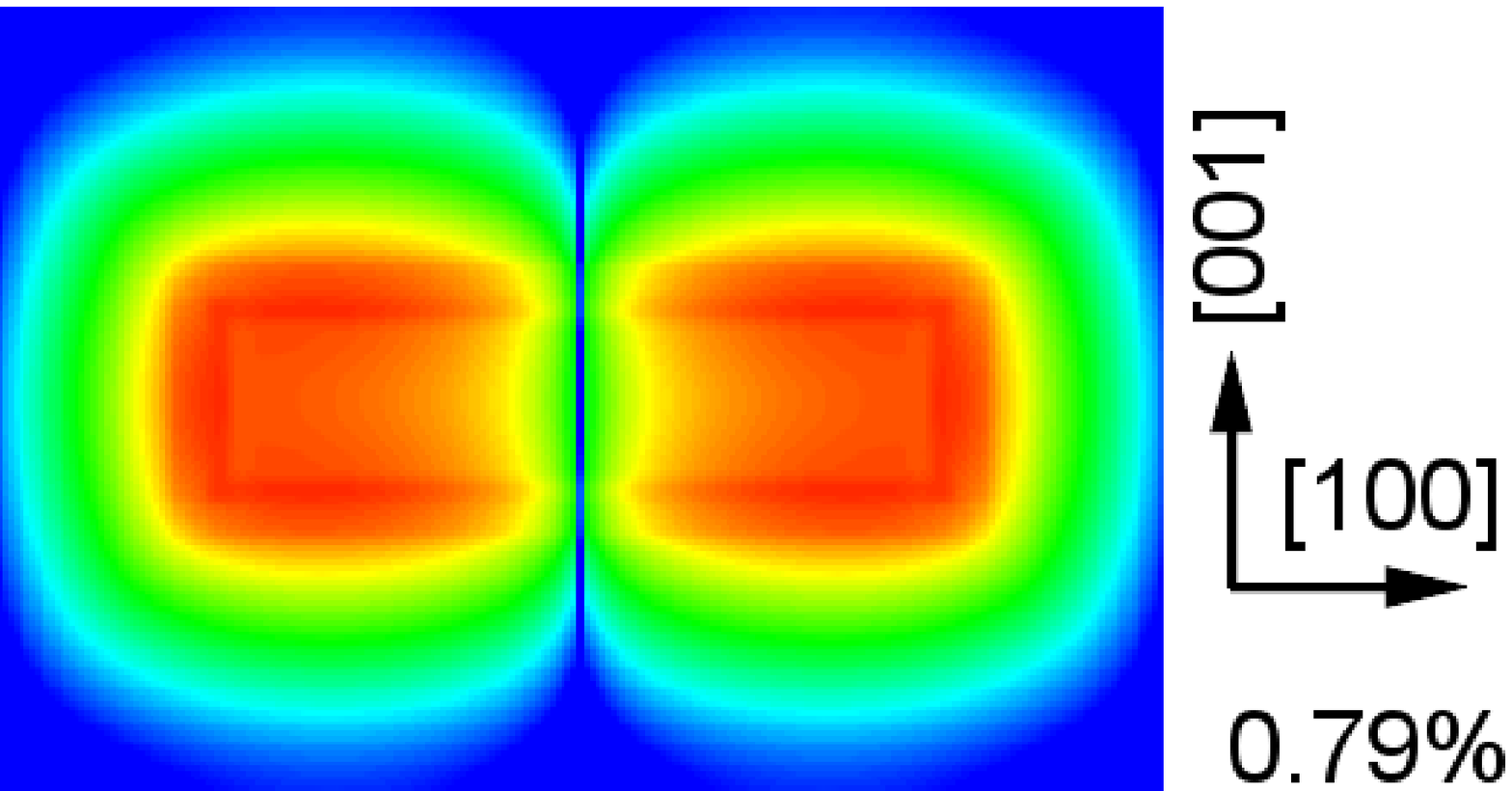}}
 & \begin{tabular}{l} $|\frac{1}{2},-\frac{1}{2}\rangle$ \\ \vspace{3mm} \\ $|\frac{1}{2},+\frac{1}{2}\rangle$ \end{tabular}
 & \parbox{0.2\columnwidth}{\vspace{1mm}\includegraphics[width=0.21\columnwidth]{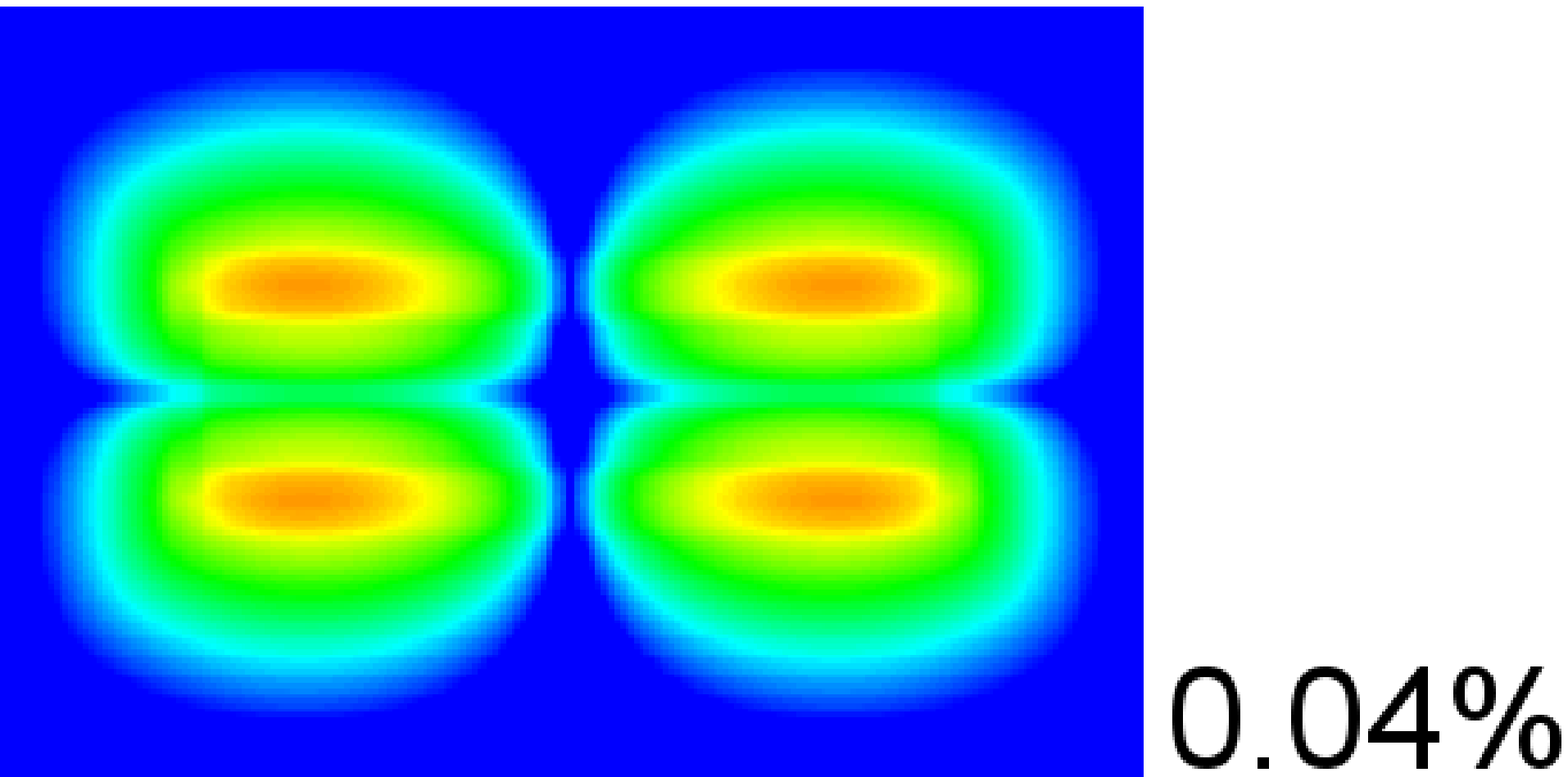} \\ \vspace{1mm}\includegraphics[width=0.21\columnwidth]{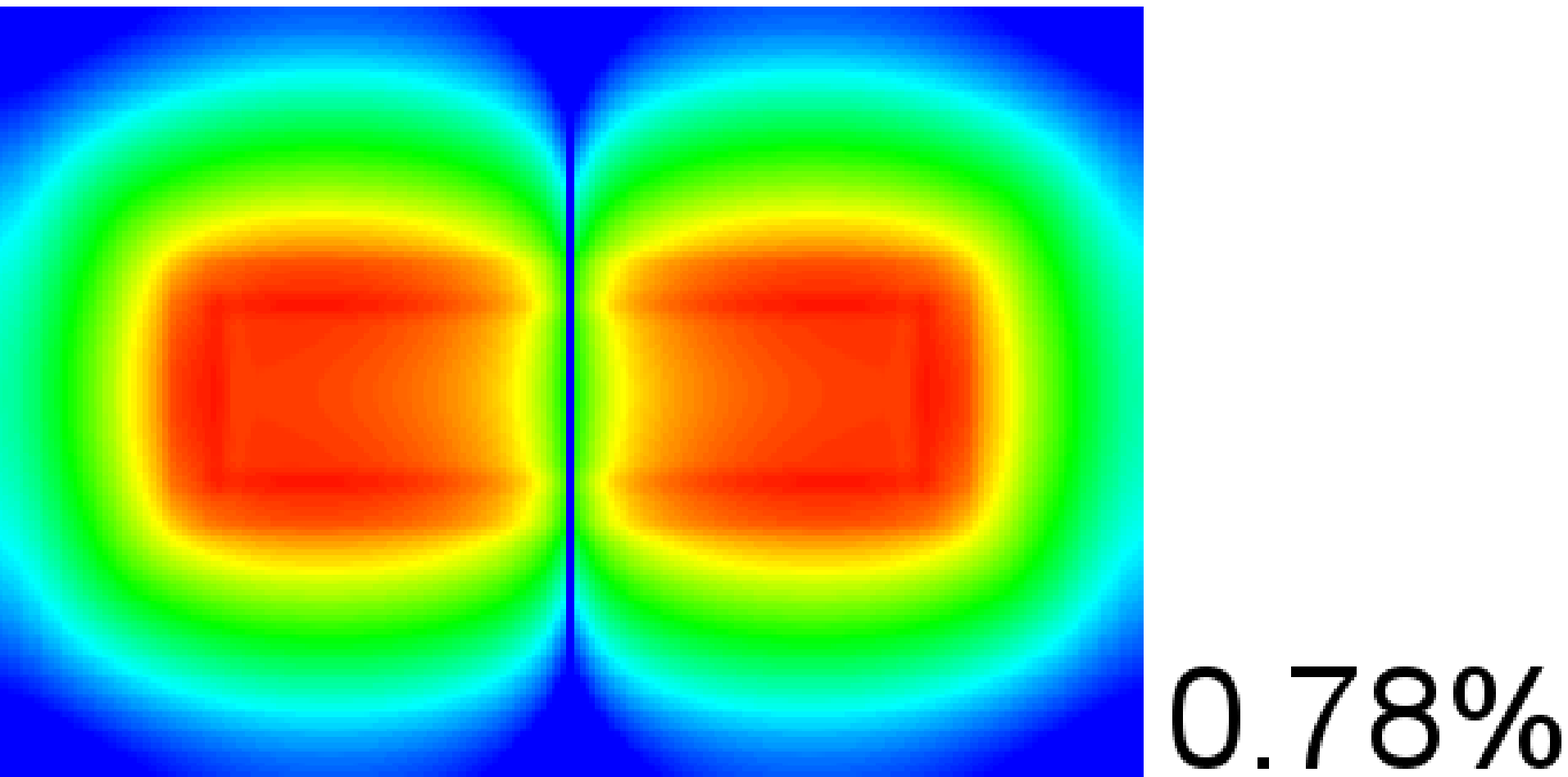}}
 & \parbox{0.2\columnwidth}{\vspace{1mm}\includegraphics[width=0.21\columnwidth]{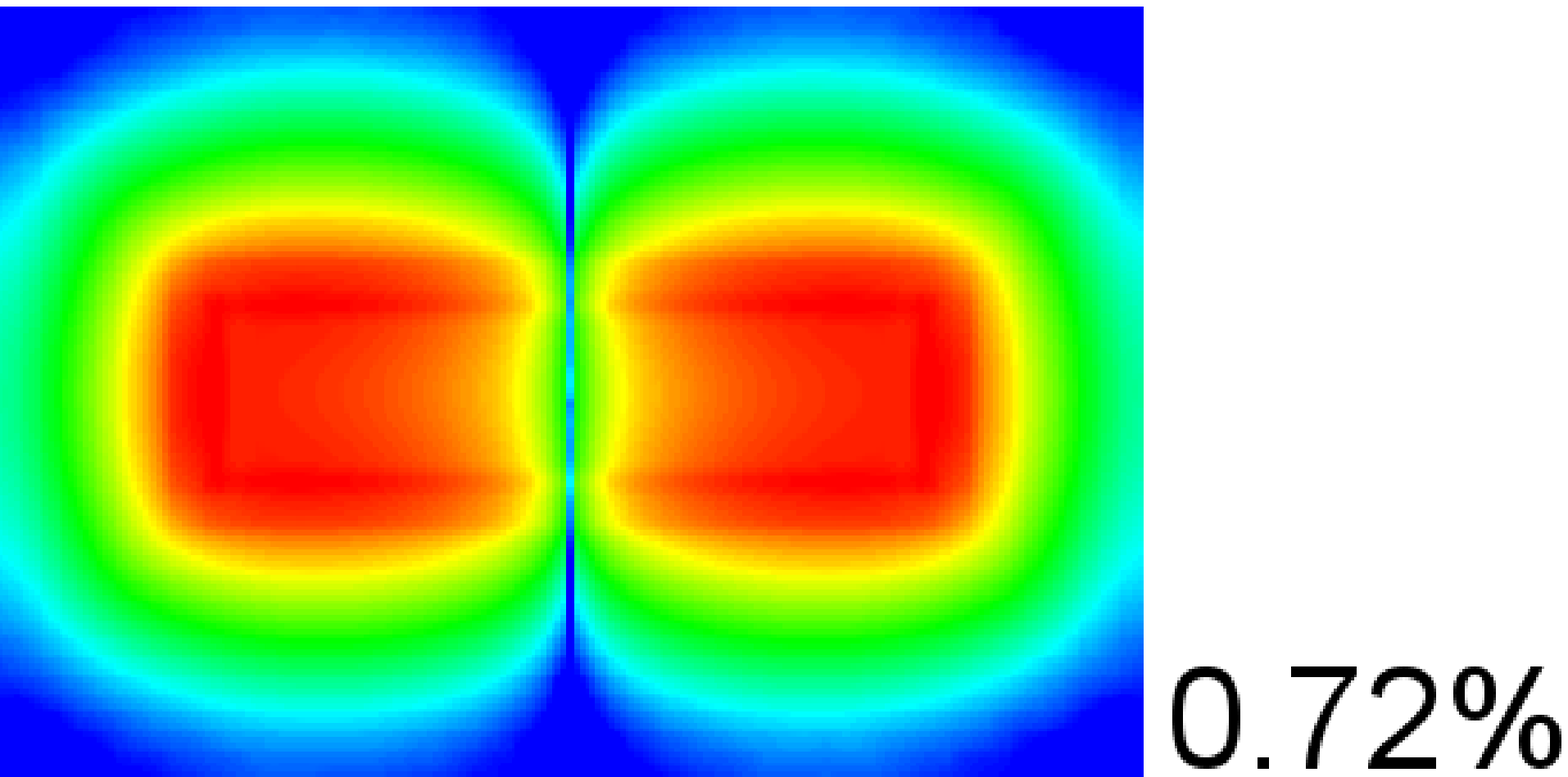} \\ \vspace{1mm}\includegraphics[width=0.21\columnwidth]{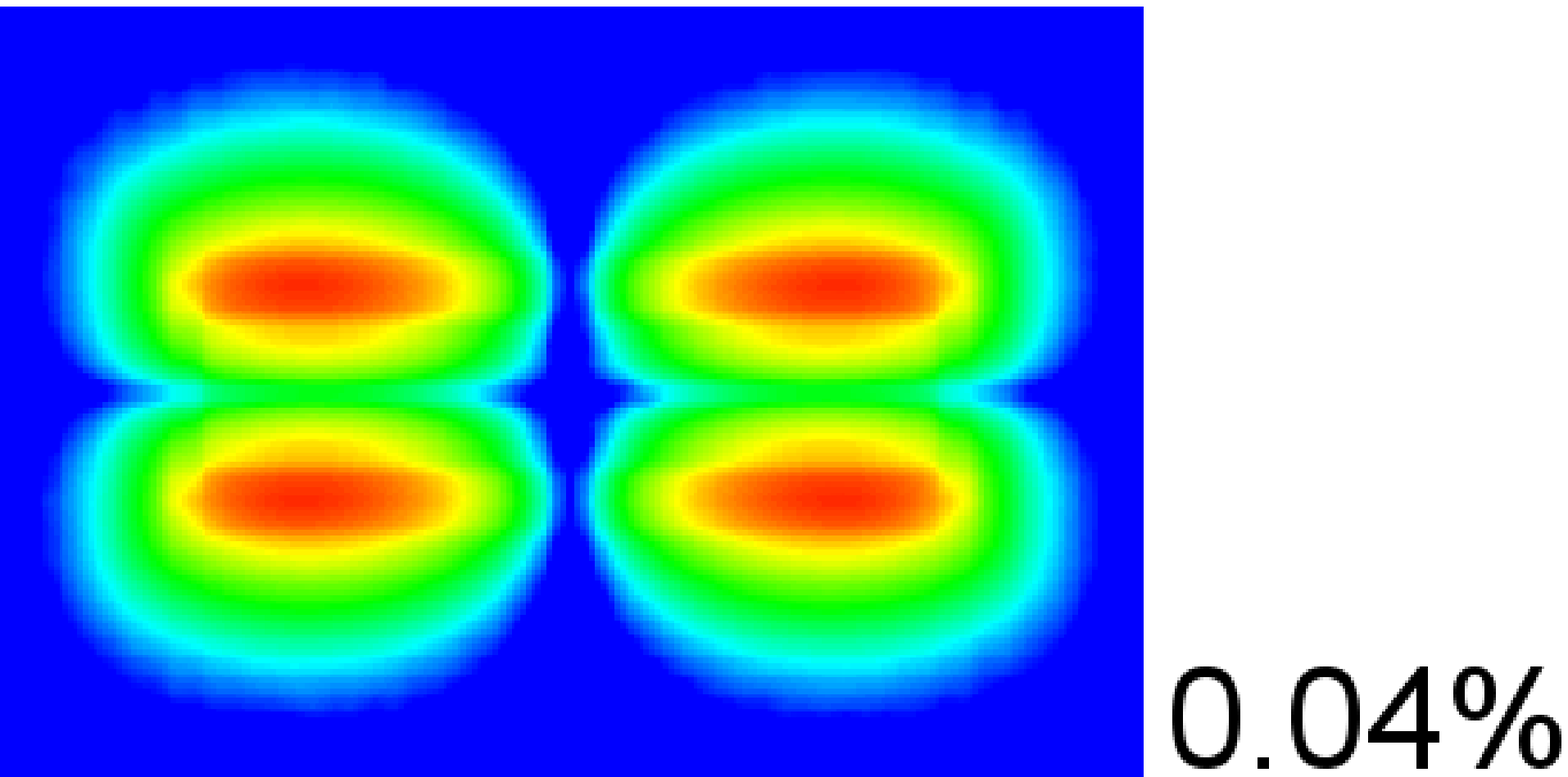}} \\
\hline \vspace{1mm}
HH 
 &\begin{tabular}{l}
$|\frac{3}{2},\pm\frac{3}{2},0,0\rangle$ \\
$|\frac{3}{2},\pm\frac{3}{2},2,0\rangle$
\end{tabular}
 & \parbox{0.3\columnwidth}{\vspace{1mm}\includegraphics[width=0.4\columnwidth]{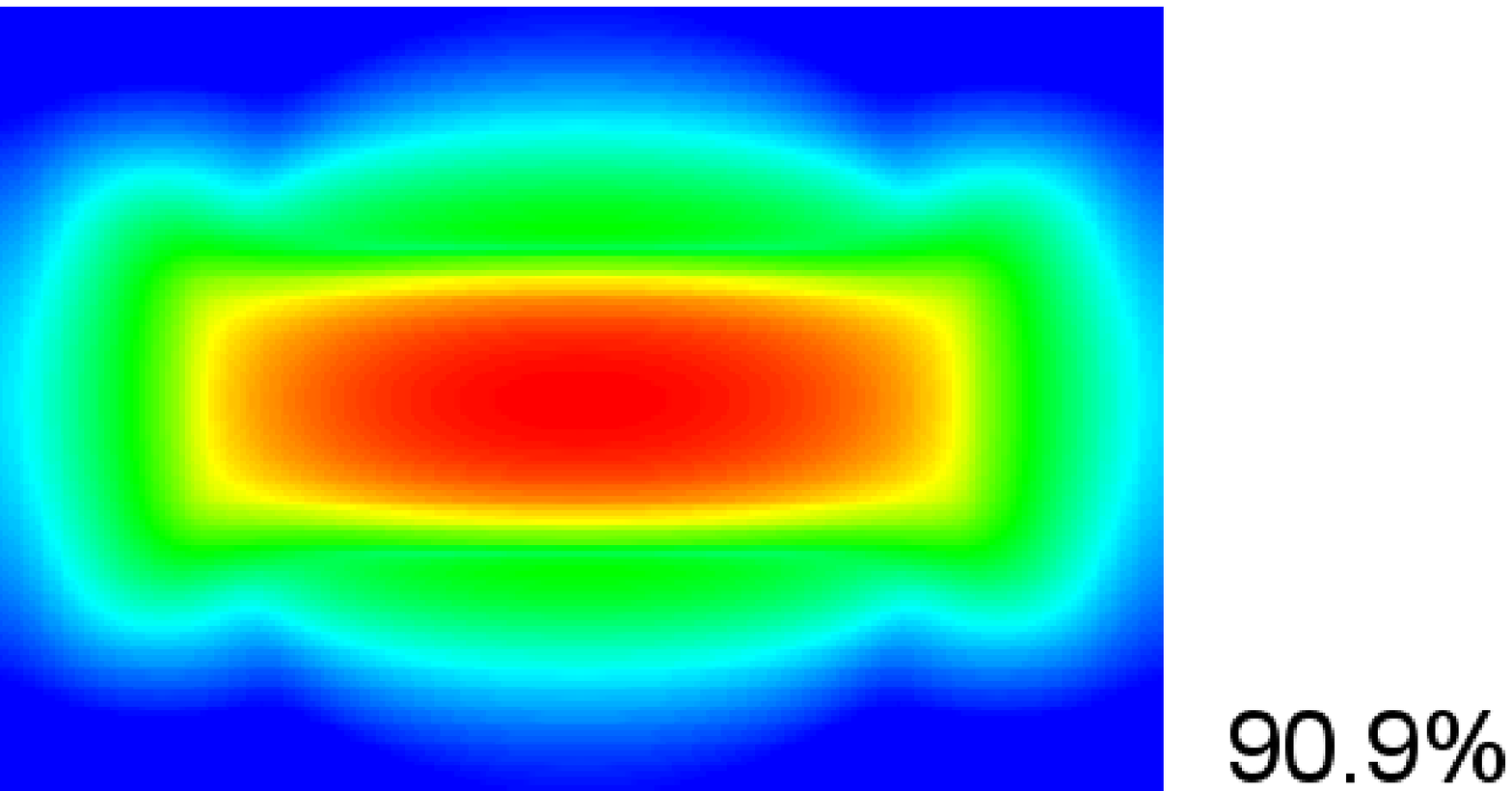}}
 & \begin{tabular}{l} $|\frac{3}{2},+\frac{3}{2}\rangle$ \\ \vspace{3mm} \\ $|\frac{3}{2},-\frac{3}{2}\rangle$ \end{tabular}
 & \parbox{0.2\columnwidth}{\vspace{1mm}\includegraphics[width=0.21\columnwidth]{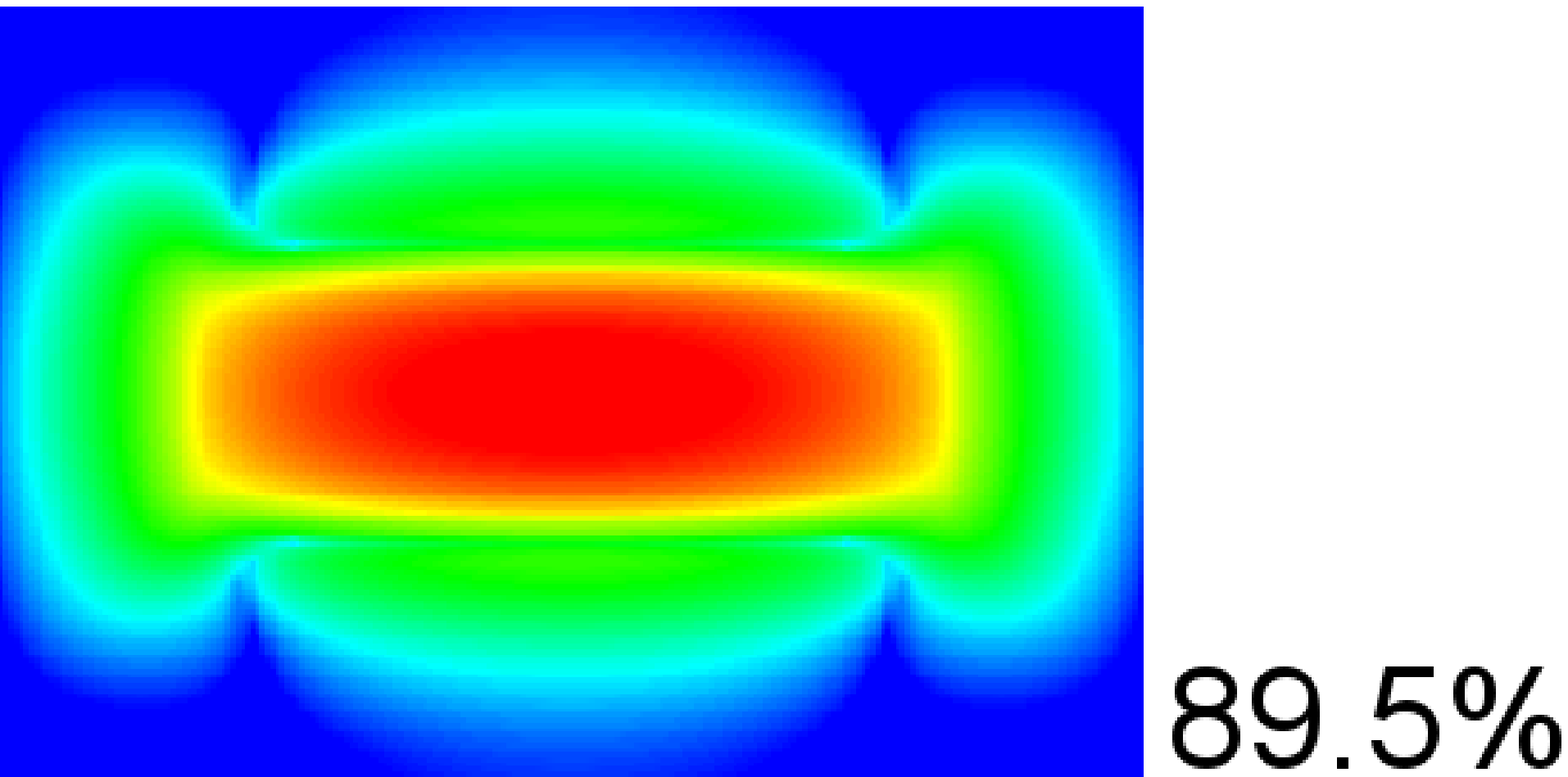} \\ \vspace{1mm}\includegraphics[width=0.21\columnwidth]{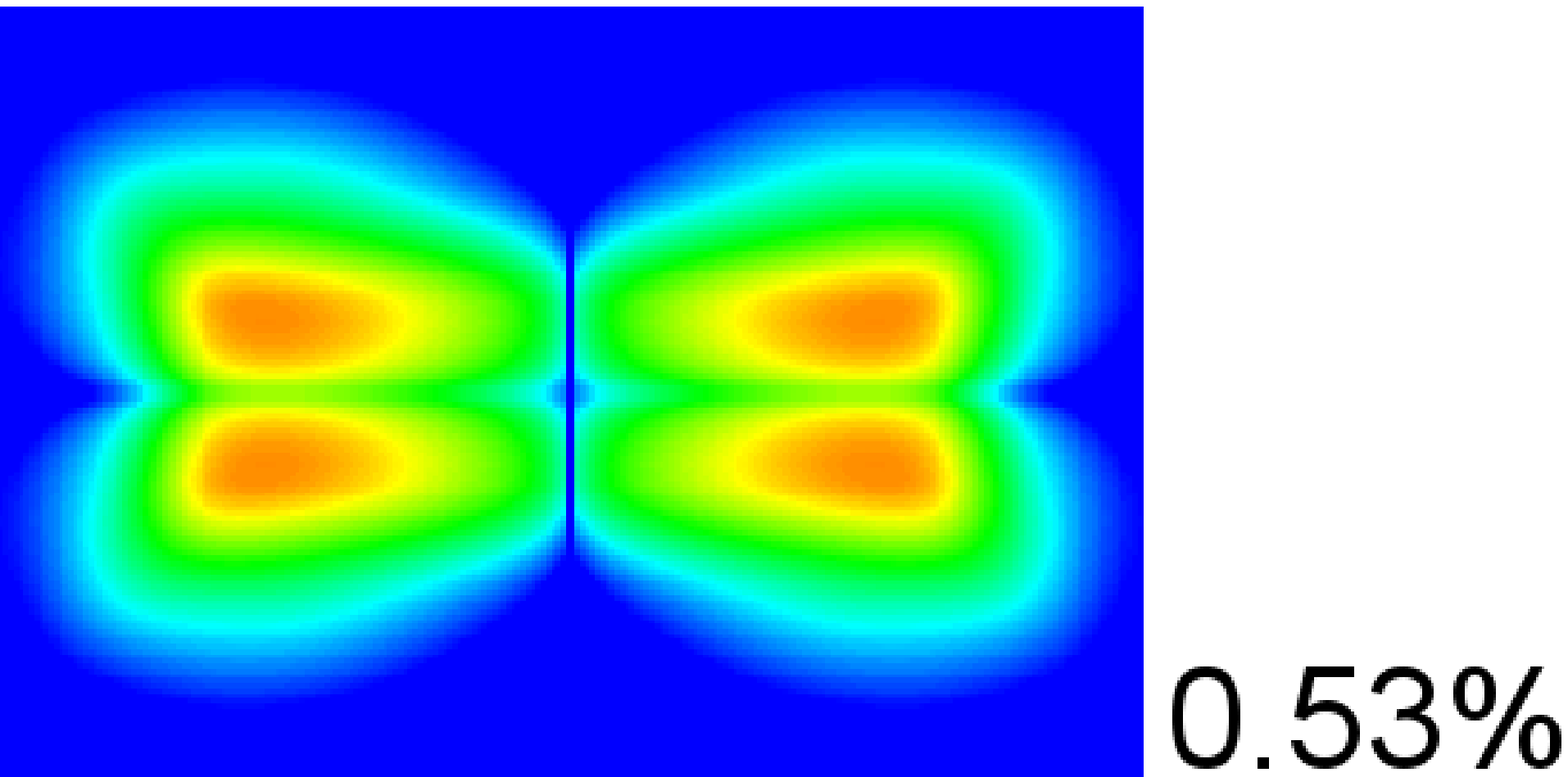}}
 & \parbox{0.2\columnwidth}{\vspace{1mm}\includegraphics[width=0.21\columnwidth]{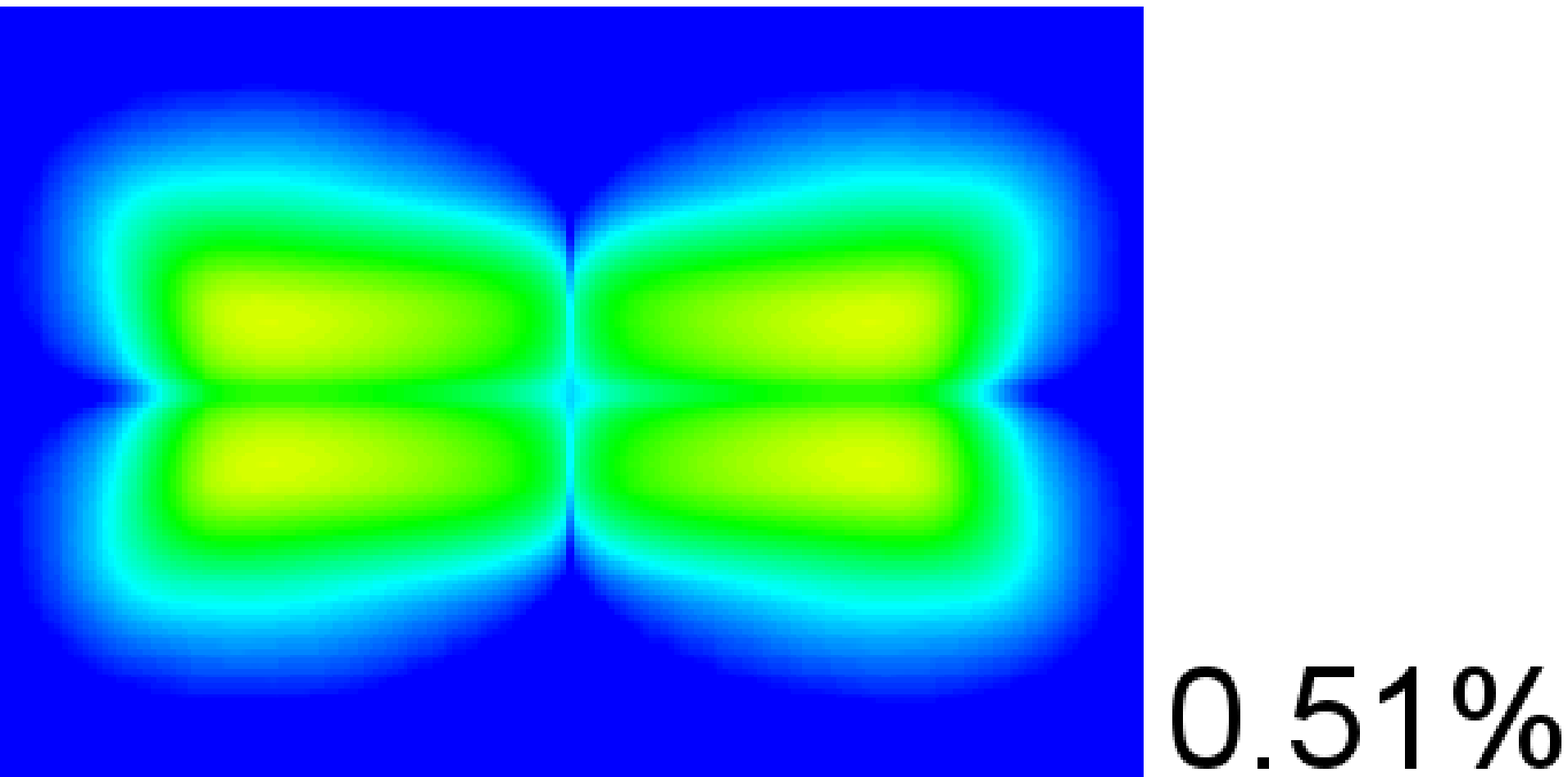} \\ \vspace{1mm}\includegraphics[width=0.21\columnwidth]{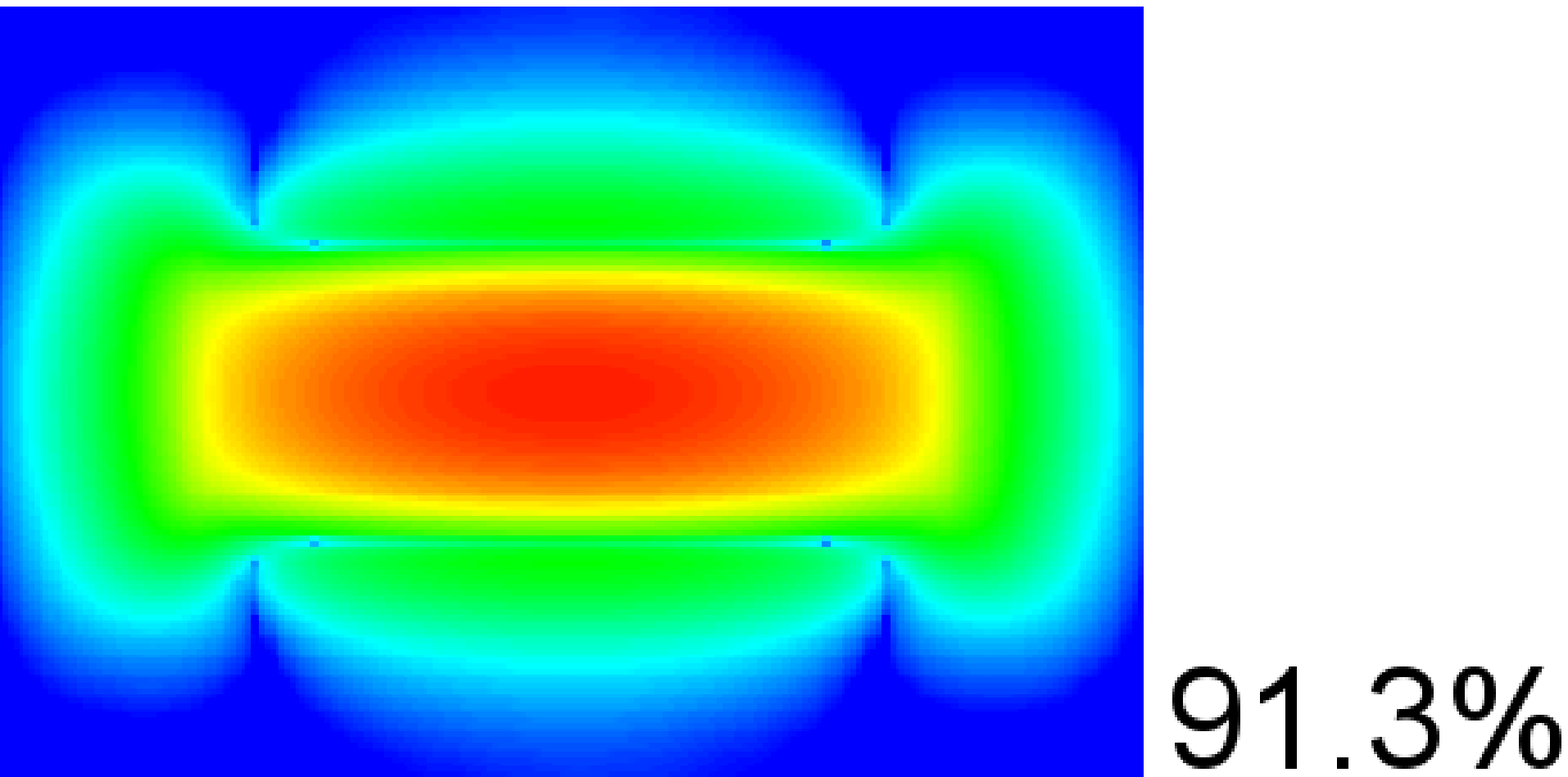}} \\
\hline \vspace{1mm}
LH
 & \begin{tabular}{l}
$|\frac{3}{2},\mp\frac{1}{2},2,\pm 2\rangle$ \\
$|\frac{3}{2},\pm\frac{1}{2},2,\pm 1\rangle$
\end{tabular}
 & \parbox{0.3\columnwidth}{\vspace{1mm}\includegraphics[width=0.4\columnwidth]{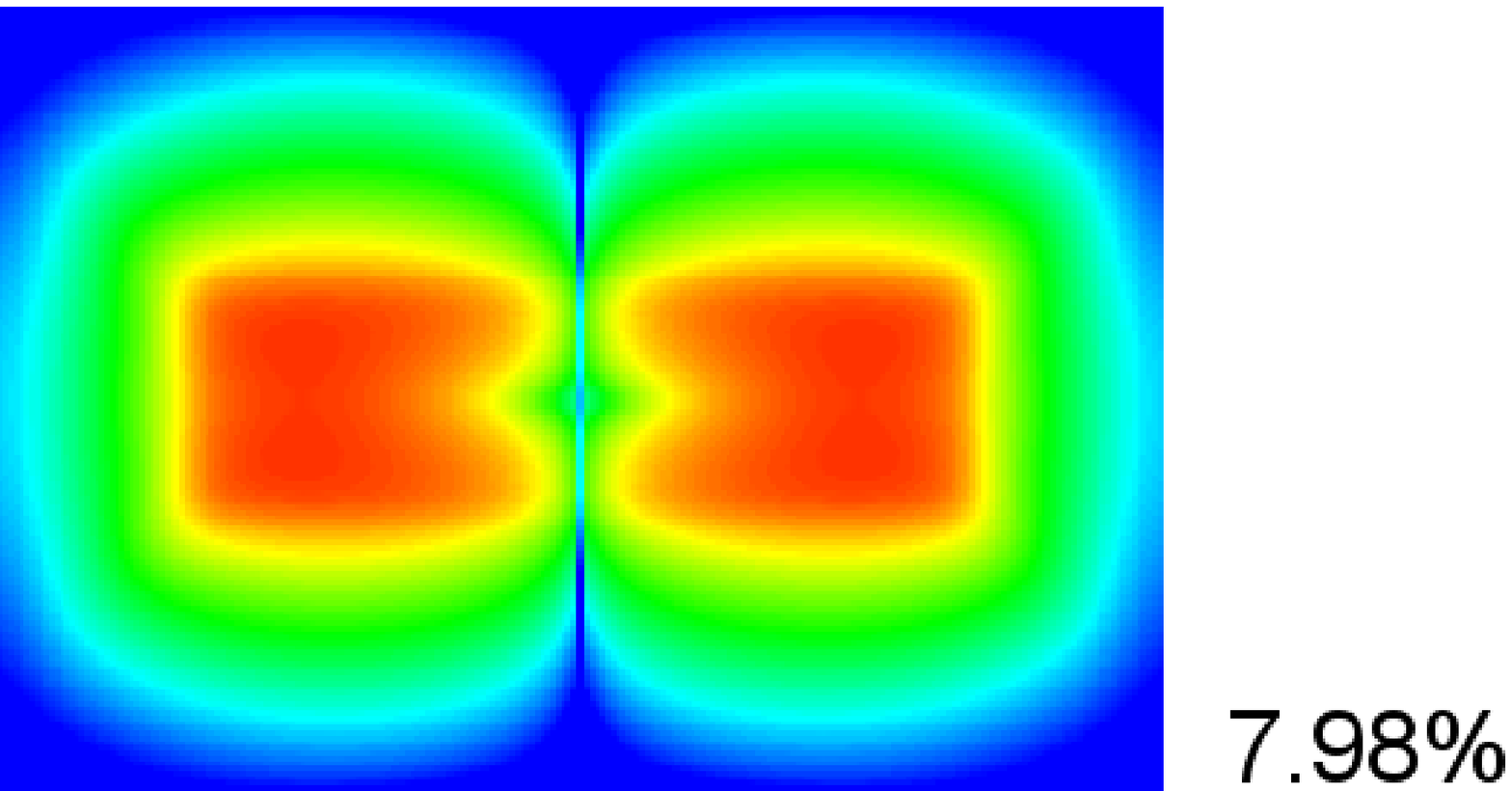}}
 & \begin{tabular}{l} $|\frac{3}{2},+\frac{1}{2}\rangle$ \\ \vspace{3mm} \\ $|\frac{3}{2},-\frac{1}{2}\rangle$ \end{tabular}
 & \parbox{0.2\columnwidth}{\vspace{1mm}\includegraphics[width=0.21\columnwidth]{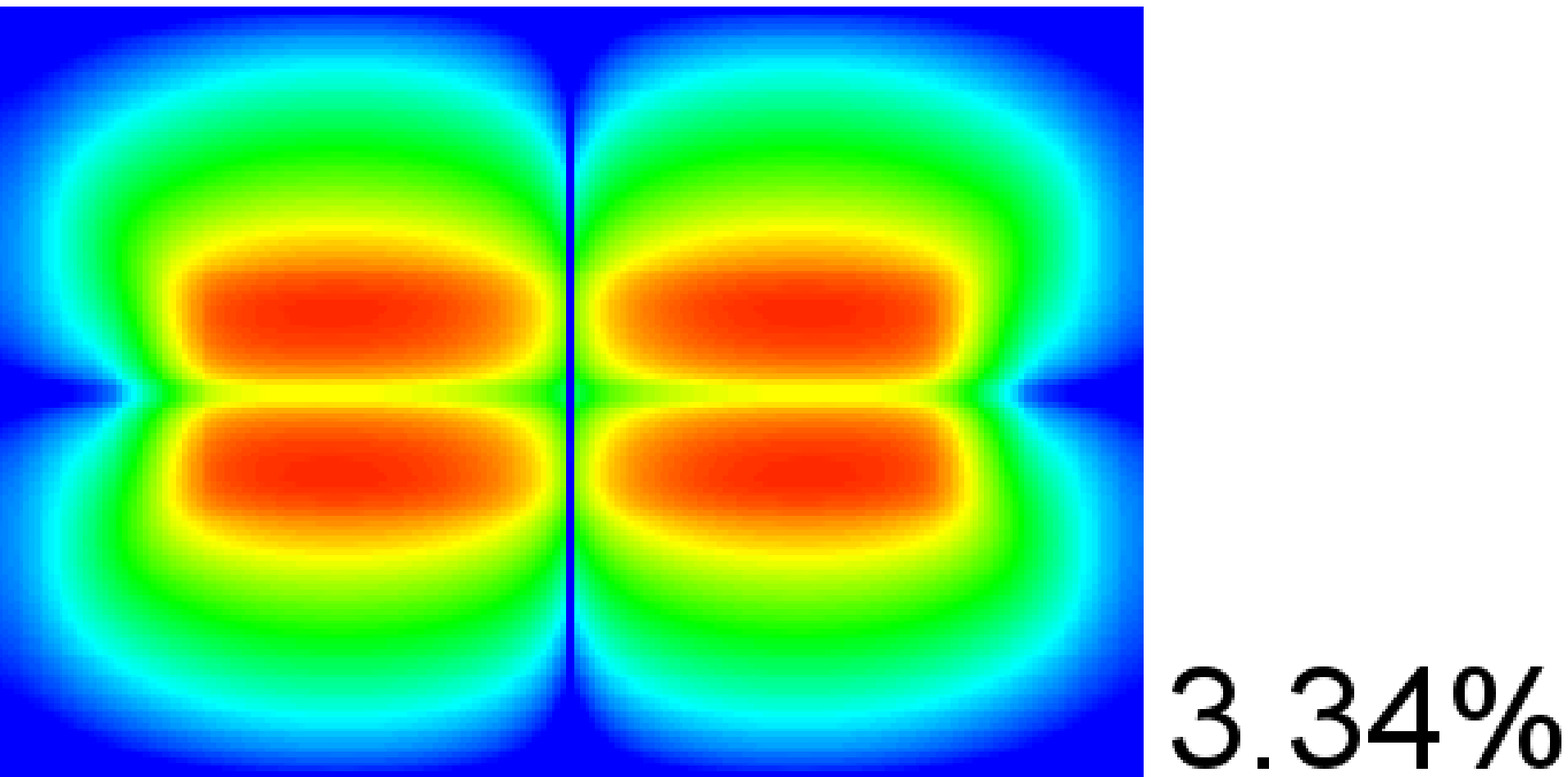} \\ \vspace{1mm}\includegraphics[width=0.21\columnwidth]{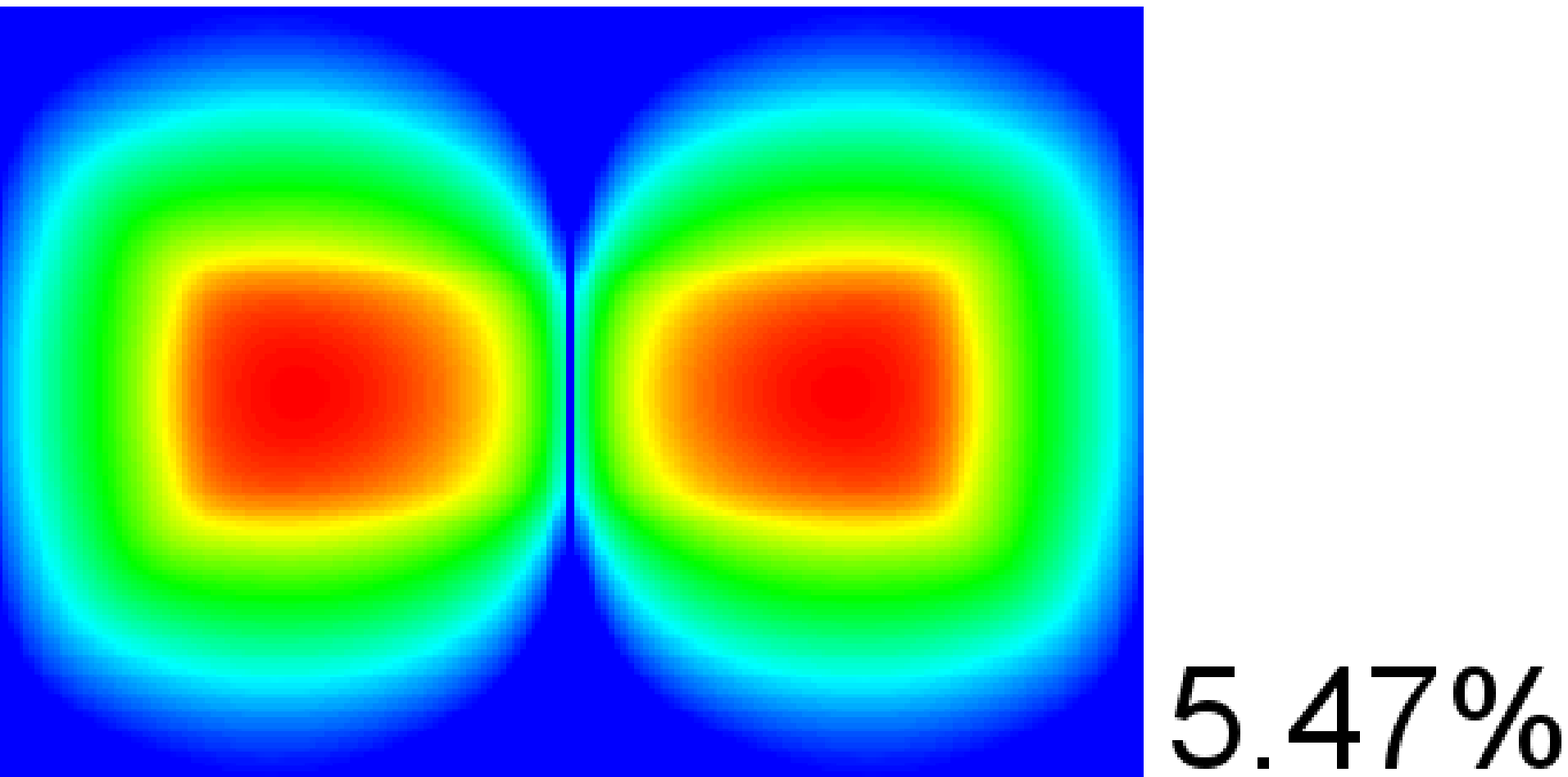}}
 & \parbox{0.2\columnwidth}{\vspace{1mm}\includegraphics[width=0.21\columnwidth]{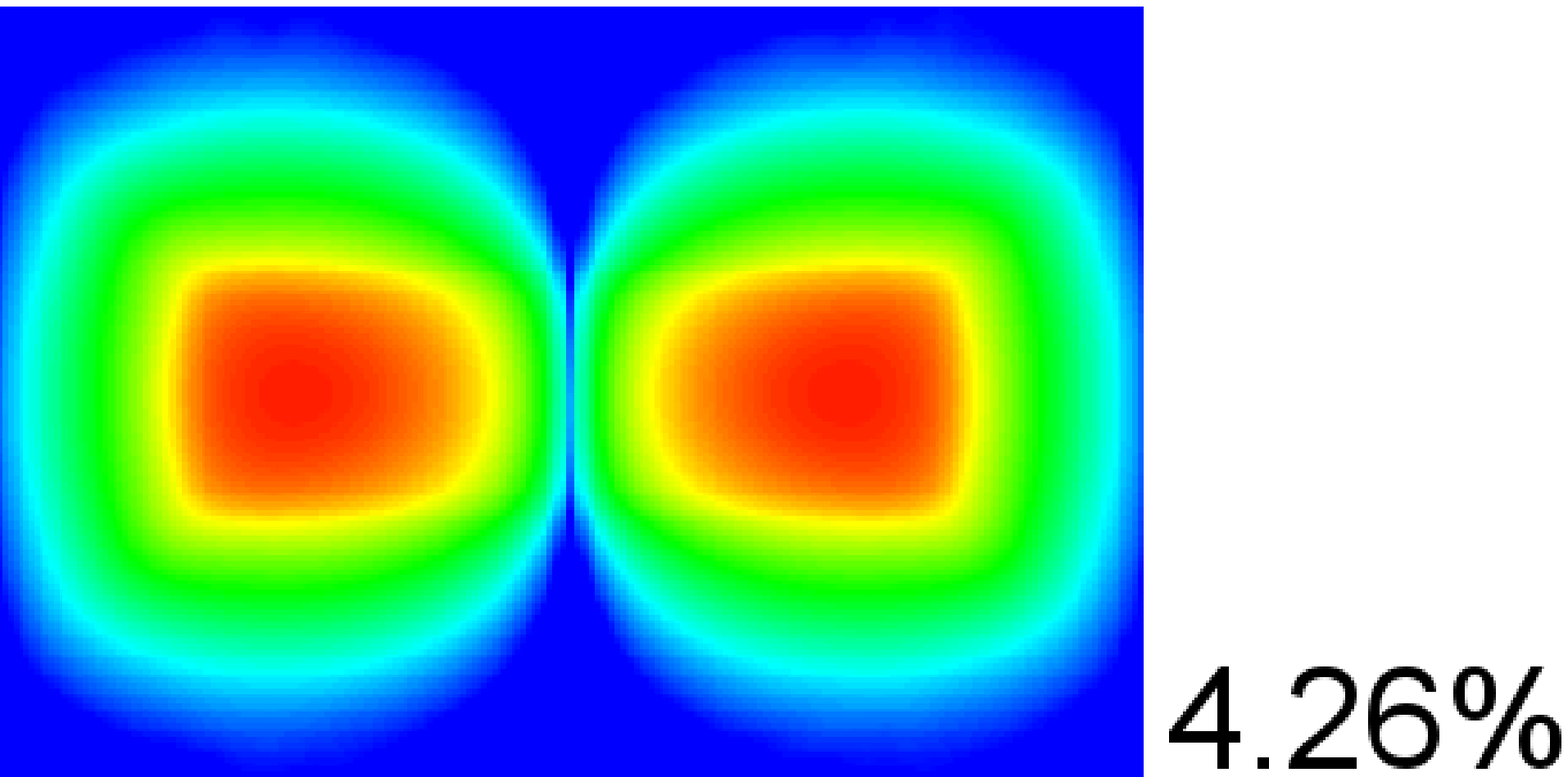} \\ \vspace{1mm}\includegraphics[width=0.21\columnwidth]{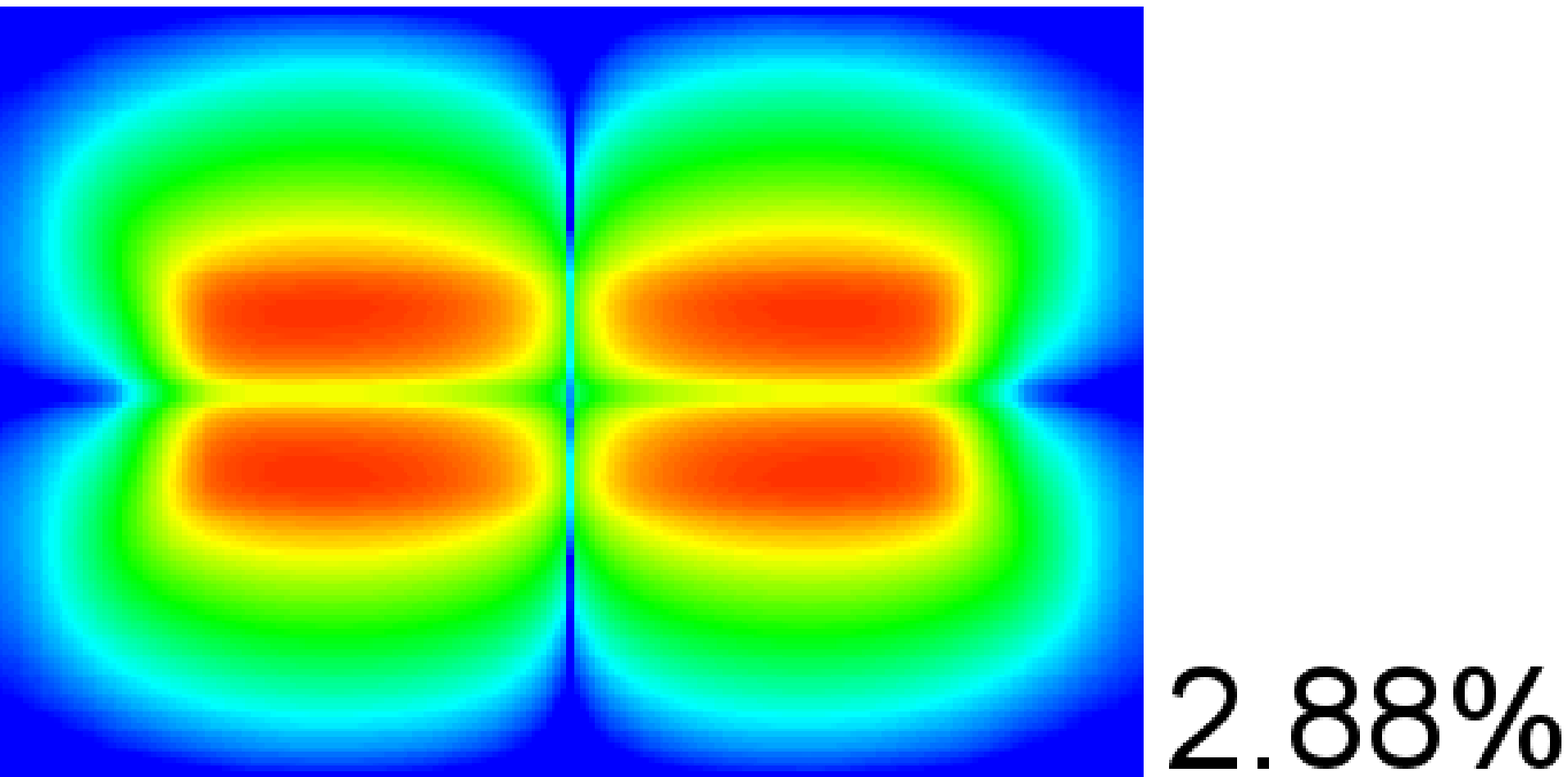}} \\
\hline \vspace{1mm}
SO
 & \begin{tabular}{l}
$|\frac{1}{2},\mp\frac{1}{2},2,\pm 2\rangle$ \\
$|\frac{1}{2},\pm\frac{1}{2},2,\pm 1\rangle$
\end{tabular}
 & \parbox{0.3\columnwidth}{\vspace{1mm}\includegraphics[width=0.4\columnwidth]{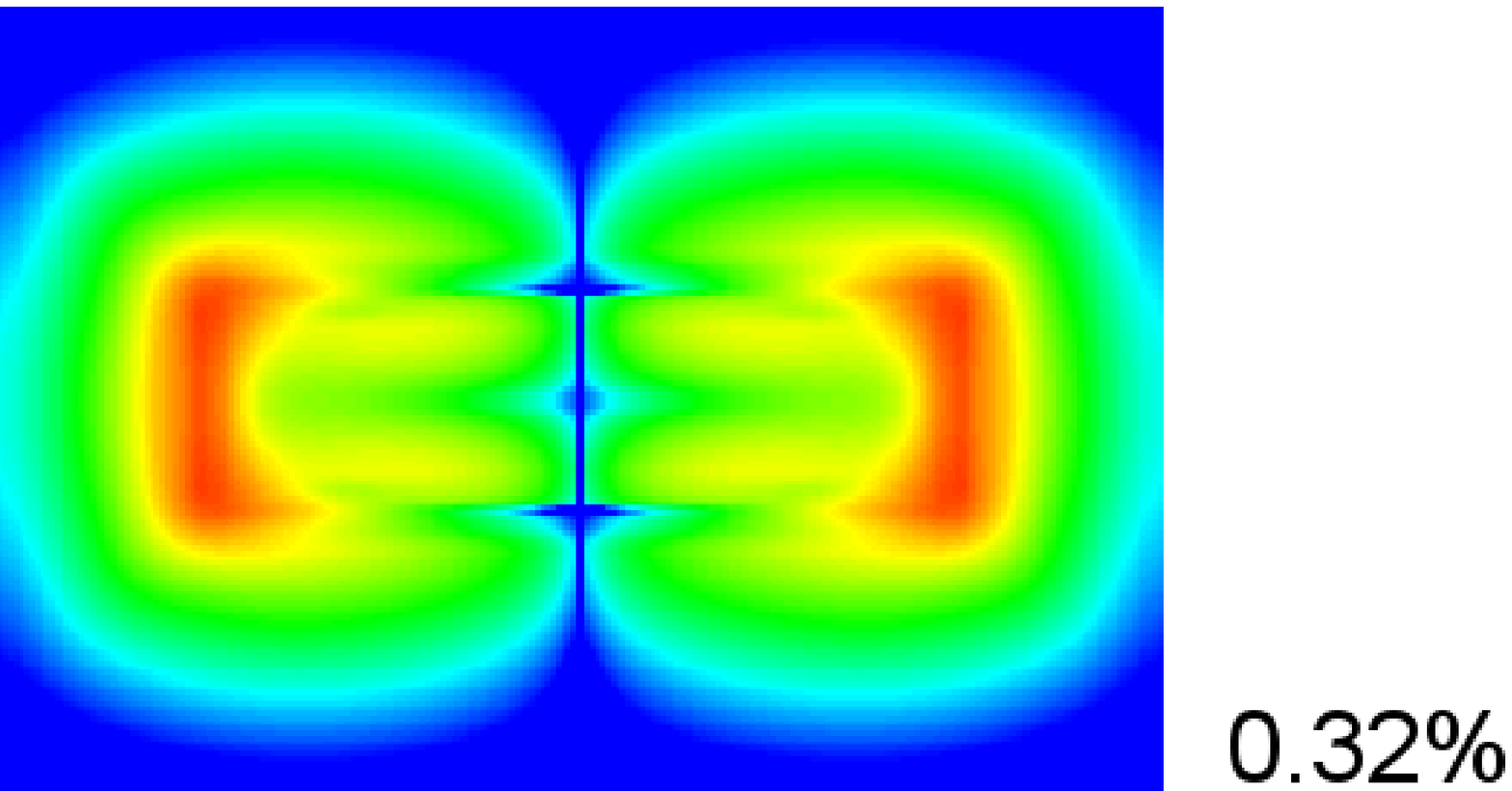}}
 & \begin{tabular}{l} $|\frac{1}{2},-\frac{1}{2}\rangle$ \\ \vspace{3mm} \\ $|\frac{1}{2},+\frac{1}{2}\rangle$ \end{tabular}
 & \parbox{0.2\columnwidth}{\vspace{1mm}\includegraphics[width=0.21\columnwidth]{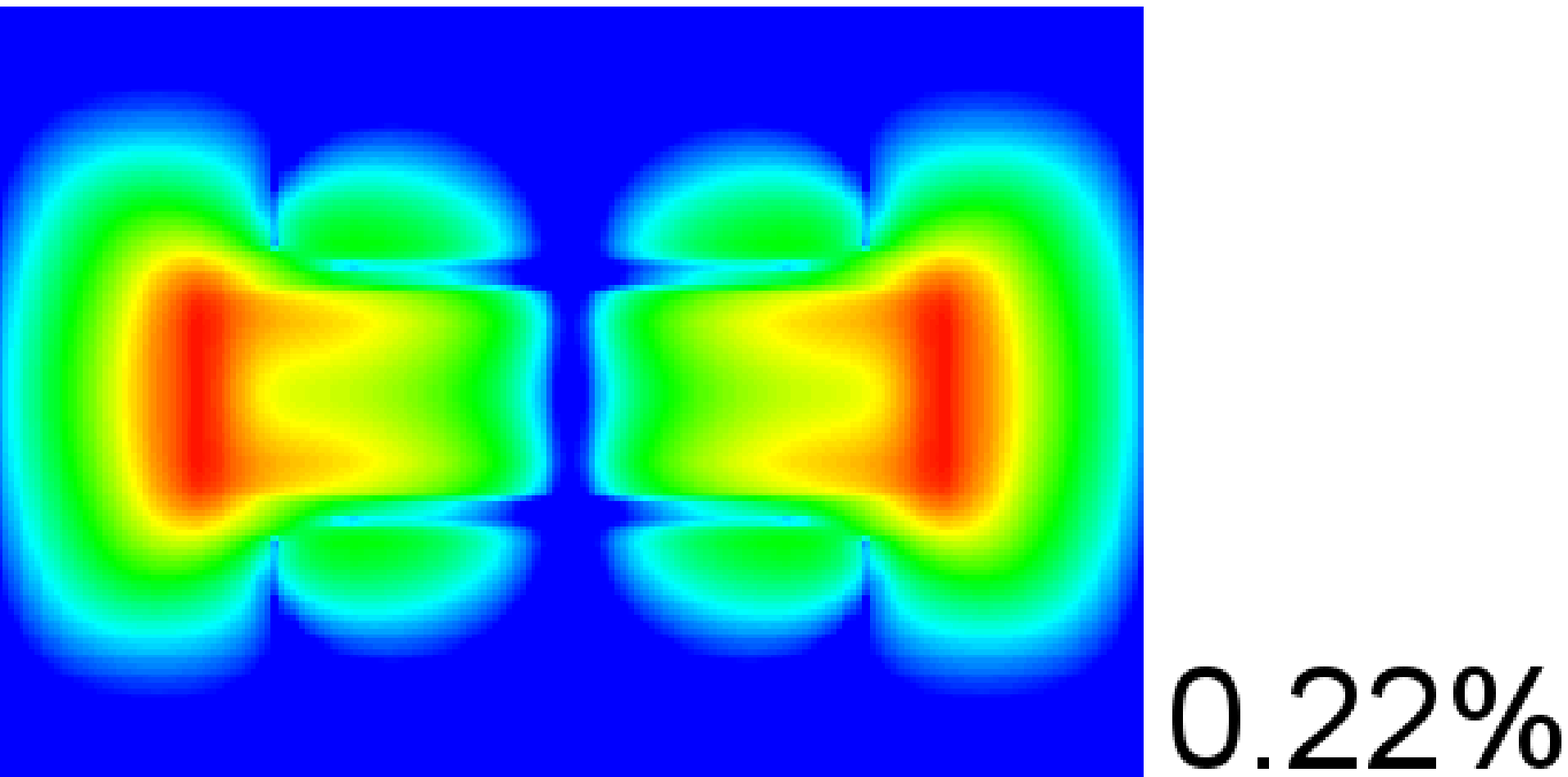} \\ \vspace{1mm}\includegraphics[width=0.21\columnwidth]{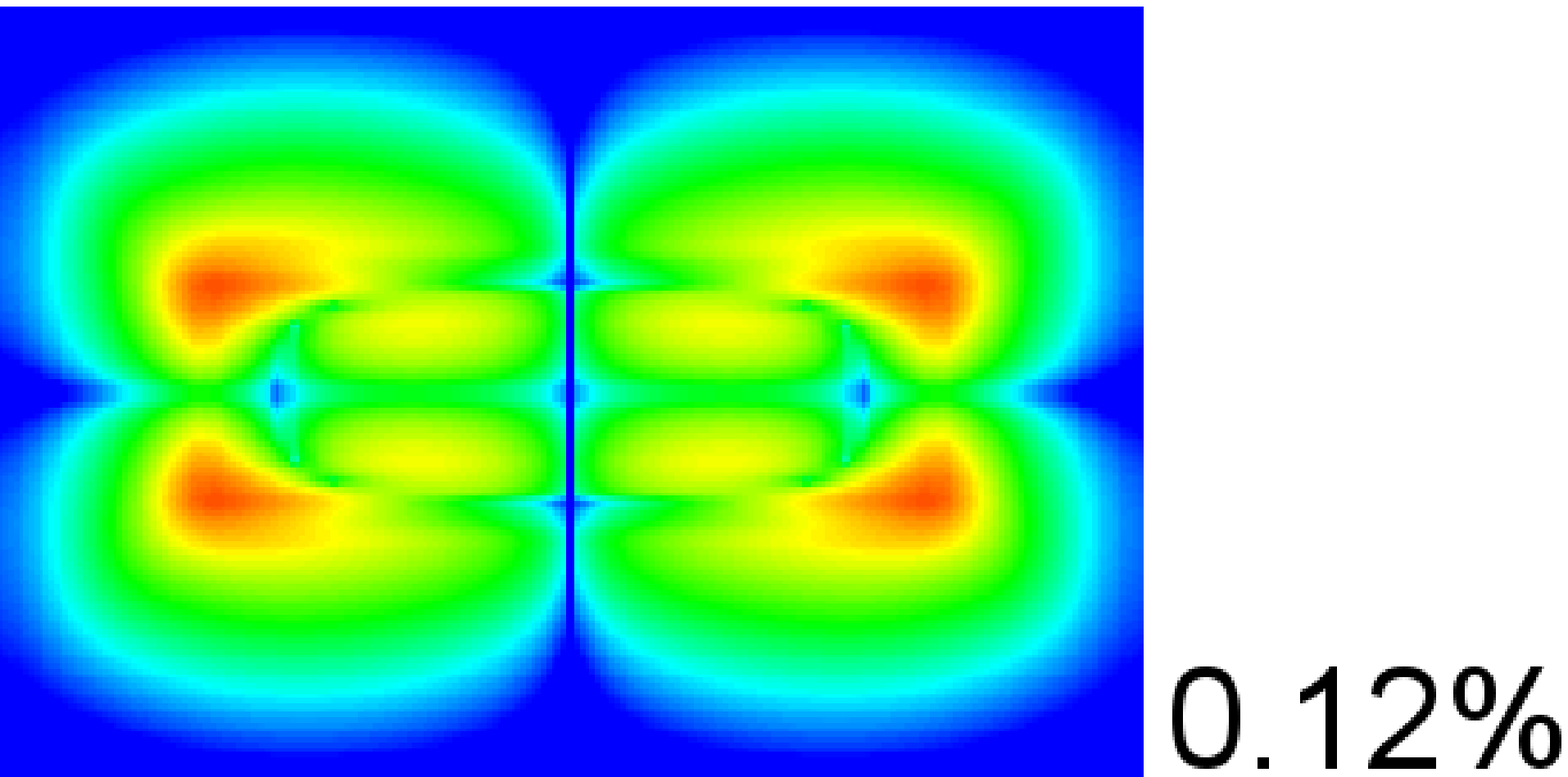}}
 & \parbox{0.2\columnwidth}{\vspace{1mm}\includegraphics[width=0.21\columnwidth]{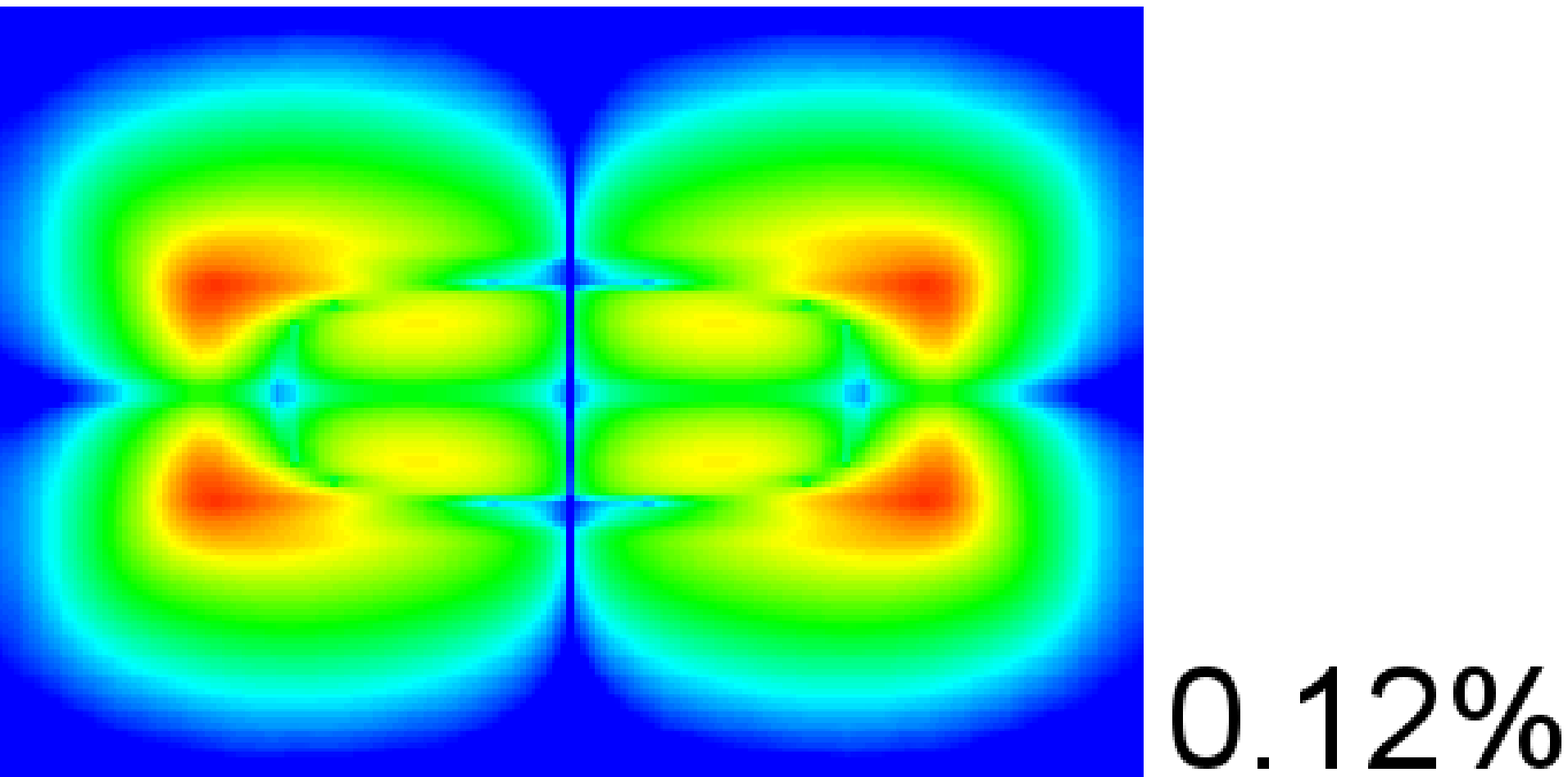} \\ \vspace{1mm}\includegraphics[width=0.21\columnwidth]{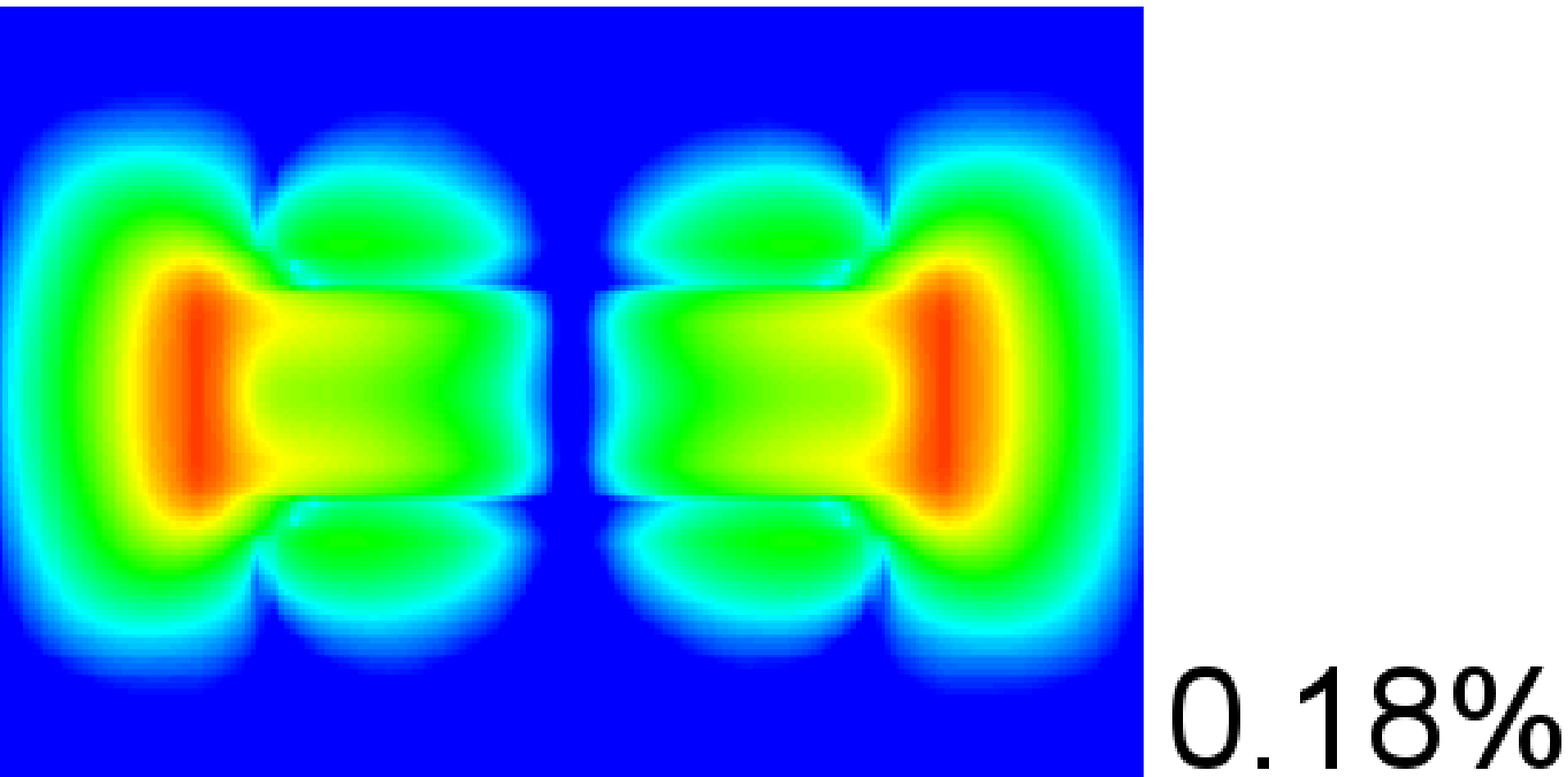}}
\end{tabular}
\end{ruledtabular}
\end{table*}

For the electron state, we have computed two cases: without strain, Fig.~\ref{fig:g_c}(a), and with strain, Fig.~\ref{fig:g_c}(c). In both cases the $g_c$-factor has a similar trend both in terms of height and radius; as confinement increases by decreasing either height or radius, the $g$-factor tends towards the free electron $g$-factor value of $+2$ due to orbital angular momentum quenching. As explained in section \ref{sec:intro}, the envelope orbital momentum causes the deviation of the electron $g$-factor from the Land{\'e}-factor. It is therefore necessary to analyze the origin of the envelope orbital angular momentum in more detail. We first focus  on the results for unstrained quantum dots.

Within the framework of eight-band ${\bf k}\cdot{\bf p}$-theory, any quantum state consists of four different band components: conduction band (CB), heavy-hole (HH), light-hole (LH), and split-off (SO). Each of these components has a Bloch and envelope wave function with corresponding (total) orbital angular momenta $J^{\text{Bloch}}$ and $L^{\text{env}}$. Due to strong spin-orbit coupling these momenta couple with each other giving rise to a combined total momentum $F$:
\begin{eqnarray}
|F,F_z\rangle =  |J^{\text{Bloch}},J^{\text{Bloch}}_z\rangle \otimes |L^{\text{env}},L^{\text{env}}_z\rangle
\end{eqnarray}
The calculated electron ground state has an almost completely $F=\frac{1}{2}$ character, which can be formed from the four different band components by taking the proper combinations of $J^{\text{Bloch}}$ and $L^{\text{env}}$. However, not all $L^{\text{env}}$ are allowed; the parity of the effective mass equations requires that only either even or odd $L^{\text{env}}$ can be mixed into the state \cite{Schechter1962}. All allowable combinations have been listed in table \ref{table:F12}, together with the probability density of the different components along the $[001]$-direction.

Table \ref{table:F12} shows the CB component has a $L^{\text{env}}=0$ and therefore does not contribute to the envelope orbital angular momentum and thus to the deviation of the $g_c$-factor from $+2$. However, the valence band portions of the wave function have components with $|L^{\text{env}}, L^{\text{env}}_z\rangle=|1,\pm1\rangle$, which do lead to a finite envelope orbital angular momentum directed along the $[001]$-direction. In Fig.~\ref{fig:g_c}(b) we have plotted the $g_c$-factor against the sum of all components of the wave function having $L^{\text{env}}_z=\pm1$. For fixed radius, there is a nearly linear relation between the $g_c$-factor and the amount of the wave function which has an envelope orbital momentum $L^{\text{env}}_z=\pm1$. Less amplitude in the band components having $L^{\text{env}}_z=\pm1$ leads to a $g_c$-factor tending to $+2$. In other words, the value of the electron $g_c$-factor for a particular quantum dot having a certain height and radius is solely determined by how the confinement has affected the composition of the wave function.

To investigate this further, we have plotted in Fig.~\ref{fig:g_c}(d) the sum of all components having either $L^{\text{env}}_z=\pm1$ (dots) and $L^{\text{env}}_z=0$ (squares) as function of height. The observed dependencies can be straightforwardly understood by examining the underlying Hamiltonian\cite{Pryor1998}. The coupling between the CB-component and the valence band components having $L^{\text{env}}_z=\pm1$ is proportional to $k_x$ and $k_y$, whereas the coupling with the components having $L^{\text{env}}_z=0$ is proportional to $k_z$. The allowed $k$-vectors for our cylindrical quantum dots having height $h$ and radius $r$ are roughly proportional to $k_z \sim 1/h$ and $k_x,k_y \sim 1/r$, and thus for larger height the fraction of $L^{\text{env}}_z=0$ decreases. Although not directly linked to the height, the fraction of $L^{\text{env}}_z=\pm1$ increases simultaneously. This is merely due to the relative increase of the $1/r$ coupling ($L^{\text{env}}_z=\pm1$) compared to the directly affected $1/h$ coupling ($L^{\text{env}}_z=0$). In a similar fashion the radial dependence can be understood. This analysis shows that quantum confinement is only affecting the composition of the wave function via the allowed $k$-vectors in the quantum dot. Thus the effect of quantum confinement can be understood as probing a bulk material at a specific $k$-vector determined by the size of the quantum dot. This is analogous to earlier work on quantum wells which found the non-parabolicity of the electronic effective mass by varying the quantum well thickness. 

As can be inferred by comparing the unstrained and strained $g_c$-factors in Figs.~\ref{fig:g_c}(a) and (c), strain is changing the $g_c$-factor only quantitatively and not qualitatively. The strain Hamiltonian has the same functional form as the kinetic part of the Hamiltonian\cite{Bahder1990}, meaning that its effect will only be able to modify the size of the couplings and energy gaps between the different bands. From this point of view strain can be regarded as a perturbation onto the system, not affecting the mechanisms determining the $g_c$-factor. In particular, compressive strain inside the QD will cause the effective band gap to become larger compared to unstrained QDs. This gives rise to an overall smaller fraction of the valence band mixed into the electron state, and so for similar sizes, the strained $g_c$-factors have values closer to $+2$ than the unstrained ones.

\subsection{Hole state}
The hole state presents additional complications versus the electron state. We show calculations without strain, Fig.~\ref{fig:g_v_strain}(a), and with strain, Fig.~\ref{fig:g_v_strain}(c). As for the electron state, the envelope orbital angular momenta of the four different band components of the wave function are essential to understand the behavior of the $g_v$-factor. As the hole ground state has a strong $F_z=\pm\frac{3}{2}$ character, different envelope orbital momenta $L^{\text{env}}$ are mixed into the state. The proper combinations of $J^{\text{Bloch}}$ and $L^{\text{env}}$ are listed in table \ref{table:F32}, together with probability density of the different components along the $[001]$-direction. As can be seen from the probability densities, higher orbital angular momenta are mixed into the hole state, since the energy gaps between the valence bands are rather small. There are a number of reasons why an analysis of the envelope orbital momenta is more complicated than for the electron state:

\textit{No clear limit at infinite confinement}. The HH-component has both  $L^{\text{env}}=0$ and $L^{\text{env}}=2$ components, each of which has $L^{\text{env}}_z=0$ and thus does not contribute to $g_v^0$-factor. However, the $L^{\text{env}}=0$ and $L^{\text{env}}=2$ parts of the wave function will have a different Land{\'e}-factor. Depending on the relative weight of these orbital angular momenta, the $g_v$-factor at infinite confinement (in the completely quenched limit) will vary. Thus we do not observe the clear limit at small heights like was obtained for the electron state.

\textit{Strain is no longer only perturbative}. The electron $g_c$-factor is determined by the admixture of valence bands into the electron state, a mechanism upon which strain  only acts perturbatively. For holes the strain  plays a more dominant role, and changes energy gaps and couplings to a much greater extent. Indeed by comparing the unstrained and strained $g_v^0$-factors [Fig.~\ref{fig:g_v_strain}(a) and (c)], it can be seen that even the qualitative trends are not the same.

\textit{Smaller ground-excited state energy splitting}. At heights larger than $\sim 15$~nm an excited state crosses the ground state, leading to mixing between the two. This means that also envelope orbital angular momenta contributions get mixed, complicating the analysis.

Due to these complications a single driving mechanism for the $g_v^0$-factor cannot be extracted. However, some general trends  can still be observed. There is a very different trend of the  $g_v^0$-factor as a function of height for  different radii. At large heights there seems to be an overall converging trend. This converging trend is more dominant in unstrained quantum dots [Fig.~\ref{fig:g_v_strain}(c)]. The divergence, more pronounced for large radii quantum dots, is  enhanced by strain
[Fig.~\ref{fig:g_v_strain}(b)].

\begin{figure*}
\begin{center}
\begingroup
  \makeatletter
  \providecommand\color[2][]{%
    \GenericError{(gnuplot) \space\space\space\@spaces}{%
      Package color not loaded in conjunction with
      terminal option `colourtext'%
    }{See the gnuplot documentation for explanation.%
    }{Either use 'blacktext' in gnuplot or load the package
      color.sty in LaTeX.}%
    \renewcommand\color[2][]{}%
  }%
  \providecommand\includegraphics[2][]{%
    \GenericError{(gnuplot) \space\space\space\@spaces}{%
      Package graphicx or graphics not loaded%
    }{See the gnuplot documentation for explanation.%
    }{The gnuplot epslatex terminal needs graphicx.sty or graphics.sty.}%
    \renewcommand\includegraphics[2][]{}%
  }%
  \providecommand\rotatebox[2]{#2}%
  \@ifundefined{ifGPcolor}{%
    \newif\ifGPcolor
    \GPcolortrue
  }{}%
  \@ifundefined{ifGPblacktext}{%
    \newif\ifGPblacktext
    \GPblacktexttrue
  }{}%
  \let\gplgaddtomacro\g@addto@macro
  \gdef\gplbacktext{}%
  \gdef\gplfronttext{}%
  \makeatother
  \ifGPblacktext
    \def\colorrgb#1{}%
    \def\colorgray#1{}%
  \else
    \ifGPcolor
      \def\colorrgb#1{\color[rgb]{#1}}%
      \def\colorgray#1{\color[gray]{#1}}%
      \expandafter\def\csname LTw\endcsname{\color{white}}%
      \expandafter\def\csname LTb\endcsname{\color{black}}%
      \expandafter\def\csname LTa\endcsname{\color{black}}%
      \expandafter\def\csname LT0\endcsname{\color[rgb]{1,0,0}}%
      \expandafter\def\csname LT1\endcsname{\color[rgb]{0,1,0}}%
      \expandafter\def\csname LT2\endcsname{\color[rgb]{0,0,1}}%
      \expandafter\def\csname LT3\endcsname{\color[rgb]{1,0,1}}%
      \expandafter\def\csname LT4\endcsname{\color[rgb]{0,1,1}}%
      \expandafter\def\csname LT5\endcsname{\color[rgb]{1,1,0}}%
      \expandafter\def\csname LT6\endcsname{\color[rgb]{0,0,0}}%
      \expandafter\def\csname LT7\endcsname{\color[rgb]{1,0.3,0}}%
      \expandafter\def\csname LT8\endcsname{\color[rgb]{0.5,0.5,0.5}}%
    \else
      \def\colorrgb#1{\color{black}}%
      \def\colorgray#1{\color[gray]{#1}}%
      \expandafter\def\csname LTw\endcsname{\color{white}}%
      \expandafter\def\csname LTb\endcsname{\color{black}}%
      \expandafter\def\csname LTa\endcsname{\color{black}}%
      \expandafter\def\csname LT0\endcsname{\color{black}}%
      \expandafter\def\csname LT1\endcsname{\color{black}}%
      \expandafter\def\csname LT2\endcsname{\color{black}}%
      \expandafter\def\csname LT3\endcsname{\color{black}}%
      \expandafter\def\csname LT4\endcsname{\color{black}}%
      \expandafter\def\csname LT5\endcsname{\color{black}}%
      \expandafter\def\csname LT6\endcsname{\color{black}}%
      \expandafter\def\csname LT7\endcsname{\color{black}}%
      \expandafter\def\csname LT8\endcsname{\color{black}}%
    \fi
  \fi
  \setlength{\unitlength}{0.0500bp}%
  \begin{picture}(10368.00,3880.80)%
    \gplgaddtomacro\gplbacktext{%
      \csname LTb\endcsname%
      \put(688,512){\makebox(0,0)[r]{\strut{} 0}}%
      \put(688,902){\makebox(0,0)[r]{\strut{} 5}}%
      \put(688,1291){\makebox(0,0)[r]{\strut{} 10}}%
      \put(688,1681){\makebox(0,0)[r]{\strut{} 15}}%
      \put(688,2070){\makebox(0,0)[r]{\strut{} 20}}%
      \put(688,2460){\makebox(0,0)[r]{\strut{} 25}}%
      \put(688,2849){\makebox(0,0)[r]{\strut{} 30}}%
      \put(688,3239){\makebox(0,0)[r]{\strut{} 35}}%
      \put(688,3628){\makebox(0,0)[r]{\strut{} 40}}%
      \put(784,352){\makebox(0,0){\strut{} 0}}%
      \put(1788,352){\makebox(0,0){\strut{} 5}}%
      \put(2792,352){\makebox(0,0){\strut{} 10}}%
      \put(3795,352){\makebox(0,0){\strut{} 15}}%
      \put(4799,352){\makebox(0,0){\strut{} 20}}%
      \csname LTb\endcsname%
      \put(224,2070){\rotatebox{-270}{\makebox(0,0){\strut{}Diamagnetic coefficient $\alpha_c$ [$\mu$eV/$T^2$]}}}%
      \put(2791,112){\makebox(0,0){\strut{}Height [nm]}}%
    }%
    \gplgaddtomacro\gplfronttext{%
      \csname LTb\endcsname%
      \put(4498,3348){\makebox(0,0)[r]{\strut{}Electron (A)}}%
      \put(4217,1603){\makebox(0,0){\strut{}Radius}}%
      \csname LT0\endcsname%
      \put(4458,1416){\makebox(0,0)[r]{\strut{}7 nm}}%
      \csname LT1\endcsname%
      \put(4458,1260){\makebox(0,0)[r]{\strut{}9 nm}}%
      \csname LT2\endcsname%
      \put(4458,1104){\makebox(0,0)[r]{\strut{}11 nm}}%
      \csname LT3\endcsname%
      \put(4458,948){\makebox(0,0)[r]{\strut{}13 nm}}%
      \csname LT4\endcsname%
      \put(4458,792){\makebox(0,0)[r]{\strut{}15 nm}}%
    }%
    \gplgaddtomacro\gplbacktext{%
      \csname LTb\endcsname%
      \put(5776,512){\makebox(0,0)[r]{\strut{}-9}}%
      \put(5776,858){\makebox(0,0)[r]{\strut{}-8}}%
      \put(5776,1204){\makebox(0,0)[r]{\strut{}-7}}%
      \put(5776,1551){\makebox(0,0)[r]{\strut{}-6}}%
      \put(5776,1897){\makebox(0,0)[r]{\strut{}-5}}%
      \put(5776,2243){\makebox(0,0)[r]{\strut{}-4}}%
      \put(5776,2589){\makebox(0,0)[r]{\strut{}-3}}%
      \put(5776,2936){\makebox(0,0)[r]{\strut{}-2}}%
      \put(5776,3282){\makebox(0,0)[r]{\strut{}-1}}%
      \put(5776,3628){\makebox(0,0)[r]{\strut{} 0}}%
      \put(5872,352){\makebox(0,0){\strut{} 0}}%
      \put(6900,352){\makebox(0,0){\strut{} 5}}%
      \put(7927,352){\makebox(0,0){\strut{} 10}}%
      \put(8955,352){\makebox(0,0){\strut{} 15}}%
      \put(9982,352){\makebox(0,0){\strut{} 20}}%
      \csname LTb\endcsname%
      \put(5408,2070){\rotatebox{-270}{\makebox(0,0){\strut{}Diamagnetic coefficient $\alpha_v$ [$\mu$eV/$T^2$]}}}%
      \put(7927,112){\makebox(0,0){\strut{}Height [nm]}}%
    }%
    \gplgaddtomacro\gplfronttext{%
      \csname LTb\endcsname%
      \put(9674,3348){\makebox(0,0)[r]{\strut{}Hole (B)}}%
      \put(9386,1603){\makebox(0,0){\strut{}Radius}}%
      \csname LT0\endcsname%
      \put(9633,1416){\makebox(0,0)[r]{\strut{}7 nm}}%
      \csname LT1\endcsname%
      \put(9633,1260){\makebox(0,0)[r]{\strut{}9 nm}}%
      \csname LT2\endcsname%
      \put(9633,1104){\makebox(0,0)[r]{\strut{}11 nm}}%
      \csname LT3\endcsname%
      \put(9633,948){\makebox(0,0)[r]{\strut{}13 nm}}%
      \csname LT4\endcsname%
      \put(9633,792){\makebox(0,0)[r]{\strut{}15 nm}}%
    }%
    \gplbacktext
    \put(0,0){\includegraphics{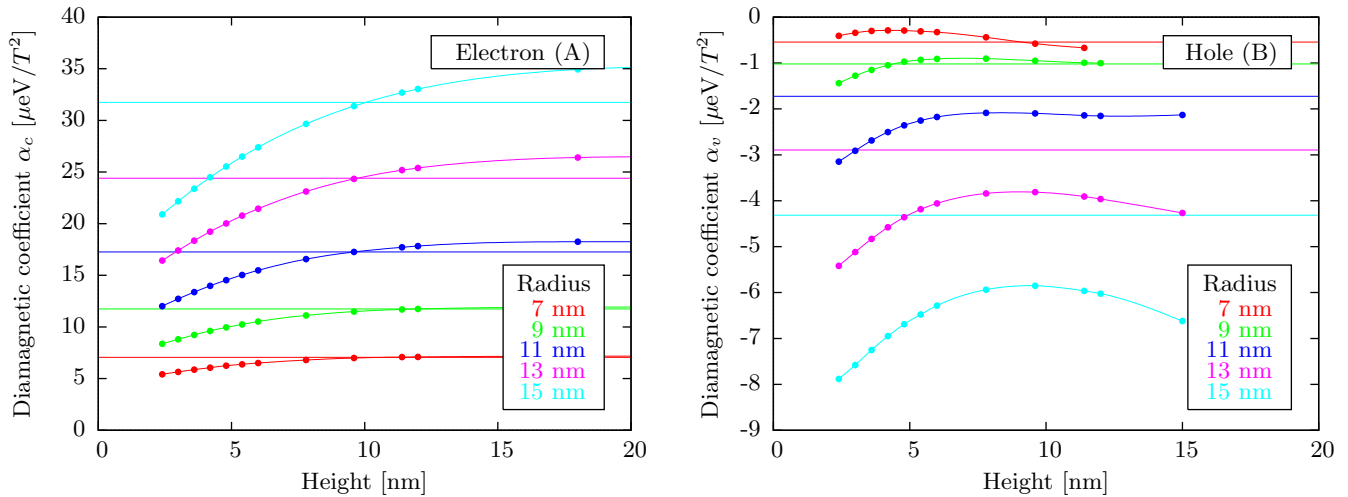}}%
    \gplfronttext
  \end{picture}%
\endgroup

\caption{(Color online) The height dependence of the (a) electron diamagnetic coefficient $\alpha_c$ and (b) hole diamagnetic coefficient $\alpha_v$ of strained InAs/InP quantum dots, respectively. The different colors indicate different radii of QDs. The horizontal lines are the $\alpha$'s of quantum wires of different radii.}
\label{fig:alpha}
\end{center}
\end{figure*}

Besides the usual zero-order $g_v^0$-factor, we have observed that there is a considerable second-order $g_v^2$-factor affecting the Zeeman energy of the hole states [Fig.~\ref{fig:g_v_strain}(d)]. Note that this quantity has per definition a dimension (see Eq.~\ref{eq:g_v_def}) and is thus dependent on the magnitude of the applied magnetic field. For example a QD with a height of $10$~nm and radius of $13$~nm has a $g_v^2 \sim 6 \cdot 10^{-3}/\text{T}^2$ leading to a deviation of the zero-order $g_v^0$-factor from $\sim 4$ to $\sim 4.6$ at a magnetic field of $10$~T. Such deviations are certainly large enough to be noticed in experiments. To the best of our knowledge, however, such non-linearities have only been seen in very tall dots\cite{Babinski2006}.  In quantum wells, non-linearities of the hole Zeeman energy have been observed experimentally and were reproduced theoretically\cite{Warburton1993,Traynor1995,Traynor1997}. In these studies the non-linearity was attributed to the admixture between HH- and LH-subbands. In that case mass reversal is induced by quantum confinement, leading to an anti-crossing of the HH- and LH-subbands as function of magnetic field. This anti-crossing only occurs for one of the two spin split subbands, leading to a non-linear Zeeman energy. However, in our calculations both spin split ground states exhibit a non-linear dependence as function of magnetic field. Therefore the mechanism as explained for quantum wells cannot explain the origin of the $g_v^2$-factor in quantum dots. We attribute the non-linearity to mixing between the HH-like ground and LH-like excited states, and have verified that the magnitude of $g_v^2$ decreases with increasing strain (larger energy splitting) and decreasing $\gamma_2$ and $\gamma_3$ parameters (smaller coupling), both resulting in a smaller degree of mixing. Furthermore, we observe that the non-linearity has a strong size dependence: increasing either height or radius decreases the energy splitting between the ground and excited states, leading to a larger $g_v^2$-factor. It remains unclear why this effect is strongly present in the calculation and has not been observed experimentally.

\begin{figure*}
\begin{center}
\begingroup
  \makeatletter
  \providecommand\color[2][]{%
    \GenericError{(gnuplot) \space\space\space\@spaces}{%
      Package color not loaded in conjunction with
      terminal option `colourtext'%
    }{See the gnuplot documentation for explanation.%
    }{Either use 'blacktext' in gnuplot or load the package
      color.sty in LaTeX.}%
    \renewcommand\color[2][]{}%
  }%
  \providecommand\includegraphics[2][]{%
    \GenericError{(gnuplot) \space\space\space\@spaces}{%
      Package graphicx or graphics not loaded%
    }{See the gnuplot documentation for explanation.%
    }{The gnuplot epslatex terminal needs graphicx.sty or graphics.sty.}%
    \renewcommand\includegraphics[2][]{}%
  }%
  \providecommand\rotatebox[2]{#2}%
  \@ifundefined{ifGPcolor}{%
    \newif\ifGPcolor
    \GPcolortrue
  }{}%
  \@ifundefined{ifGPblacktext}{%
    \newif\ifGPblacktext
    \GPblacktexttrue
  }{}%
  \let\gplgaddtomacro\g@addto@macro
  \gdef\gplbacktext{}%
  \gdef\gplfronttext{}%
  \makeatother
  \ifGPblacktext
    \def\colorrgb#1{}%
    \def\colorgray#1{}%
  \else
    \ifGPcolor
      \def\colorrgb#1{\color[rgb]{#1}}%
      \def\colorgray#1{\color[gray]{#1}}%
      \expandafter\def\csname LTw\endcsname{\color{white}}%
      \expandafter\def\csname LTb\endcsname{\color{black}}%
      \expandafter\def\csname LTa\endcsname{\color{black}}%
      \expandafter\def\csname LT0\endcsname{\color[rgb]{1,0,0}}%
      \expandafter\def\csname LT1\endcsname{\color[rgb]{0,1,0}}%
      \expandafter\def\csname LT2\endcsname{\color[rgb]{0,0,1}}%
      \expandafter\def\csname LT3\endcsname{\color[rgb]{1,0,1}}%
      \expandafter\def\csname LT4\endcsname{\color[rgb]{0,1,1}}%
      \expandafter\def\csname LT5\endcsname{\color[rgb]{1,1,0}}%
      \expandafter\def\csname LT6\endcsname{\color[rgb]{0,0,0}}%
      \expandafter\def\csname LT7\endcsname{\color[rgb]{1,0.3,0}}%
      \expandafter\def\csname LT8\endcsname{\color[rgb]{0.5,0.5,0.5}}%
    \else
      \def\colorrgb#1{\color{black}}%
      \def\colorgray#1{\color[gray]{#1}}%
      \expandafter\def\csname LTw\endcsname{\color{white}}%
      \expandafter\def\csname LTb\endcsname{\color{black}}%
      \expandafter\def\csname LTa\endcsname{\color{black}}%
      \expandafter\def\csname LT0\endcsname{\color{black}}%
      \expandafter\def\csname LT1\endcsname{\color{black}}%
      \expandafter\def\csname LT2\endcsname{\color{black}}%
      \expandafter\def\csname LT3\endcsname{\color{black}}%
      \expandafter\def\csname LT4\endcsname{\color{black}}%
      \expandafter\def\csname LT5\endcsname{\color{black}}%
      \expandafter\def\csname LT6\endcsname{\color{black}}%
      \expandafter\def\csname LT7\endcsname{\color{black}}%
      \expandafter\def\csname LT8\endcsname{\color{black}}%
    \fi
  \fi
  \setlength{\unitlength}{0.0500bp}%
  \begin{picture}(10368.00,7761.60)%
    \gplgaddtomacro\gplbacktext{%
      \csname LTb\endcsname%
      \put(688,4392){\makebox(0,0)[r]{\strut{}-4}}%
      \put(688,4837){\makebox(0,0)[r]{\strut{}-2}}%
      \put(688,5282){\makebox(0,0)[r]{\strut{} 0}}%
      \put(688,5727){\makebox(0,0)[r]{\strut{} 2}}%
      \put(688,6173){\makebox(0,0)[r]{\strut{} 4}}%
      \put(688,6618){\makebox(0,0)[r]{\strut{} 6}}%
      \put(688,7063){\makebox(0,0)[r]{\strut{} 8}}%
      \put(688,7508){\makebox(0,0)[r]{\strut{} 10}}%
      \put(784,4232){\makebox(0,0){\strut{} 600}}%
      \put(1286,4232){\makebox(0,0){\strut{} 700}}%
      \put(1788,4232){\makebox(0,0){\strut{} 800}}%
      \put(2290,4232){\makebox(0,0){\strut{} 900}}%
      \put(2792,4232){\makebox(0,0){\strut{} 1000}}%
      \put(3293,4232){\makebox(0,0){\strut{} 1100}}%
      \put(3795,4232){\makebox(0,0){\strut{} 1200}}%
      \put(4297,4232){\makebox(0,0){\strut{} 1300}}%
      \put(4799,4232){\makebox(0,0){\strut{} 1400}}%
      \csname LTb\endcsname%
      \put(224,5950){\rotatebox{-270}{\makebox(0,0){\strut{}g-factor $g_{ex}$}}}%
      \put(2791,3992){\makebox(0,0){\strut{}Emission energy [meV]}}%
    }%
    \gplgaddtomacro\gplfronttext{%
      \csname LTb\endcsname%
      \put(4498,7228){\makebox(0,0)[r]{\strut{}Exciton (A)}}%
      \put(4217,5483){\makebox(0,0){\strut{}Radius}}%
      \csname LT0\endcsname%
      \put(4458,5296){\makebox(0,0)[r]{\strut{}7 nm}}%
      \csname LT1\endcsname%
      \put(4458,5140){\makebox(0,0)[r]{\strut{}9 nm}}%
      \csname LT2\endcsname%
      \put(4458,4984){\makebox(0,0)[r]{\strut{}11 nm}}%
      \csname LT3\endcsname%
      \put(4458,4828){\makebox(0,0)[r]{\strut{}13 nm}}%
      \csname LT4\endcsname%
      \put(4458,4672){\makebox(0,0)[r]{\strut{}15 nm}}%
    }%
    \gplgaddtomacro\gplbacktext{%
      \csname LTb\endcsname%
      \put(5872,4392){\makebox(0,0)[r]{\strut{} 0}}%
      \put(5872,4837){\makebox(0,0)[r]{\strut{} 5}}%
      \put(5872,5282){\makebox(0,0)[r]{\strut{} 10}}%
      \put(5872,5727){\makebox(0,0)[r]{\strut{} 15}}%
      \put(5872,6173){\makebox(0,0)[r]{\strut{} 20}}%
      \put(5872,6618){\makebox(0,0)[r]{\strut{} 25}}%
      \put(5872,7063){\makebox(0,0)[r]{\strut{} 30}}%
      \put(5872,7508){\makebox(0,0)[r]{\strut{} 35}}%
      \put(5968,4232){\makebox(0,0){\strut{} 600}}%
      \put(6470,4232){\makebox(0,0){\strut{} 700}}%
      \put(6972,4232){\makebox(0,0){\strut{} 800}}%
      \put(7473,4232){\makebox(0,0){\strut{} 900}}%
      \put(7975,4232){\makebox(0,0){\strut{} 1000}}%
      \put(8477,4232){\makebox(0,0){\strut{} 1100}}%
      \put(8979,4232){\makebox(0,0){\strut{} 1200}}%
      \put(9480,4232){\makebox(0,0){\strut{} 1300}}%
      \put(9982,4232){\makebox(0,0){\strut{} 1400}}%
      \csname LTb\endcsname%
      \put(5408,5950){\rotatebox{-270}{\makebox(0,0){\strut{}Diamagnetic coefficient $\alpha_{ex}$ [$\mu$eV/$T^2$]}}}%
      \put(7975,3992){\makebox(0,0){\strut{}Emission energy [meV]}}%
    }%
    \gplgaddtomacro\gplfronttext{%
      \csname LTb\endcsname%
      \put(9681,7228){\makebox(0,0)[r]{\strut{}Exciton (C)}}%
      \put(9400,5483){\makebox(0,0){\strut{}Radius}}%
      \csname LT0\endcsname%
      \put(9641,5296){\makebox(0,0)[r]{\strut{}7 nm}}%
      \csname LT1\endcsname%
      \put(9641,5140){\makebox(0,0)[r]{\strut{}9 nm}}%
      \csname LT2\endcsname%
      \put(9641,4984){\makebox(0,0)[r]{\strut{}11 nm}}%
      \csname LT3\endcsname%
      \put(9641,4828){\makebox(0,0)[r]{\strut{}13 nm}}%
      \csname LT4\endcsname%
      \put(9641,4672){\makebox(0,0)[r]{\strut{}15 nm}}%
    }%
    \gplgaddtomacro\gplbacktext{%
      \csname LTb\endcsname%
      \put(688,512){\makebox(0,0)[r]{\strut{}-4}}%
      \put(688,957){\makebox(0,0)[r]{\strut{}-2}}%
      \put(688,1402){\makebox(0,0)[r]{\strut{} 0}}%
      \put(688,1847){\makebox(0,0)[r]{\strut{} 2}}%
      \put(688,2293){\makebox(0,0)[r]{\strut{} 4}}%
      \put(688,2738){\makebox(0,0)[r]{\strut{} 6}}%
      \put(688,3183){\makebox(0,0)[r]{\strut{} 8}}%
      \put(688,3628){\makebox(0,0)[r]{\strut{} 10}}%
      \put(784,352){\makebox(0,0){\strut{} 600}}%
      \put(1286,352){\makebox(0,0){\strut{} 700}}%
      \put(1788,352){\makebox(0,0){\strut{} 800}}%
      \put(2290,352){\makebox(0,0){\strut{} 900}}%
      \put(2792,352){\makebox(0,0){\strut{} 1000}}%
      \put(3293,352){\makebox(0,0){\strut{} 1100}}%
      \put(3795,352){\makebox(0,0){\strut{} 1200}}%
      \put(4297,352){\makebox(0,0){\strut{} 1300}}%
      \put(4799,352){\makebox(0,0){\strut{} 1400}}%
      \csname LTb\endcsname%
      \put(224,2070){\rotatebox{-270}{\makebox(0,0){\strut{}g-factor $g_{ex}$}}}%
      \put(2791,112){\makebox(0,0){\strut{}Emission energy [meV]}}%
    }%
    \gplgaddtomacro\gplfronttext{%
      \csname LTb\endcsname%
      \put(4498,3348){\makebox(0,0)[r]{\strut{}Exciton - Coulomb interaction (B)}}%
      \put(4217,1603){\makebox(0,0){\strut{}Radius}}%
      \csname LT0\endcsname%
      \put(4458,1416){\makebox(0,0)[r]{\strut{}7 nm}}%
      \csname LT1\endcsname%
      \put(4458,1260){\makebox(0,0)[r]{\strut{}9 nm}}%
      \csname LT2\endcsname%
      \put(4458,1104){\makebox(0,0)[r]{\strut{}11 nm}}%
      \csname LT3\endcsname%
      \put(4458,948){\makebox(0,0)[r]{\strut{}13 nm}}%
      \csname LT4\endcsname%
      \put(4458,792){\makebox(0,0)[r]{\strut{}15 nm}}%
    }%
    \gplgaddtomacro\gplbacktext{%
      \csname LTb\endcsname%
      \put(5872,512){\makebox(0,0)[r]{\strut{} 0}}%
      \put(5872,957){\makebox(0,0)[r]{\strut{} 5}}%
      \put(5872,1402){\makebox(0,0)[r]{\strut{} 10}}%
      \put(5872,1847){\makebox(0,0)[r]{\strut{} 15}}%
      \put(5872,2293){\makebox(0,0)[r]{\strut{} 20}}%
      \put(5872,2738){\makebox(0,0)[r]{\strut{} 25}}%
      \put(5872,3183){\makebox(0,0)[r]{\strut{} 30}}%
      \put(5872,3628){\makebox(0,0)[r]{\strut{} 35}}%
      \put(5968,352){\makebox(0,0){\strut{} 600}}%
      \put(6470,352){\makebox(0,0){\strut{} 700}}%
      \put(6972,352){\makebox(0,0){\strut{} 800}}%
      \put(7473,352){\makebox(0,0){\strut{} 900}}%
      \put(7975,352){\makebox(0,0){\strut{} 1000}}%
      \put(8477,352){\makebox(0,0){\strut{} 1100}}%
      \put(8979,352){\makebox(0,0){\strut{} 1200}}%
      \put(9480,352){\makebox(0,0){\strut{} 1300}}%
      \put(9982,352){\makebox(0,0){\strut{} 1400}}%
      \csname LTb\endcsname%
      \put(5408,2070){\rotatebox{-270}{\makebox(0,0){\strut{}Diamagnetic coefficient $\alpha_{ex}$ [$\mu$eV/$T^2$]}}}%
      \put(7975,112){\makebox(0,0){\strut{}Emission energy [meV]}}%
    }%
    \gplgaddtomacro\gplfronttext{%
      \csname LTb\endcsname%
      \put(9681,3348){\makebox(0,0)[r]{\strut{}Exciton - Coulomb interaction (D)}}%
      \put(9400,1603){\makebox(0,0){\strut{}Radius}}%
      \csname LT0\endcsname%
      \put(9641,1416){\makebox(0,0)[r]{\strut{}7 nm}}%
      \csname LT1\endcsname%
      \put(9641,1260){\makebox(0,0)[r]{\strut{}9 nm}}%
      \csname LT2\endcsname%
      \put(9641,1104){\makebox(0,0)[r]{\strut{}11 nm}}%
      \csname LT3\endcsname%
      \put(9641,948){\makebox(0,0)[r]{\strut{}13 nm}}%
      \csname LT4\endcsname%
      \put(9641,792){\makebox(0,0)[r]{\strut{}15 nm}}%
    }%
    \gplbacktext
    \put(0,0){\includegraphics{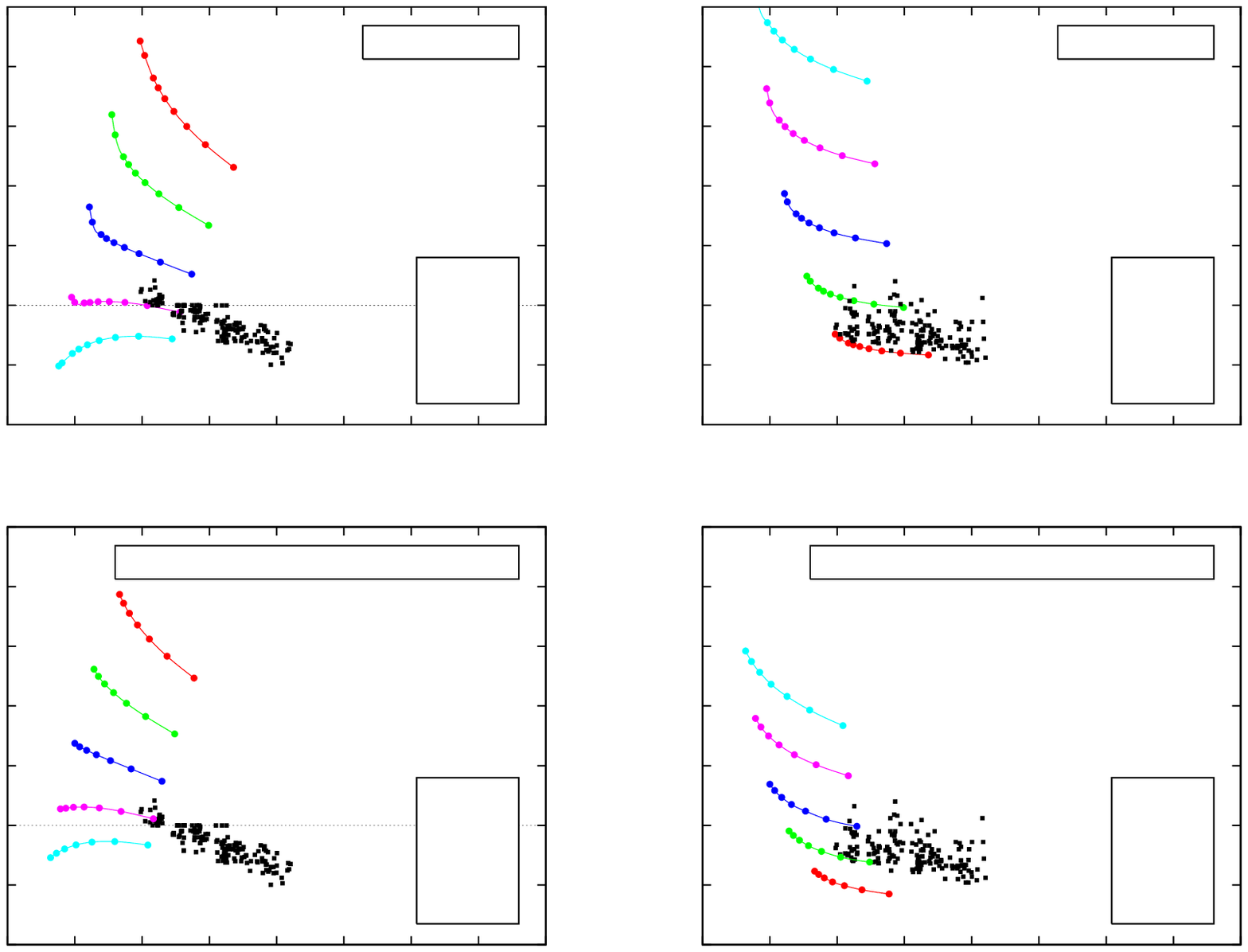}}%
    \gplfronttext
  \end{picture}%
\endgroup

\caption{(Color online) The emission energy dependence of the (a) exciton $g_{ex}$-factor and (c) diamagnetic coefficient $\alpha_{ex}$ of InAs/InP quantum dots, respectively. The emission energy dependence of (b) $g_{ex}$ and (d)  $\alpha_{ex}$ with Coulomb interaction taken into account, respectively. The different colors indicate different radii of QDs.}
\label{fig:E_ex}
\end{center}
\end{figure*}

\section{Diamagnetic coefficients} \label{sec:alpha}

We will now discuss the diamagnetic coefficient for the electron and hole state (see Fig.~\ref{fig:alpha}). Within the framework of the ${\bf k}\cdot{\bf p}$-approximation, in first order perturbation the diamagnetic coefficient of a carrier in a semiconductor is proportional to\cite{NASH1989}:
\begin{eqnarray}
\alpha \sim \frac{\langle r^2\rangle}{m^*} \label{eq:alpha}
\end{eqnarray}
where $\langle r^2\rangle$ is the average lateral extension of the wave function perpendicular to the applied magnetic field (and can therefore be directly associated with the radius), and $m^*$ the effective mass of the carrier. As a result there is a strong radial dependence of the diamagnetic coefficient, both for the electron and hole state: a large radius quantum dot has a large diamagnetic coefficient. The strong height dependence is unexpected; for the electron state the diamagnetic coefficient becomes smaller at smaller heights, whereas for the hole state it shifts both to larger and smaller values depending on the height. We will start by discussing the electron state.

The height dependence should arise either from details in the average lateral extension $\langle r^2\rangle$ or the effective mass $m^*$, as can be inferred from Eq.~(\ref{eq:alpha}). It has been verified that $\langle r^2\rangle$ is only weakly dependent on the height ($\leq10\%$ for the most affected 15~nm radius quantum dots), and hence cannot be responsible for the strong deviations in $\alpha_c$ of almost a factor of 2. We therefore attribute the strong height dependence to the energy-dependence of the effective mass. As mentioned in previous sections, quantum confinement is effectively probing the bulk dispersion at specific $k$-vectors. At the small quantum dot heights considered here, the dispersion is sampled at large $k$-vectors, where the conduction band is no longer parabolic and corrections to the effective mass have to be made, i.e. $m^* = m^*(h)$. For InAs these corrections lead to a more heavier mass at large $k$-vectors, causing the diamagnetic coefficient to become smaller at smaller heights, as can be observed in Fig.~\ref{fig:alpha}. We want to note that we have verified that strain is not an important mechanism in determining $\alpha_c$, as was the case for the electron $g_c$-factor. The sole effect of strain is to modify the magnitude of the diamagnetic coefficient (mainly via $m^*$), the qualitative trends stay the same.

The hole state diamagnetic coefficient has an even more peculiar height dependence: at intermediate heights it decreases, whereas at large heights it increases. The nature of the hole state induces a number of ways by which the diamagnetic coefficient can be modified. First of all the lateral extension can be influenced by size dependent HH-LH mixing. We have verified that $\langle r^2\rangle$ varies by $\leq15\%$ for the 15~nm radius quantum dots (and varies less for smaller radii). Contrary to the electron state, this variation is of the same order as the deviations in $\alpha_v$. However, the effective mass will also vary with height. It is a priori not possible to disentangle these effects and study their respective influence. Note that we also calculated the diamagnetic coefficients including the $\kappa$-parameter. Similar to the $g_v$-factors, to first order the $\kappa$-parameter does not modify the diamagnetic coefficients (see appendix~\ref{app:remote} for a discussion about the remote band parameters).


\section{Comparison with experiment} \label{sec:comparison}
Using the separate electron and hole $g$-factors and diamagnetic coefficients, one can compute the exciton $g_{ex}$-factor and exciton diamagnetic coefficient $\alpha_{ex}$, as defined by Eqs.~(\ref{eq:g_ex})-(\ref{eq:E_ex}). Since in the experiment the Zeeman energy was found to be linear with magnetic field, we have only taken the $g_v^0$-factor into account, which is exactly valid for small magnetic fields. Since experimentally only the emission energy $E_{em}$ is accessible, we have plotted the $g_{ex}$-factor and $\alpha_{ex}$ as function of $E_{em}$ in Fig.~\ref{fig:E_ex}. Again the different colors represent different radii QDs. Note that the different dots along a line of fixed radius represent QDs with roughly a ML height difference. The black squares are the experimental data points from Ref. \onlinecite{Kleemans2009}. We will first discuss the $g_{ex}$-factor.

It was experimentally established \cite{Kleemans2009} that the general trend of the $g_{ex}$-factor with emission energy originates from the dependence on the height of the QDs. This is  confirmed by the theoretical calculation, in which irrespective of the exact radius of the QDs the height dependence matches with the experimentally observed trend. Furthermore, from cross-sectional scanning tunneling microscopy analysis it was inferred that the maximum radius of the QDs was $\sim 15$~nm, meaning that the $13$~nm radius theoretical calculations that overlap the experimental data are a good match. Experimentally the spread around the trend was related to variations in the lateral size: QDs below the trend had on average a smaller $\alpha_{ex}$ than QDs above the trend. The opposite trend is observed in the calculations. However, the assumptions made about the shape and composition of the quantum dot may be too simplistic to reproduce this detailed trend; for example, a size-dependent composition profile might alleviate the discrepancy. Note that it is mainly the hole $g_v^0$-factor which determines the $g_{ex}$-factor;  the electron $g_c$-factor is nearly independent of the radius when plotted as a function of $E_{em}$. We want to stress that, since the $g_{ex}$-factor is a combination of the $g_c$-factor and $g_v^0$-factor, it is difficult to extract relevant information from it to identify the physics determining the $g$-factors of the individual carriers. It is far more useful to have experimental access to the $g_c$-factor and $g_v^0$-factor separately, preferably in a system where the size of the quantum dot can be highly controlled.

The diamagnetic coefficient $\alpha_{ex}$ is plotted against $E_{em}$ in Fig.~\ref{fig:E_ex}(c). Experimentally it was established that the emission energy is dominated by the height dependence. Assuming that $\alpha_{ex}$ is then governed by the radius (see Eq.~\ref{eq:alpha}), there should only be a weak dependence between $\alpha_{ex}$ and $E_{em}$. This has been observed experimentally, and confirmed the interpretation of the height-dominated $E_{em}$. The theoretical calculations support this analysis; there is only a rather weak dependence between $\alpha_{ex}$ and $E_{em}$. The calculations, however, show that the electron and hole diamagnetic coefficients separately are strongly dependent on the height (Fig.~\ref{fig:alpha}). It is the coincidental height dependence of the emission energy and the (partially) opposing height dependence of the electron and hole diamagnetic coefficients, which conceal the true size dependency. Experimental access to the separate electron and hole energy levels would be the only way to verify this. The match in terms of absolute size of the quantum dot  is less satisfactory than for the $g$-factor; the calculations with a radius of $\sim8$~nm  fit the diamagnetic coefficient measurements best.

To improve on the match for $\alpha_{ex}$, we included the Coulomb interaction between the electron and hole states via a self-consistent iterative Hartree approximation\cite{Pryor1998}. After two iterations more than $99\%$ of the total modification to the emission energy is taken into account. In graphs (b) and (d) of Fig.~\ref{fig:E_ex} the influence of the Coulomb interaction to respectively $g_{ex}$ and $\alpha_{ex}$ are displayed. Except for corrections to the emission energy ($\leq7\%$), there are only small modifications to $g_{ex}$. This was not too surprising, as the Coulomb interaction might modify the spatial extension of the wave function, but not as much the composition of the states. Supporting that picture, the diamagnetic coefficient is much more strongly modified: it becomes smaller, such that there is a better agreement between the experimentally found radius ($\sim13$~nm) and the theoretical match ($\sim 11$~nm). The Coulomb interaction pulls the electron and hole wave functions together, such that the exciton state has a smaller lateral extension (and therefore $\alpha_{ex}$) compared to the non-interacting states for a similar radius quantum dot.

\section{Conclusions}
We have calculated the $g$-factors and diamagnetic coefficients of electron and hole states in InAs/InP quantum dots. The electron $g_c$-factor is determined by how quantum confinement affects the amount of envelope orbital angular momentum mixed in through the admixture of valence-band contributions. In the limit of infinite barriers and a dot of vanishing size the electron $g_c$-factor will approach the Land{\'e}-factor of $+2$. Strain acts merely a perturbative mechanism on the $g_c$-factor. A similar analysis is less straightforward for the hole $g_v$-factor. We attribute the observed non-linear hole Zeeman energy to ground-excited state mixing. The electron and hole diamagnetic coefficients are determined both by the lateral extension of the wave function and the effective mass. The height dependence of the electron diamagnetic coefficient is due to a strongly height- (or energy-) dependent effective mass. Using the single-particle states, the calculated exciton $g$-factor and diamagnetic coefficient agree well with the experimental data. Including the Coulomb interaction between the electron and hole states improves the agreement, especially by decreasing the diamagnetic coefficient due to reductions in the lateral extent of the wave function. 

\appendix

\begin{figure*}
\begin{center}
\begingroup
  \makeatletter
  \providecommand\color[2][]{%
    \GenericError{(gnuplot) \space\space\space\@spaces}{%
      Package color not loaded in conjunction with
      terminal option `colourtext'%
    }{See the gnuplot documentation for explanation.%
    }{Either use 'blacktext' in gnuplot or load the package
      color.sty in LaTeX.}%
    \renewcommand\color[2][]{}%
  }%
  \providecommand\includegraphics[2][]{%
    \GenericError{(gnuplot) \space\space\space\@spaces}{%
      Package graphicx or graphics not loaded%
    }{See the gnuplot documentation for explanation.%
    }{The gnuplot epslatex terminal needs graphicx.sty or graphics.sty.}%
    \renewcommand\includegraphics[2][]{}%
  }%
  \providecommand\rotatebox[2]{#2}%
  \@ifundefined{ifGPcolor}{%
    \newif\ifGPcolor
    \GPcolortrue
  }{}%
  \@ifundefined{ifGPblacktext}{%
    \newif\ifGPblacktext
    \GPblacktexttrue
  }{}%
  \let\gplgaddtomacro\g@addto@macro
  \gdef\gplbacktext{}%
  \gdef\gplfronttext{}%
  \makeatother
  \ifGPblacktext
    \def\colorrgb#1{}%
    \def\colorgray#1{}%
  \else
    \ifGPcolor
      \def\colorrgb#1{\color[rgb]{#1}}%
      \def\colorgray#1{\color[gray]{#1}}%
      \expandafter\def\csname LTw\endcsname{\color{white}}%
      \expandafter\def\csname LTb\endcsname{\color{black}}%
      \expandafter\def\csname LTa\endcsname{\color{black}}%
      \expandafter\def\csname LT0\endcsname{\color[rgb]{1,0,0}}%
      \expandafter\def\csname LT1\endcsname{\color[rgb]{0,1,0}}%
      \expandafter\def\csname LT2\endcsname{\color[rgb]{0,0,1}}%
      \expandafter\def\csname LT3\endcsname{\color[rgb]{1,0,1}}%
      \expandafter\def\csname LT4\endcsname{\color[rgb]{0,1,1}}%
      \expandafter\def\csname LT5\endcsname{\color[rgb]{1,1,0}}%
      \expandafter\def\csname LT6\endcsname{\color[rgb]{0,0,0}}%
      \expandafter\def\csname LT7\endcsname{\color[rgb]{1,0.3,0}}%
      \expandafter\def\csname LT8\endcsname{\color[rgb]{0.5,0.5,0.5}}%
    \else
      \def\colorrgb#1{\color{black}}%
      \def\colorgray#1{\color[gray]{#1}}%
      \expandafter\def\csname LTw\endcsname{\color{white}}%
      \expandafter\def\csname LTb\endcsname{\color{black}}%
      \expandafter\def\csname LTa\endcsname{\color{black}}%
      \expandafter\def\csname LT0\endcsname{\color{black}}%
      \expandafter\def\csname LT1\endcsname{\color{black}}%
      \expandafter\def\csname LT2\endcsname{\color{black}}%
      \expandafter\def\csname LT3\endcsname{\color{black}}%
      \expandafter\def\csname LT4\endcsname{\color{black}}%
      \expandafter\def\csname LT5\endcsname{\color{black}}%
      \expandafter\def\csname LT6\endcsname{\color{black}}%
      \expandafter\def\csname LT7\endcsname{\color{black}}%
      \expandafter\def\csname LT8\endcsname{\color{black}}%
    \fi
  \fi
  \setlength{\unitlength}{0.0500bp}%
  \begin{picture}(10368.00,3880.80)%
    \gplgaddtomacro\gplbacktext{%
      \csname LTb\endcsname%
      \put(592,512){\makebox(0,0)[r]{\strut{}-8}}%
      \put(592,902){\makebox(0,0)[r]{\strut{}-6}}%
      \put(592,1291){\makebox(0,0)[r]{\strut{}-4}}%
      \put(592,1681){\makebox(0,0)[r]{\strut{}-2}}%
      \put(592,2070){\makebox(0,0)[r]{\strut{} 0}}%
      \put(592,2460){\makebox(0,0)[r]{\strut{} 2}}%
      \put(592,2849){\makebox(0,0)[r]{\strut{} 4}}%
      \put(592,3239){\makebox(0,0)[r]{\strut{} 6}}%
      \put(592,3628){\makebox(0,0)[r]{\strut{} 8}}%
      \put(688,352){\makebox(0,0){\strut{} 0}}%
      \put(1716,352){\makebox(0,0){\strut{} 5}}%
      \put(2744,352){\makebox(0,0){\strut{} 10}}%
      \put(3771,352){\makebox(0,0){\strut{} 15}}%
      \put(4799,352){\makebox(0,0){\strut{} 20}}%
      \csname LTb\endcsname%
      \put(224,2070){\rotatebox{-270}{\makebox(0,0){\strut{}Zero-order g-factor $g_v^0$}}}%
      \put(2743,112){\makebox(0,0){\strut{}Height [nm]}}%
    }%
    \gplgaddtomacro\gplfronttext{%
      \csname LTb\endcsname%
      \put(4491,3348){\makebox(0,0)[r]{\strut{}Hole - without $\kappa$ (A)}}%
      \put(4203,1603){\makebox(0,0){\strut{}Radius}}%
      \csname LT0\endcsname%
      \put(4450,1416){\makebox(0,0)[r]{\strut{}7 nm}}%
      \csname LT1\endcsname%
      \put(4450,1260){\makebox(0,0)[r]{\strut{}9 nm}}%
      \csname LT2\endcsname%
      \put(4450,1104){\makebox(0,0)[r]{\strut{}11 nm}}%
      \csname LT3\endcsname%
      \put(4450,948){\makebox(0,0)[r]{\strut{}13 nm}}%
      \csname LT4\endcsname%
      \put(4450,792){\makebox(0,0)[r]{\strut{}15 nm}}%
    }%
    \gplgaddtomacro\gplbacktext{%
      \csname LTb\endcsname%
      \put(5872,512){\makebox(0,0)[r]{\strut{}-32}}%
      \put(5872,902){\makebox(0,0)[r]{\strut{}-30}}%
      \put(5872,1291){\makebox(0,0)[r]{\strut{}-28}}%
      \put(5872,1681){\makebox(0,0)[r]{\strut{}-26}}%
      \put(5872,2070){\makebox(0,0)[r]{\strut{}-24}}%
      \put(5872,2460){\makebox(0,0)[r]{\strut{}-22}}%
      \put(5872,2849){\makebox(0,0)[r]{\strut{}-20}}%
      \put(5872,3239){\makebox(0,0)[r]{\strut{}-18}}%
      \put(5872,3628){\makebox(0,0)[r]{\strut{}-16}}%
      \put(5968,352){\makebox(0,0){\strut{} 0}}%
      \put(6972,352){\makebox(0,0){\strut{} 5}}%
      \put(7975,352){\makebox(0,0){\strut{} 10}}%
      \put(8979,352){\makebox(0,0){\strut{} 15}}%
      \put(9982,352){\makebox(0,0){\strut{} 20}}%
      \csname LTb\endcsname%
      \put(5408,2070){\rotatebox{-270}{\makebox(0,0){\strut{}Zero-order g-factor $g_v^0$}}}%
      \put(7975,112){\makebox(0,0){\strut{}Height [nm]}}%
    }%
    \gplgaddtomacro\gplfronttext{%
      \csname LTb\endcsname%
      \put(9681,3348){\makebox(0,0)[r]{\strut{}Hole - with $\kappa$ (B)}}%
      \put(9400,1603){\makebox(0,0){\strut{}Radius}}%
      \csname LT0\endcsname%
      \put(9641,1416){\makebox(0,0)[r]{\strut{}7 nm}}%
      \csname LT1\endcsname%
      \put(9641,1260){\makebox(0,0)[r]{\strut{}9 nm}}%
      \csname LT2\endcsname%
      \put(9641,1104){\makebox(0,0)[r]{\strut{}11 nm}}%
      \csname LT3\endcsname%
      \put(9641,948){\makebox(0,0)[r]{\strut{}13 nm}}%
      \csname LT4\endcsname%
      \put(9641,792){\makebox(0,0)[r]{\strut{}15 nm}}%
    }%
    \gplbacktext
    \put(0,0){\includegraphics{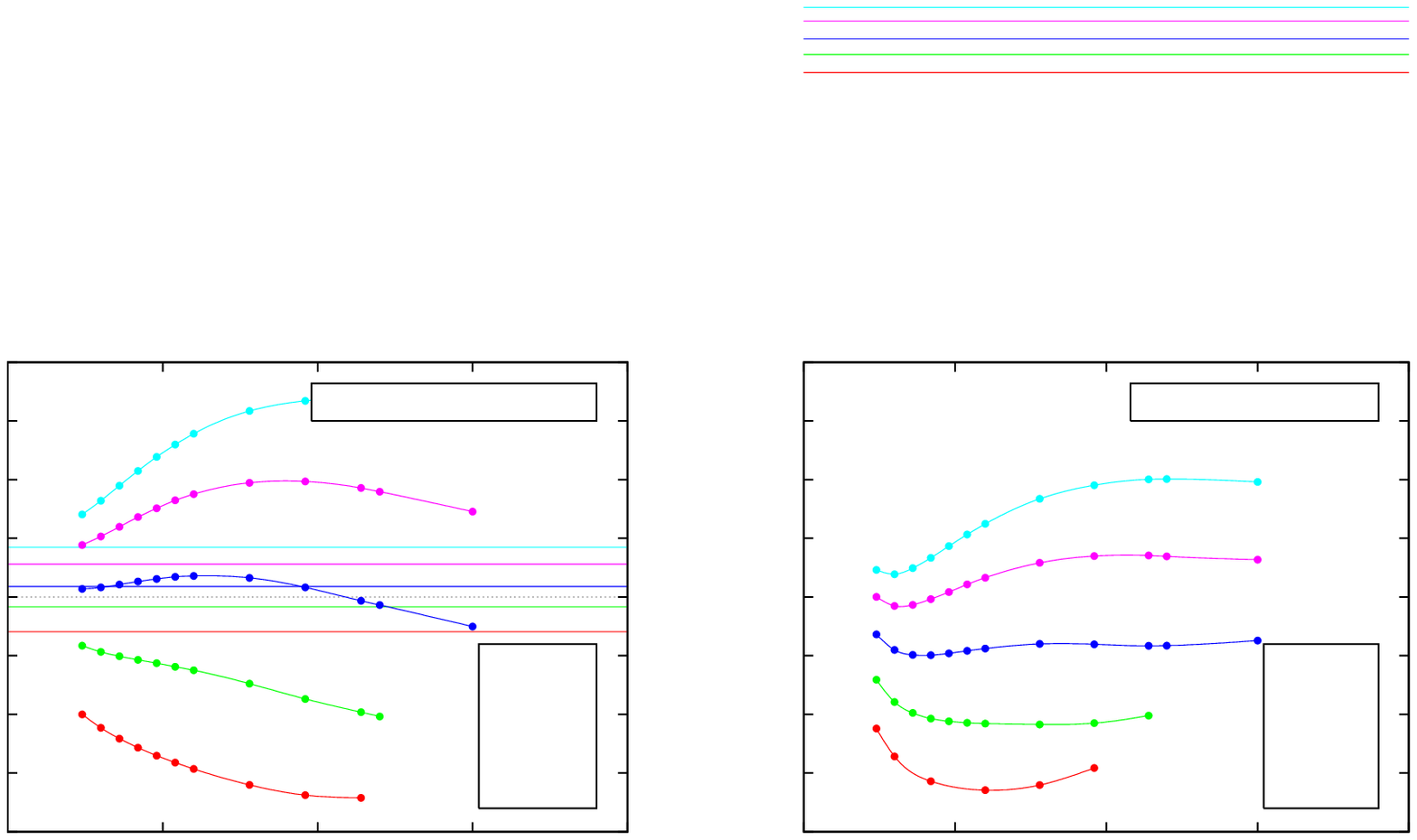}}%
    \gplfronttext
  \end{picture}%
\endgroup

\caption{(Color online) Height dependence of the zero-order hole $g_v$-factor of InAs/InP quantum dots (a) without and (b) with the $\kappa$-parameter included, respectively. Different colors indicate different radii of QDs. The horizontal lines are the $g_v^0$-factors of quantum wires of different radii.}
\label{fig:g_v_kappa}
\end{center}
\end{figure*}

\section{Remote band contributions} \label{app:remote}
A ${\bf k}\cdot{\bf p}$-calculation takes only a finite number of bands into account, and the influence of  remote bands can be incorporated via phenomenological parameters. These remote band parameters then do not change as the confinement experienced by an electron or hole in the dot changes. As explained in section \ref{sec:intro}, orbital angular momentum quenching strongly influences the $g$-factor and diamagnetic coefficient. The influence of remote bands, considered perturbatively in a ${\bf k}\cdot{\bf p}$-calculation, is not quenched with increasing confinement. The quantum dots considered in this work are in the strongly confined limit, so it is reasonable to assume that the contribution of the remote bands has been completely quenched. In the calculations presented above we have  left out the remote band parameters for the $g$-factors and used the Land{\'e}-factors in the Hamiltonian. This is certainly a very good approximation for the conduction band $g$-factor, since in InAs remote bands contribute only about $4\%$ to the bulk electron $g_c$-factor \cite{Winkler2003}. For holes the situation is more complicated, since the remote band contribution is non-negligible and one should consider including the phenomenological $\kappa$-parameter\cite{Luttinger1956} in a bulk calculation.

To view the effect of $\kappa$, the calculations are performed both with and without it [Figs.~ \ref{fig:g_v_kappa}(a, b)].  The dependence of the $g_v^0$-factor with height is not very different for both cases, however including the $\kappa$-parameter means that a large offset is introduced (the unquenched effect of these remote bands). This offset is very large compared to experimental results (see Sec.~\ref{sec:comparison}), and we therefore conclude that including the $\kappa$-parameter is not only poorly grounded theoretically, but also produces worse quantitative results than excluding it. For these dots  the dot size  is small enough to have quenched the influence of the remote bands sufficiently to exclude the remote bands from the calculations.

\section{Averaging methods} \label{app:average}
To avoid the complexity of fully  quantum mechanical calculations, one might consider using simpler averaging schemes to calculate the $g$-factor. The rationale behind such schemes is that an electron or hole experiences a spatially varying environment with spatially-dependent bulk-like properties. A property $\xi$ of this state can be calculated by weighing the bulk-like property $\xi_{\text{bulk}}\left({\bf r}\right)$ with the probability density:
\begin{eqnarray*}
\langle \xi \rangle = \int_V |\Psi\left({\bf r}\right)|^2 \xi_{\text{bulk}}\left({\bf r}\right)  d{\bf r}
\end{eqnarray*}
where $\Psi\left({\bf r}\right)$ is the wave function of the state. Such averaging schemes have no validity from a quantum-mechanical point of view; perturbing the environment locally will alter the whole quantum mechanical state and not merely the value of $\langle \xi \rangle$. Furthermore, there is not a unique choice for  $\xi_{\text{bulk}}\left({\bf r}\right)$. Some choices might give accurate results in particular cases, but there is not a single choice that would be generally valid. The minimal approach to calculate the properties of the electronic state properly for these QDs is  eight-band ${\bf k}\cdot{\bf p}$-theory. However, to stress the differences between the methods, we have calculated the electron $g$-factor using two different averaging schemes.

\begin{figure*}
\begin{center}
\begingroup
  \makeatletter
  \providecommand\color[2][]{%
    \GenericError{(gnuplot) \space\space\space\@spaces}{%
      Package color not loaded in conjunction with
      terminal option `colourtext'%
    }{See the gnuplot documentation for explanation.%
    }{Either use 'blacktext' in gnuplot or load the package
      color.sty in LaTeX.}%
    \renewcommand\color[2][]{}%
  }%
  \providecommand\includegraphics[2][]{%
    \GenericError{(gnuplot) \space\space\space\@spaces}{%
      Package graphicx or graphics not loaded%
    }{See the gnuplot documentation for explanation.%
    }{The gnuplot epslatex terminal needs graphicx.sty or graphics.sty.}%
    \renewcommand\includegraphics[2][]{}%
  }%
  \providecommand\rotatebox[2]{#2}%
  \@ifundefined{ifGPcolor}{%
    \newif\ifGPcolor
    \GPcolortrue
  }{}%
  \@ifundefined{ifGPblacktext}{%
    \newif\ifGPblacktext
    \GPblacktexttrue
  }{}%
  \let\gplgaddtomacro\g@addto@macro
  \gdef\gplbacktext{}%
  \gdef\gplfronttext{}%
  \makeatother
  \ifGPblacktext
    \def\colorrgb#1{}%
    \def\colorgray#1{}%
  \else
    \ifGPcolor
      \def\colorrgb#1{\color[rgb]{#1}}%
      \def\colorgray#1{\color[gray]{#1}}%
      \expandafter\def\csname LTw\endcsname{\color{white}}%
      \expandafter\def\csname LTb\endcsname{\color{black}}%
      \expandafter\def\csname LTa\endcsname{\color{black}}%
      \expandafter\def\csname LT0\endcsname{\color[rgb]{1,0,0}}%
      \expandafter\def\csname LT1\endcsname{\color[rgb]{0,1,0}}%
      \expandafter\def\csname LT2\endcsname{\color[rgb]{0,0,1}}%
      \expandafter\def\csname LT3\endcsname{\color[rgb]{1,0,1}}%
      \expandafter\def\csname LT4\endcsname{\color[rgb]{0,1,1}}%
      \expandafter\def\csname LT5\endcsname{\color[rgb]{1,1,0}}%
      \expandafter\def\csname LT6\endcsname{\color[rgb]{0,0,0}}%
      \expandafter\def\csname LT7\endcsname{\color[rgb]{1,0.3,0}}%
      \expandafter\def\csname LT8\endcsname{\color[rgb]{0.5,0.5,0.5}}%
    \else
      \def\colorrgb#1{\color{black}}%
      \def\colorgray#1{\color[gray]{#1}}%
      \expandafter\def\csname LTw\endcsname{\color{white}}%
      \expandafter\def\csname LTb\endcsname{\color{black}}%
      \expandafter\def\csname LTa\endcsname{\color{black}}%
      \expandafter\def\csname LT0\endcsname{\color{black}}%
      \expandafter\def\csname LT1\endcsname{\color{black}}%
      \expandafter\def\csname LT2\endcsname{\color{black}}%
      \expandafter\def\csname LT3\endcsname{\color{black}}%
      \expandafter\def\csname LT4\endcsname{\color{black}}%
      \expandafter\def\csname LT5\endcsname{\color{black}}%
      \expandafter\def\csname LT6\endcsname{\color{black}}%
      \expandafter\def\csname LT7\endcsname{\color{black}}%
      \expandafter\def\csname LT8\endcsname{\color{black}}%
    \fi
  \fi
  \setlength{\unitlength}{0.0500bp}%
  \begin{picture}(10368.00,3880.80)%
    \gplgaddtomacro\gplbacktext{%
      \csname LTb\endcsname%
      \put(688,720){\makebox(0,0)[r]{\strut{}-12}}%
      \put(688,1135){\makebox(0,0)[r]{\strut{}-10}}%
      \put(688,1551){\makebox(0,0)[r]{\strut{}-8}}%
      \put(688,1966){\makebox(0,0)[r]{\strut{}-6}}%
      \put(688,2382){\makebox(0,0)[r]{\strut{}-4}}%
      \put(688,2797){\makebox(0,0)[r]{\strut{}-2}}%
      \put(688,3213){\makebox(0,0)[r]{\strut{} 0}}%
      \put(688,3628){\makebox(0,0)[r]{\strut{} 2}}%
      \put(784,352){\makebox(0,0){\strut{} 0}}%
      \put(1358,352){\makebox(0,0){\strut{} 5}}%
      \put(1931,352){\makebox(0,0){\strut{} 10}}%
      \put(2505,352){\makebox(0,0){\strut{} 15}}%
      \put(3078,352){\makebox(0,0){\strut{} 20}}%
      \put(3652,352){\makebox(0,0){\strut{} 25}}%
      \put(4225,352){\makebox(0,0){\strut{} 30}}%
      \put(4799,352){\makebox(0,0){\strut{} 35}}%
      \csname LTb\endcsname%
      \put(224,2070){\rotatebox{-270}{\makebox(0,0){\strut{}g-factor $g_c$}}}%
      \put(2791,112){\makebox(0,0){\strut{}Height [nm]}}%
    }%
    \gplgaddtomacro\gplfronttext{%
      \csname LTb\endcsname%
      \put(4498,3348){\makebox(0,0)[r]{\strut{}Electron - unstrained (A)}}%
      \put(4217,1603){\makebox(0,0){\strut{}Radius}}%
      \csname LT0\endcsname%
      \put(4458,1416){\makebox(0,0)[r]{\strut{}7 nm}}%
      \csname LT1\endcsname%
      \put(4458,1260){\makebox(0,0)[r]{\strut{}9 nm}}%
      \csname LT2\endcsname%
      \put(4458,1104){\makebox(0,0)[r]{\strut{}11 nm}}%
      \csname LT3\endcsname%
      \put(4458,948){\makebox(0,0)[r]{\strut{}13 nm}}%
      \csname LT4\endcsname%
      \put(4458,792){\makebox(0,0)[r]{\strut{}15 nm}}%
    }%
    \gplgaddtomacro\gplbacktext{%
      \csname LTb\endcsname%
      \put(5872,720){\makebox(0,0)[r]{\strut{}-12}}%
      \put(5872,1135){\makebox(0,0)[r]{\strut{}-10}}%
      \put(5872,1551){\makebox(0,0)[r]{\strut{}-8}}%
      \put(5872,1966){\makebox(0,0)[r]{\strut{}-6}}%
      \put(5872,2382){\makebox(0,0)[r]{\strut{}-4}}%
      \put(5872,2797){\makebox(0,0)[r]{\strut{}-2}}%
      \put(5872,3213){\makebox(0,0)[r]{\strut{} 0}}%
      \put(5872,3628){\makebox(0,0)[r]{\strut{} 2}}%
      \put(5968,352){\makebox(0,0){\strut{} 0}}%
      \put(6526,352){\makebox(0,0){\strut{} 5}}%
      \put(7083,352){\makebox(0,0){\strut{} 10}}%
      \put(7641,352){\makebox(0,0){\strut{} 15}}%
      \put(8198,352){\makebox(0,0){\strut{} 20}}%
      \put(8756,352){\makebox(0,0){\strut{} 25}}%
      \put(9313,352){\makebox(0,0){\strut{} 30}}%
      \put(9871,352){\makebox(0,0){\strut{} 35}}%
      \csname LTb\endcsname%
      \put(5408,2070){\rotatebox{-270}{\makebox(0,0){\strut{}g-factor $g_c$}}}%
      \put(7975,112){\makebox(0,0){\strut{}Height [nm]}}%
    }%
    \gplgaddtomacro\gplfronttext{%
      \csname LTb\endcsname%
      \put(9681,3348){\makebox(0,0)[r]{\strut{}Electron - strained (B)}}%
      \put(9400,1603){\makebox(0,0){\strut{}Radius}}%
      \csname LT0\endcsname%
      \put(9641,1416){\makebox(0,0)[r]{\strut{}7 nm}}%
      \csname LT1\endcsname%
      \put(9641,1260){\makebox(0,0)[r]{\strut{}9 nm}}%
      \csname LT2\endcsname%
      \put(9641,1104){\makebox(0,0)[r]{\strut{}11 nm}}%
      \csname LT3\endcsname%
      \put(9641,948){\makebox(0,0)[r]{\strut{}13 nm}}%
      \csname LT4\endcsname%
      \put(9641,792){\makebox(0,0)[r]{\strut{}15 nm}}%
    }%
    \gplbacktext
    \put(0,0){\includegraphics{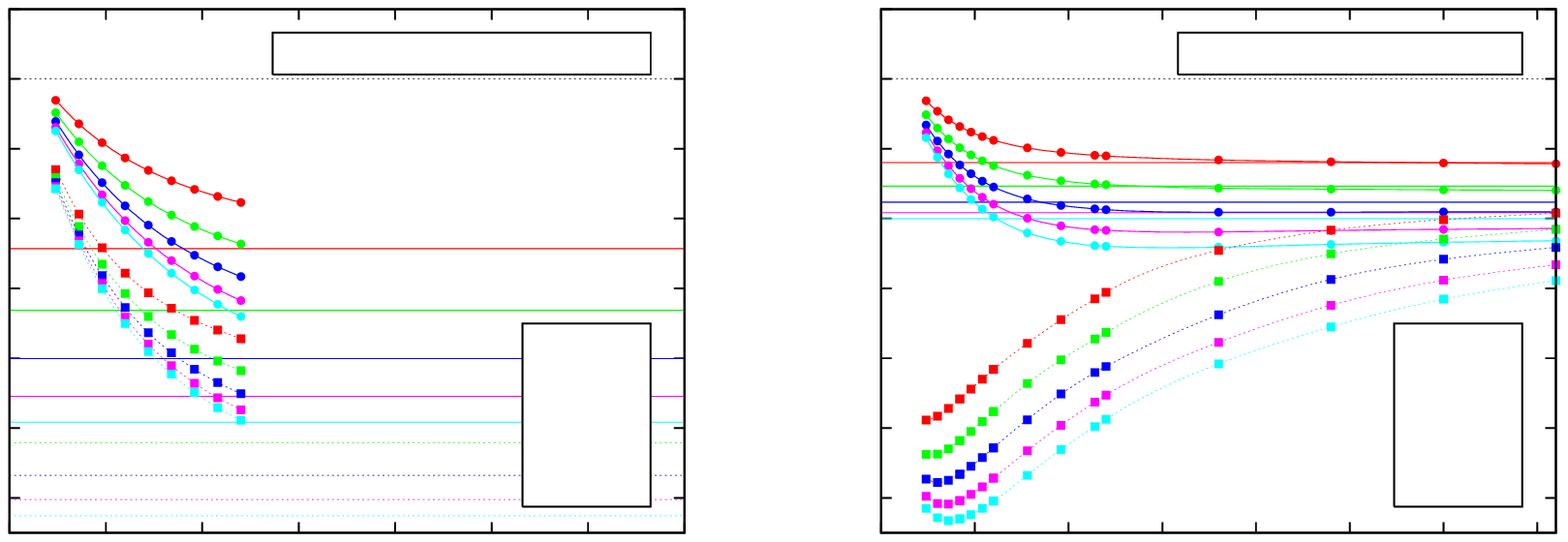}}%
    \gplfronttext
  \end{picture}%
\endgroup

\caption{(Color online) Height dependence of the electron $g_c$-factor  of (a) unstrained and (b) strained InAs/InP quantum dots. The circles and solid horizontal lines result from the eight band ${\bf k} \cdot {\bf p}$ model. The squares and dotted horizontal lines result from averaging models: (a) the adapted Roth formula is used [Eq.~(\ref{eq:Roth})], (b) the average effective mass is used [Eqs.~(\ref{eq:strain})]. Different colors indicate different radii of QDs. The horizontal lines are the electron $g_c$-factors of quantum wires of different radii.}
\label{fig:avg_c}
\end{center}
\end{figure*}

For the unstrained electron $g$-factors case, we have used an adapted Roth-formula\cite{Roth1959}:
\begin{eqnarray}
g_{\text{Roth}}^{\text{X}} = 2 - \frac{2E_p^{\text{X}}\Delta^{\text{X}}}{3E_{\text{gap}}\left(E_{\text{gap}}+\Delta^{\text{X}}\right)} \label{eq:Roth}
\end{eqnarray}
where $E_p^{\text{X}}$ is the Kane energy and $\Delta^{\text{X}}$ is the spin-orbit coupling of material $\text{X}$, and $E_{\text{gap}}$ the energy gap between the ground electron and hole state. The use of $E_{\text{gap}}$ takes some effects of confinement into account. Penetration into the barrier material is taken into account by weighting with the probability density $|\Psi\left({\bf r}\right)|^2$:
\begin{eqnarray*}
\langle g_{\text{Roth}} \rangle &=& \int_V |\Psi\left({\bf r}\right)|^2 g_{\text{Roth}}^{\text{X}} \left({\bf r}\right)d{\bf r}, \\
&=& w_{\text{QD}} g_{\text{Roth}}^{\text{InAs}} + w_{\text{B}} g_{\text{Roth}}^{\text{InP}},
\end{eqnarray*}
where $w_{\text{QD,B}}$ and $g_{\text{Roth}}^{\text{In(As,P)}}$ are the fraction of the probability density and the Roth $g$-factor inside the InAs QD and in the InP barrier material, respectively. Note that this information is taken from the ${\bf k}\cdot{\bf p}$ wave functions at zero magnetic field. The $g_c$-factors obtained via this method are indicated by the squares and dotted lines in Fig.~ \ref{fig:avg_c}(a). The trend from this averaging method is similar to the one from the ${\bf k} \cdot {\bf p}$-calculation. However, it underestimates the value of the $g_c$-factor, since orbital angular momentum quenching is not taken into account. This is even more clearly observed from the discrepancy with the quantum wire limits; the lack of orbital angular momentum quenching causes the quantum wire limits to be much closer to the bulk InAs electron $g$-factor of $\sim-14.4$.

To show the necessity of including confinement, the strained $g_c$-factor case has also been calculated using a different averaging scheme. To fully incorporate the effects of the strain $\epsilon\left({\bf r}\right)$, one needs to include band mixing. The valence bands are mixed due to strain, leading to a mixing of the split-off and light-hole states that is linear in the strain\cite{Hendorfer1991}:
\begin{eqnarray*}
|\text{A}\rangle &=& \left| \frac{3}{2}, \pm \frac{3}{2} \right>, \nonumber \\
|\text{B}\rangle &=& \left| \frac{3}{2}, \pm \frac{1}{2} \right> + \frac{\alpha_0\left(\epsilon\right)}{\sqrt{2}} \left| \frac{1}{2}, \pm \frac{1}{2} \right>,\nonumber \\
|\text{C}\rangle &=& \left| \frac{1}{2}, \pm \frac{1}{2} \right> - \frac{\alpha_0\left(\epsilon\right)}{\sqrt{2}} \left| \frac{3}{2}, \pm \frac{1}{2} \right>. \nonumber
\end{eqnarray*}
The parameter $\alpha_0\left(\epsilon\right)$ determines the mixture. The $g_c$-factor for a strained bulk crystal is then, to first order in strain \cite{Hendorfer1991},
\begin{widetext}
\begin{eqnarray}
g_{\text{strained}}^{\text{X}}\left(\epsilon\right) = 2-\frac{2 E_p^{\text{X}}}{3}\bigg\{\frac{3}{2\left[E_{\text{CB}}\left(\epsilon\right)-E_{\text{A}}\left(\epsilon\right)\right]}-\frac{1}{2\left[E_{\text{CB}}\left(\epsilon\right)-E_{\text{B}}\left(\epsilon\right)\right]}-\frac{1}{E_{\text{CB}}\left(\epsilon\right)-E_{\text{C}}\left(\epsilon\right)} \label{eq:strain} \\ \nonumber
-\alpha_0\left(\epsilon\right)\left(\frac{1}{E_{\text{CB}}\left(\epsilon\right)-E_{\text{B}}\left(\epsilon\right)}-\frac{1}{E_{\text{CB}}\left(\epsilon\right)-E_{\text{C}}\left(\epsilon\right)}\right)\bigg\}
\end{eqnarray}
\end{widetext}
where $E_{\text{CB}}$, $E_{\text{A}}$, $E_{\text{B}}$, and $E_{\text{C}}$ are respectively the conduction band (CB), the heavy-hole like band (A), the light-hole like band (B), and the split-off like band (C) edge energies. These energies and the mixing parameter $\alpha_0\left(\epsilon\right)$ depend on strain. The $g_c$-factor can then be determined by weighting the locally varying $g_{\text{strained}}^{\text{X}}\left(\epsilon\left({\bf r}\right)\right)$ by the probability density:
\begin{eqnarray*}
\langle g_{\text{strained}} \rangle = \int_V |\Psi\left({\bf r}\right)|^2 g_{\text{strained}}\left(\epsilon\left({\bf r}\right)\right)  d{\bf r}.
\end{eqnarray*}
The resulting calculated $g_c$-factor is indicated in Fig.~ \ref{fig:avg_c}(b) by the squares and dotted lines. The trend is opposite to the ${\bf k}\cdot{\bf p}$-calculation, mainly due to the absence of confinement in this averaging scheme. This  shows that confinement is essential in determining the $g$-factor.

\end{document}